\UseRawInputEncoding

\documentclass[aps,prc,twocolumn,groupedaddress,floatfix,showpacs,nofootinbib]{revtex4-1}

\usepackage{graphicx}
\usepackage{array}
\usepackage{longtable}
\usepackage{braket}
\usepackage{amsmath}
\usepackage{subfigure}

\usepackage{hyperref}
\hypersetup{
colorlinks=true,
urlcolor= blue,
citecolor=blue,
linkcolor= blue,
bookmarks=true,
bookmarksopen=false,
}

\begin{document}

\title{Low-spin particle/hole-core excitations in  $^{41,47,49}$Ca isotopes \\ studied by cold-neutron capture reactions}

\author{
S.~Bottoni$^{1,2}$, N.~Cieplicka-Ory\'nczak$^{3}$, S.Leoni$^{1,2}$, B. Fornal$^{3}$, G.Col\`o$^{1,2}$, 
P.F. Bortignon$^{1,2}$, 
G.~Bocchi$^{1,2}$, D. Bazzacco$^{4}$, G. Benzoni$^{2}$, A. Blanc$^{5}$, A. Bracco$^{1,2}$, S. Ceruti$^{1,2}$, F~.C~.L.~Crespi$^{1,2}$, G.~de~ France$^{6}$, E.~R.~Gamba$^{7,2}$, \L.W. Iskra$^{2,3}$, M. Jentschel$^{5}$, U. K\"oster$^{5}$, C. Michelagnoli$^{5}$, B.~Million$^{2}$, D.~Mengoni$^{8,4}$, P.~Mutti$^{5}$, Y. Niu$^{9}$, C. Porzio$^{1,2}$, G. Simpson$^{5}$, T. Soldner$^{5}$, B. Szpak$^{3}$, A. T\"urler$^{10}$, C.A. Ur$^{11}$, W. Urban$^{12}$  
}

\affiliation{$^{1}$Dipartimento di Fisica, Universit\`a degli Studi di Milano, 20133 Milano, Italy}
\affiliation{$^{2}$INFN Sezione di Milano, 20133, Milano, Italy }
\affiliation{$^{3}$Institute of Nuclear Physics, PAN, 31-342 Krak\'ow, Poland}
\affiliation{$^{4}$INFN Sezione di Padova, 35131 Padova, Italy}
\affiliation{$^{5}$Institut Laue-Langevin, 38042 Grenoble CEDEX 9, France}
\affiliation{$^{6}$GANIL, BP 55027, 14076 Caen CEDEX 5, France}
\affiliation{$^{7}$Museo Storico della Fisica e Centro di Studi e Ricerche Enrico Fermi, 00184 Roma, Italy}
\affiliation{$^{8}$Dipartimento di Fisica e Astronomia, Università degli Studi di Padova, 35131 Padova, Italy}
\affiliation{$^{9}$School of Nuclear Science and Technology, Lanzhou University, Lanzhou 730000, China}
\affiliation{$^{10}$Universit\"at Bern and Paul Scherrer Institut, Villigen, Switzerland}
\affiliation{$^{11}$ELI-NP, Magurele-Bucharest, Romania}
\affiliation{$^{12}$Faculty of Physics, Warsaw University, 00-681 Warsaw, Poland}

%\email[]{}

%\date{\today}

\begin{abstract}

We present recent results on the structure of the one-valence-particle $^{41}$Ca and $^{49}$Ca, and one-valence-hole $^{47}$Ca, nuclei. The isotopes of interest were populated via the cold-neutron capture reactions $^{40}$Ca(n,$\gamma$), $^{48}$Ca(n,$\gamma$) and $^{46}$Ca(n,$\gamma$), respectively. The experiments were performed at the Institut Laue-Langevin, within the EXILL campaign, which employed a large array of HPGe detectors. The $\gamma$ decay and level schemes of these nuclei were investigated by $\gamma$-ray coincidence relationships, leading to the identification of 41, 10, and 6 new transitions in  $^{41}$Ca, $^{47}$Ca, and $^{49}$Ca, respectively. Branching ratios and intensities were extracted for the $\gamma$ decay from each state, and $\gamma$-ray angular correlations were performed to establish a number of transition multipolarities and mixing ratios, thus helping in the spin assignment of the states. The experimental findings are discussed along with microscopic, self-consistent beyond-mean-field calculations performed with the Hybrid Configuration Mixing model, based on a Skyrme  SkX Hamiltonian. The latter suggests that a fraction of the low-spin states of the $^{41}$Ca, $^{49}$Ca, and $^{47}$Ca nuclei is characterized by the coexistence of either 2p-1h and 1p-2h excitations, or couplings between single-particle/hole degrees of freedom and collective vibrations (phonons) of the doubly-magic ``core". 

\end{abstract}


\maketitle

\section{INTRODUCTION}
\label{intro}
The structure of calcium isotopes between the doubly-magic $^{40}$Ca (\textit{N}=20) and $^{48}$Ca (\textit{N}=28) nuclei has been the subject of many experimental studies over the past decades~\cite{Yas10,Lui11,Mon11plb,Mon11,Mon12,Gar15,Noj15,Gad16,Had16,Ril16,Cra17,Tal18,Had18}. With six stable isotopes, calcium plays a crucial role in stellar nucleosynthesis~\cite{Bod68,Tru72,Kap84}. The formation of Ca isotopes involves several astrophysical processes, such as silicon- and oxygen-burning~\cite{Woo73}, as well as s- and r-processes~\cite{Kap82,Cam79}, which generate, for example, the heaviest, symmetric \textit{N=Z} stable nucleus, i.e. $^{40}$Ca, and the lightest stable doubly-magic neutron-rich system, namely $^{48}$Ca, in the nuclide chart. Moreover, the \textit{Z}=20 isotopic chain contains a rare cosmogenic radioactive nucleus, i.e. $^{41}$Ca, produced by neutron-capture reactions on $^{40}$Ca induced by cosmic rays~\cite{Kub86}. \\
\noindent
In this context, nuclear structure studies along Ca isotopes are crucial to understand, for instance, the evolution of single-particle states and collectivity from symmetric to neutron-rich systems, which are properties significantly affecting the reaction rates in stellar environments. Moreover, new experimental results may serve as a benchmark for the most advanced theoretical models, such as state-of-the-art shell model calculations~\cite{Hon02,Hon05,Uts12,Tsu13} and \textit{ab initio} approaches, employing chiral two- and three-nucleon interactions~\cite{Heb11,Hol12,Hol14,Sim16}.  \\
\noindent
The low-lying structure of $^{40}$Ca is characterized by a 0$^+$ state at 3.4 MeV,	 as a first excited state - a clear signature of a robust double shell closure in this nucleus - and a very collective octupole, 3$^{-}$ vibration at 3.7 MeV, with a B(E3) of $\approx$ 30 W.u.~\cite{Kib02}. Moreover, in the spin range 2$\hbar$-8$\hbar$, deformed and superdeformed bands have been observed and associated to 4p-4h and 8p-8h excitations, respectively~\cite{Ide01}. These features are gradually lost in mid-shell Ca nuclei, where deformed structures take over spherical ones already at low energies, owing to neutron p-h excitations across the \textit{pfg}  energy gap. This scenario changes again in $^{48}$Ca, where the presence of a low-lying 0$^+$  state at 4.3 MeV and a 3$^{-}$ phonon with  B(E3) $\approx$ 7~W.u.~\cite{Bur06} suggests the restoration of the spherical symmetry, although weaker than in $^{40}$Ca.\\
\noindent
In this framework, Ca nuclei one-particle or one-hole away from double shell closures are of particular interest. These isotopes are ideal to investigate the interplay between fermionic and bosonic degrees of freedom, as it occurs in the coexistence and competition between pure p-h excitations and the so-called particle/hole-vibration coupling~\cite{BM}.  As a matter of fact, the low-lying structure of one-valence-particle/hole nuclei is strongly influenced by the collective phonons of the underlying ``core". On the other hand, core excitations are perturbed and damped by the single particle/hole motion and non-collective p-h excitations~\cite{Ham69,Ham74}. Therefore,  a comprehensive investigation of these mechanisms, moving along the Ca isotopic chain, may significantly advance our understanding of the emergence of complex phenomena, such as the quenching of spectroscopic factors and the anarmonicity of vibrational spectra in this mass region. \\
\noindent
In this paper, we present new experimental results in $^{41}$Ca, $^{47}$Ca, and $^{49}$Ca, populated via cold neutron-capture reactions and studied by $\gamma$-ray spectroscopy. Neutron-capture reactions induced by cold and thermal neutrons populate the corresponding\textit{ N+1} systems at the neutron separation energy S$_{n}$. The spin of the capture level depends on the ground-state spin \textit{J} of the target nucleus and can only be  \textit{J} $\pm$ 1/2,  being 1/2 the spin of the neutron. As a consequence, neutron-induced reactions on even-even nuclei always proceed through  a single 1/2$^+$ neutron-capture state. The $\gamma$-ray decay is typically dominated by high-energy, E1 primary transitions, which preferentially populate 1/2$^{-}$ and 3/2$^{-}$ states (based on $\gamma$-decay selection rules), followed by secondary electromagnetic radiation of different character and multipolarity. \\
\noindent
In this context, it is clear that the combined use of neutron-capture reactions and detectors with high energy resolution (e.g. HPGe crystals) enables to perform an almost-complete $\gamma$-ray spectroscopy from the neutron binding energy to the ground state, providing an exhaustive picture of the low-spin structure of the nuclei of interest. The present experimental results on the $^{41}$Ca, $^{47}$Ca, and $^{49}$Ca nuclei will be discussed  in the framework of the Hybrid Configuration Mixing Model~\cite{Col17,Bot19}, with particular attention to the interplay between single-particle/hole states and couplings with core excitations.\\
\noindent
The paper is organized as follows: in Sec.~\ref{exp} the experimental details will be presented along with the different reactions performed; in Sec.~\ref{Analysis} the analysis of the data will be discussed, while in Sec.\ref{hybrid} the experimental results will be outlined in connection with theoretical interpretations.    

\section{THE EXPERIMENT}
\label{exp}
\begin{figure}[t!]
 	\includegraphics[width=1\columnwidth]{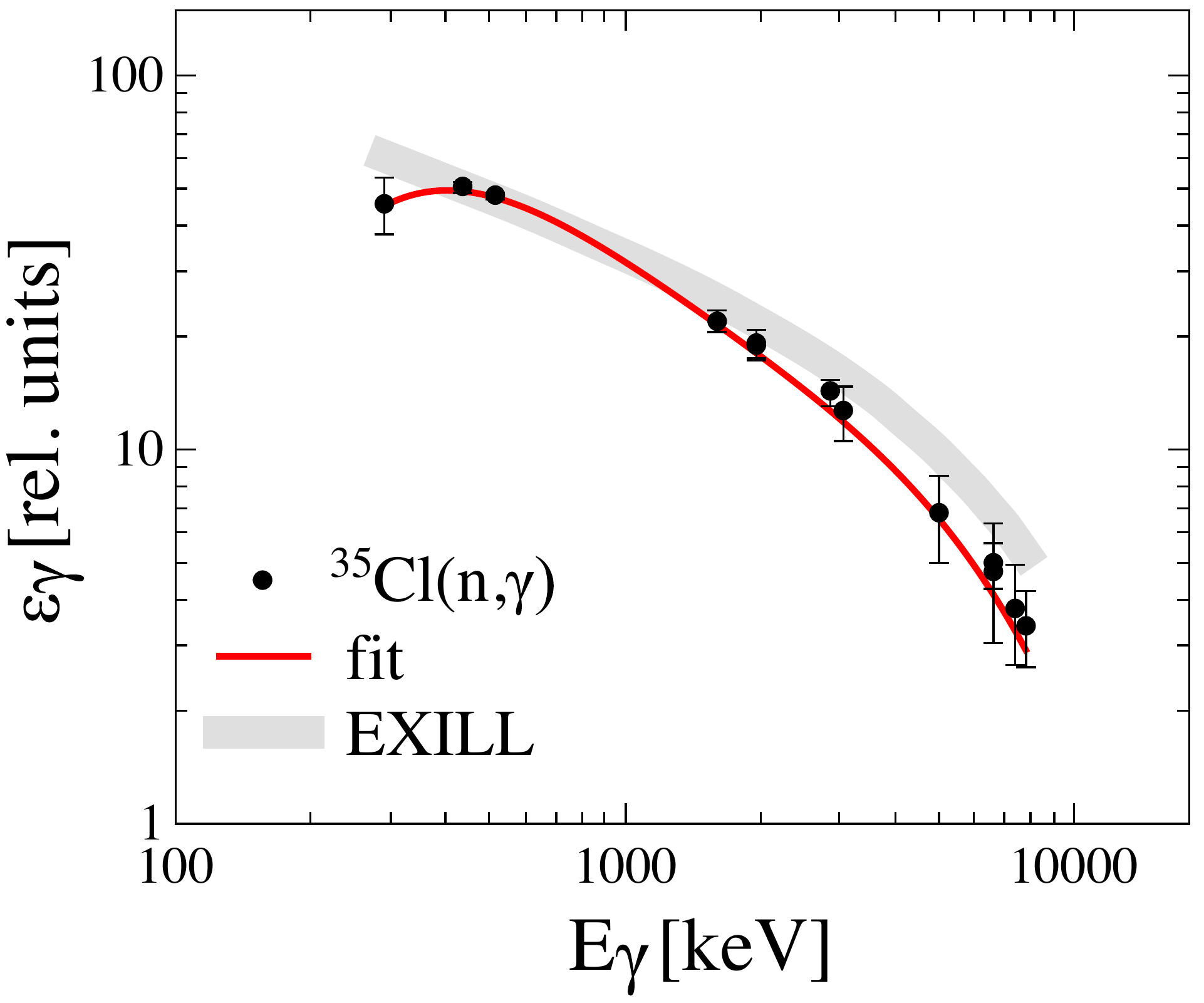}
  		\caption{(Color online) Relative $\gamma$-ray efficiency of the HPGe array used in the $^{48}$Ca(n,$\gamma$) experiment. Experimental data obtained from the $^{35}$Cl(n,$\gamma$) reaction are displayed along with the fit function (red). The efficiency of the full EXILL setup is also reported~\cite{Jen17} (see text for details).}  
 	\label{figeff}
\end{figure}
The experiments were performed at Institut Lau-Langevin (ILL) in Grenoble, within the EXILL experimental campaign~\cite{Jen17}. Neutron-capture reactions were studied at the High Flux Reactor of ILL~\cite{Mol74}, which delivers the most intense, continuous neutron beams worldwide for scientific research. \\
\noindent
In the present measurement, a high-efficiency, high-resolution  composite HPGe array was installed at the PF1B cold-neutron beam line~\cite{Abe06}, where the neutron flux was about 10$^{8}$ neutrons cm$^{-1}$ s$^{-1}$, after collimation.\\
 \noindent
The array comprised 8 clover detectors from the EXOGAM setup~\cite{Sim00}, 6 coaxial GASP detectors~\cite{Alv93} and 2 ILL clover detectors, providing a total photopeak efficiency of $\approx$ 6\% at 1.3 MeV. Apart from the ILL clovers, all the other HPGe detectors were equipped with BGO anti-Compton shields for background suppression. In the case of the $^{48}$Ca(n,$\gamma$) experiment, the GASP and ILL detectors were replaced by 16 LaBr$_3$:Ce fast scintillators from the FATIMA collaboration~\cite{Rob14} for lifetime measurements by using fast-timing techniques~\cite{Reg13}. As a consequence, the $\gamma$-ray efficiency of this HPGe detector configuration, comprising the EXOGAM clovers only, differs from the one of the full EXILL setup. In Fig.\ref{figeff}, the relative efficiency of the HPGe array used in the $^{48}$Ca(n,$\gamma$) experiment is reported, arbitrary normalized at $\approx$ 400 keV to the efficiency curve of the full EXILL configuration~\cite{Jen17}. The experimental efficiency values were determined up to 8 MeV using known $\gamma$-ray transitions in the $^{36}$Cl nucleus, populated in the  $^{35}$Cl(n,$\gamma$) reaction. At high energy, the deviation of the present efficiency from the EXILL curve  is consistent with the reduced number of HPGe detectors in the setup.\\
\noindent
The compact geometry of the 8 EXOGAM clovers, mounted in a symmetric, ring configuration around the scattering chamber, was used to study $\gamma$-ray angular correlations, with the aim of determining the multipolarity of the detected radiation, thus constraining the spin and parity of the observed states. All the possible angular combinations between crystals (11 angles from 0$^{\circ}$ to 90$^{\circ}$) were grouped into 3 angles only, i.e. 0$^{\circ}$, 45$^{\circ}$, and 90$^{\circ}$, corresponding to the angles between clover detectors with respect to the target position. This enabled to increase the statistics for $\gamma$-$\gamma$ coincidences, allowing us to perform angular-correlation studies also in cases of weak transitions. Experimental data were fitted by using the analytic function~\cite{Mor76}
\begin{equation}
 W(\theta)=1+a_{22}q_2P_2(\cos\theta)+a_{44}q_4P_4(\cos\theta),
 \end{equation}
\noindent 
where $a_{ii}$ are the multipole expansion coefficients, P$_i$(cos$\theta$) the Legendre Polynomials, and q$_{i}$ the attenuation parameters which take into account the finite size of the detectors. The latter were determined by studying $\gamma$-ray angular correlations of known transitions of the $^{152}$Eu $\gamma$-ray source, and were found to be q$_{1}$=0.87 and q$_{2}$=0.6.\\
\noindent
The $^{41}$Ca and $^{47}$Ca nuclei were populated by (n,$\gamma$) reactions on an enriched target. For this purpose, a 40.6-mg Ca(NO$_3$)$_2$ compound, enriched to 31.7\% in $^{46}$Ca, was prepared at Paul Scherrer Institute, Switzerland. The nitrate solution was directly dried in a 25 $\mu$m thin FEP (fluorinated ethylene propylene) bag, which has negligible neutron-capture cross section.  It is important to note that $^{46}$Ca is the isotope with the second-lowest natural relative isotopic abundance (only 0.004\%), after ${^3}$He. A large fraction of the target (60.5\%) was composed by $^{40}$Ca, with the latter being the most abundant Z=20 isotope. This allowed us to perform $^{46}$Ca(n,$\gamma$) and $^{40}$Ca(n,$\gamma$) reactions at the same time. On the other hand, the $^{49}$Ca nucleus was populated by neutron-capture reactions on a 350-mg CaCO$_3$ compound target, enriched to 60.5\% in $^{48}$Ca. Also in this case, traces of other Ca isotopes were present in the sample. \\
\noindent
The composition of the  targets used in the current experiments, along with the (n,$\gamma$) cross sections for thermal neutrons~\cite{Mug18}, are reported in Tab.~\ref{Tab:targets}. The isotopes of interest for this work and the corresponding targets are highlighted in bold.
\begin{table}[h!]
\centering
\caption{Isotopic composition of the Ca(NO$_3$)$_2$ and CaCO$_3$ compounds used in the current experiment and corresponding (n,$\gamma$) cross sections~\cite{Mug18}. The (n,$\gamma$) columns show the percentage of capture reactions for a given isotope. The nuclei of interest for the the present work are marked in bold. Details of the FEP bag are also reported (see text for details).}
\begin{tabular}{cccccccc}
\hline
\hline
&  $\sigma$(n,$\gamma$) && Ca(NO$_3$)$_2$& & CaCO$_3$  \\
&[barn]  && [atoms \%] & (n,$\gamma$) & [atoms \%] &(n,$\gamma$)\\
\hline
\hline
\underline{TARGET} &&&&&&\\
\textbf{$^{40}$Ca} & \textbf{0.41} && \textbf{60.5 }& \textbf{34\%} & 27.9 & 13\%\\

$^{42}$Ca & 0.68 && 0.63 & 1\% & 0.3 & 0\%\\

$^{43}$Ca & 6.2 & & 0.15 &1\%&0.1&1\%\\

$^{44}$Ca & 0.88 & &5.35 &6\%&2.5&2\%\\

\textbf{$^{46}$Ca} & \textbf{0.74}&& \textbf{31.7}& \textbf{32\%}&0.1&0\%  \\

\textbf{$^{48}$Ca }& \textbf{1.09} && 1.57 & 2\% &\textbf{69.2} & \textbf{83\%}\\

C & 3.84$\cdot 10^{-3}$ && &0\%&100&0\%\\

N & 7.47$\cdot 10^{-2}$ && 200 &20\%& &0\%\\

O & 2.24$\cdot 10^{-4}$ && 600 &0\%& 300&0\%\\
\hline
\underline{FEP} &&&&&&\\
C & 3.84$\cdot 10^{-3}$ && 89 &0\%&6&0\%\\
F & 9.51$\cdot 10^{-3}$ && 178 &2\%&13 &0\%\\
\hline
\hline
\end{tabular}
\label{Tab:targets}      
\end{table}
\noindent

\section{DATA ANALYSIS}
\label{Analysis}        
Data were acquired using fast, digital electronics in triggerless mode and the analysis was performed by considering coincident events, built within a 200-ns, prompt time window.  The good energy resolution and efficiency of the HPGe array turned out to be essential to observe, with high accuracy, very weak $\gamma$-ray decay paths, whereas the BGO Compton shields significantly suppressed  $\gamma$-ray coincidences with Compton-scattered radiation. \\
\noindent  
The level schemes and the $\gamma$-ray decays of the $^{41}$Ca, $^{47}$Ca, and $^{49}$Ca nuclei were studied by using $\gamma$-$\gamma$ and triple-$\gamma$ coincidence techniques. At first, very selective gates on primary, high-energy $\gamma$ transitions were used to identify secondary $\gamma$-ray cascades and to locate new low-lying states. The energies of the latter were determined by correcting the measured $\gamma$ energies by the recoil energy of the nucleus. This is particularly crucial for high-energy transitions, considering the relatively light mass of the isotopes studied in this work.  As a second step, gates on $\gamma$ transitions depopulating low-lying states enabled to determine new decay paths from the neutron-capture level and to measure precisely the value of the S$_{n}$ neutron separation energy for all three nuclei. The latter were obtained by considering all the possible combinations of $\gamma$ rays decaying directly from the neutron-capture level. The S$_{n}$ values obtained in this work are presented in Tab~\ref{tab:sn}, along with those ones reported in the literature~\cite{Gru67,Cra70,Arn69,Nes16,Bur07,Bur08}.
\begin{table}[h!]
\centering
\caption{Neutron separation energies (S$_n$) for the $^{41}$Ca, $^{49}$Ca, and $^{47}$Ca nuclei obtained in this work, compared with values reported in the literature~\cite{Gru67,Cra70,Arn69,Nes16,Bur07,Bur08}.}
\begin{tabular}{ccc}
\hline
\hline
Isotope & S$_{n}$ [keV] &  S$_{n}$ [keV] \\
& (this work) & (literature)\\
\hline
\hline
\textbf{$^{41}$Ca} & 8362.4(2) & 8362.8(2) \\
\\
\textbf{$^{47}$Ca} & 7275.4(2) & 7276.4(3) \\
\\
\textbf{$^{49}$Ca} & 5146.3(3) &5146.5(2) \\
\hline
\hline
\end{tabular}
\label{tab:sn}      
\end{table}

\begin{figure*}
\includegraphics[width=2\columnwidth]{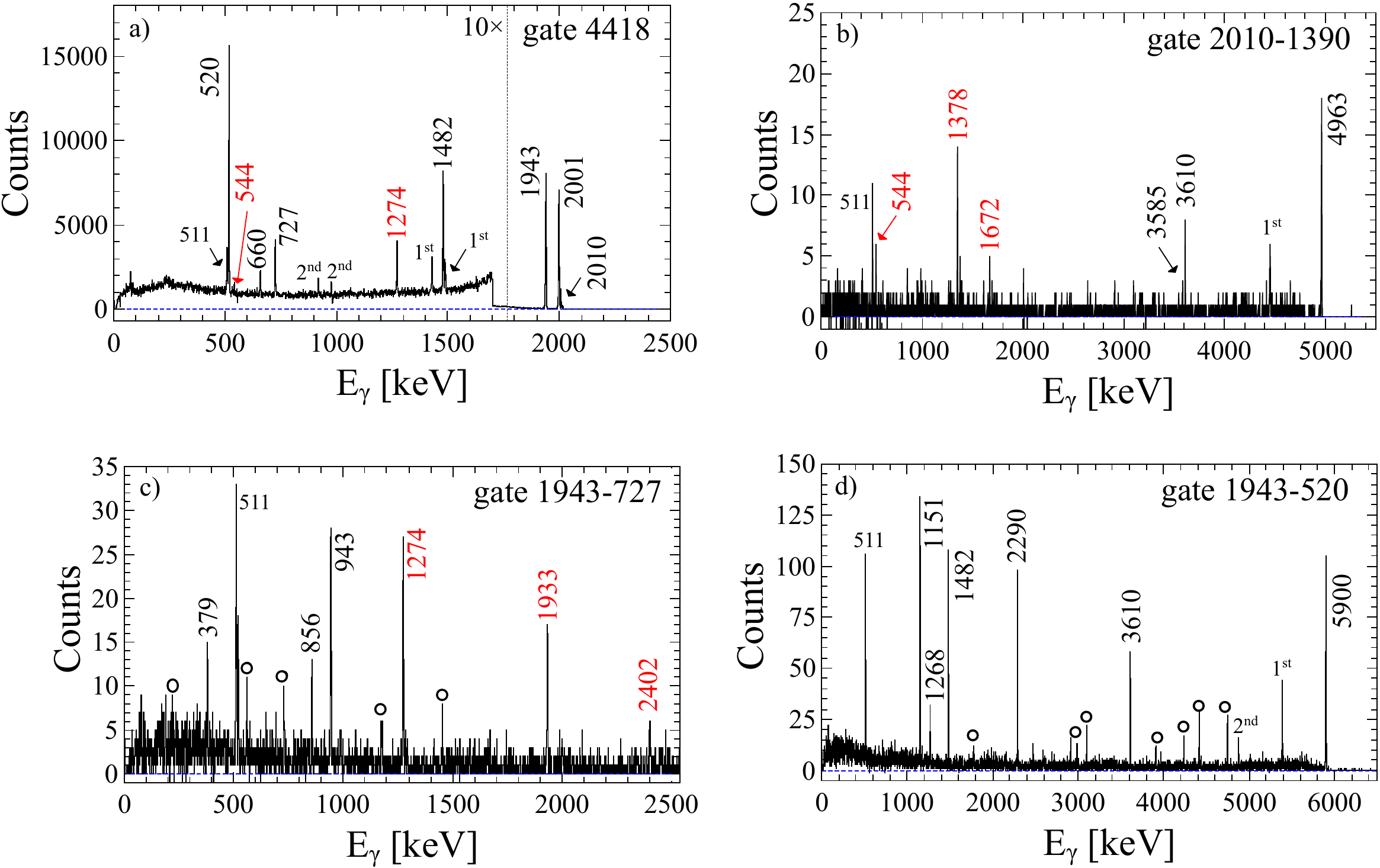}
  		\caption{(Color online) Projection of the $\gamma$-$\gamma$ coincidence matrix measured in the $^{40}$Ca(n,$\gamma$) reaction a), gated on the 4418-keV transition of the $^{41}$Ca nucleus. Projections of the triple-$\gamma$ coincidence matrix, gated on the 2010- and 1390-keV b), 1943- and 727-keV c), and 1943- and 520-keV d) transitions. New $\gamma$ rays, observed for the first time, are marked in red, while those already known in the literature~\citep{Gru67,Cra70,Arn69,Nes16,PGAA} are marked in black. Transitions associated to (n,$\gamma$) reactions on contaminants present in the target are labelled by circles. First and second escape peaks for high-energy transitions are marked by 1$^{\text{st}}$ and 2$^{\text{nd}}$, respectively. }  
 \label{fig:spectra_41Ca}
\end{figure*}

\noindent
The $\gamma$-ray intensities and branching ratios for each level were evaluated using $\gamma$-$\gamma$ matrices, constructed considering all HPGe detectors in the array. Gates were set on transitions feeding the level of interest and relative intensities of deexciting transitions, with respect to a given $\gamma$ ray in the level scheme, were extracted, taking into account efficiency corrections. Branching ratios were determined by taking the ratio of the intensity of a $\gamma$ transition to the summed intensity of all transitions deexciting a given state. Concerning primary $\gamma$ rays, all possible decay paths for each transition were considered, based on the analysis of the $\gamma$-$\gamma$ coincidence matrix, and the intensity balance was used to extract $\gamma$-ray intensities and branching ratios from the neutron-capture state. Systematic errors associated to the partial angular coverage of the detectors (i.e., $\approx$ 90\%) have been taken into account, by including a conservative 1\% systematic error on the measured $\gamma$-ray intensities. Uncertainties originating from efficiency correction (see Fig.~\ref{figeff}) were also considered. Possible uncertainties coming from self-absorption of the targets and summing effects were not included, being negligible when compared to other sources of error. In particular, the former can be excluded due to the small samples used in the current experiment (see Sec.~\ref{exp}), and the latter was estimated to be lower than 10$^{-4}$.

\subsection{$^{41}$Ca }

\label{41Ca}
\begin{center}
\begin{figure*}[htpb]
 	\includegraphics[width=2\columnwidth]{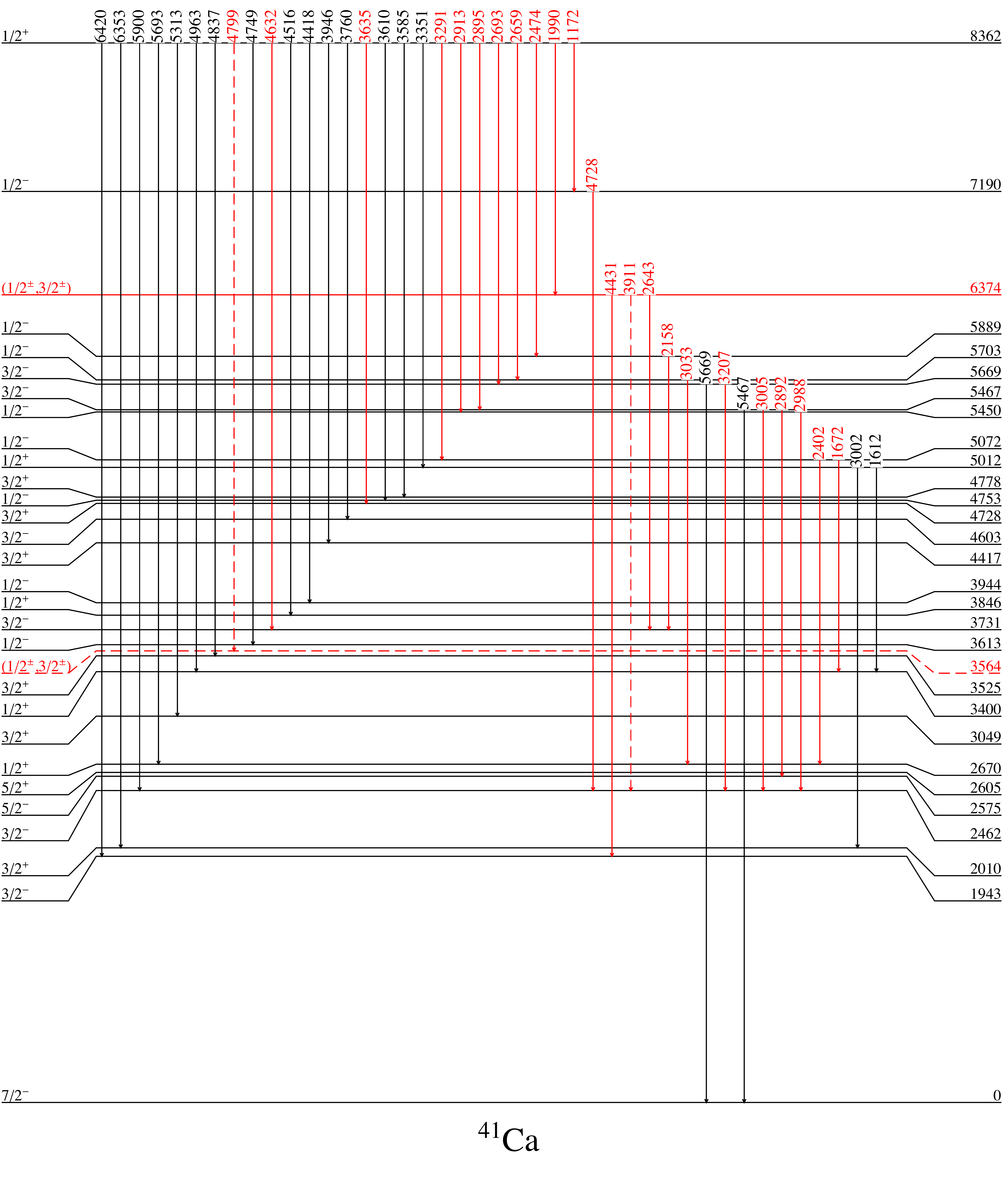}
  		\caption{(Color online) Level scheme of $^{41}$Ca, as measured in the current experiment. Newly-observed $\gamma$-ray transitions and levels are reported in red. Levels with tentative spin assignment are marked by dashed lines. Dashed arrows indicate $\gamma$ rays with no firm placement in the level scheme (i.e., the 4799- and 1622-keV cascade) or very weakly observed (see text for details).}  
 	\label{fig:41Ca}
\end{figure*}
\setcounter{figure}{2}    
\begin{figure*}[htbp]
 	\includegraphics[width=2\columnwidth]{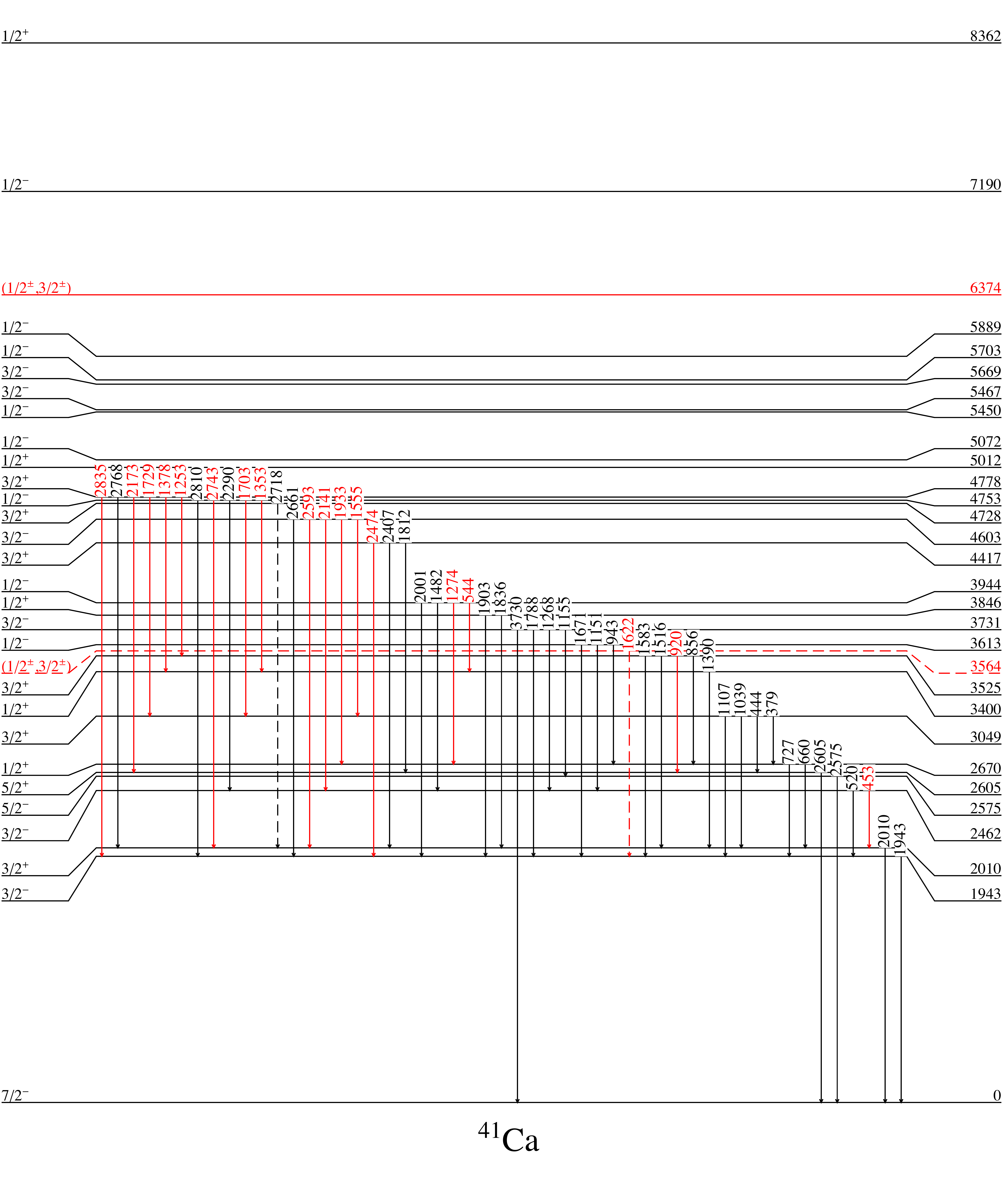}
  		\caption{-\textit{continued} }  
 	\label{fig:41Ca}
\end{figure*}
\end{center}

The level scheme of the $^{41}$Ca nucleus was built by using both $\gamma$-$\gamma$ and $\gamma$-$\gamma$-$\gamma$ ray coincidence relationships, setting gates on the most intense $\gamma$ rays. This was possible thanks to the high statistics collected and the rather high level density between the neutron-capture level and the ground state, which results in the emission of $\gamma$ rays with multiplicity greater than two.   
Fig.~\ref{fig:spectra_41Ca} presents examples of $\gamma$-ray spectra measured in the current experiment, for the  $^{40}$Ca(n,$\gamma$) reaction. The $\gamma$ rays observed here, for the first time, are marked in red, while all other correspond to transitions reported in the literature~\cite{Gru67,Cra70,Arn69,Nes16,PGAA}. Peaks coming from transitions associated to (n,$\gamma$) reactions on contaminants present in the target are labelled by circles. Panel a) presents the projection of the $\gamma$-$\gamma$ coincidence matrix, obtained by setting a gate on the known 4418-keV, primary $\gamma$-ray transition, populating the 1/2$^-$ state at 3944 keV. Two new $\gamma$ rays with energies 544 keV and 1274 keV, depopulating the 3944-keV, 1/2$^-$ state can be seen. Panels b), c), and d) show the projections of the $\gamma$-$\gamma$-$\gamma$ coincidence matrix, gated on the 2010-1390-, 1943-727-, and 1943-520-keV combinations of $\gamma$ rays. In these cases, all the spectra are almost background-free, enabling the identification of very weak $\gamma$ rays. In particular, it is worth noting the 1672-keV b) and the 2402-keV c) lines ($\approx$ 5 counts only), depopulating the 1/2$^-$ state at 5072 keV. 

\begin{figure*}[t!]
\includegraphics[width=2\columnwidth]{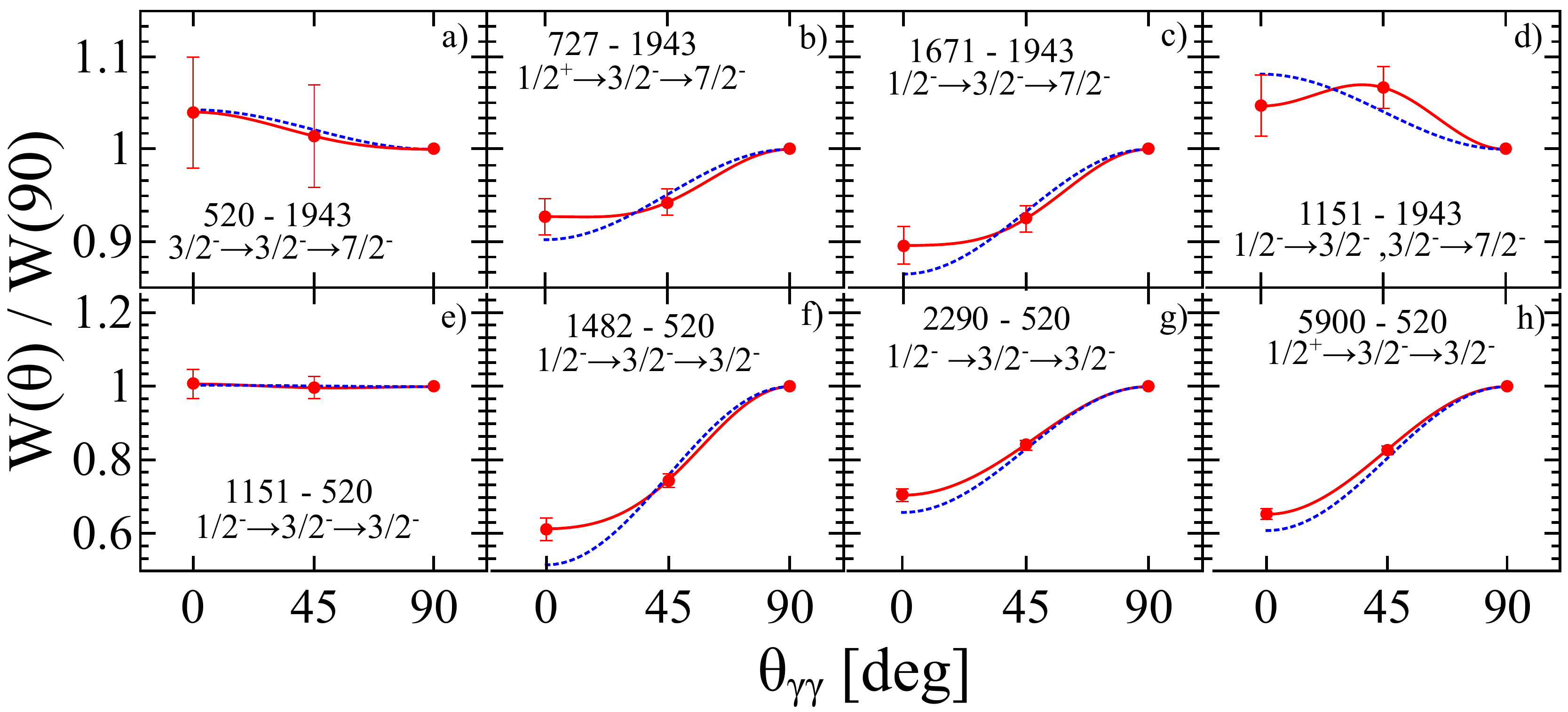}
  		\caption{(Color online) Panels a)-d): $\gamma$-ray angular correlation between the 3/2$^-$ $\rightarrow$ 7/2$^-$, 1943-keV transition and the 520-, 720-, 1671-, and 1151-keV $\gamma$ rays, depopulating the 3/2$^-$, 1/2$^+$, 1/2$^-$ states at 2462, 2670, and 3613 keV, respectively. Panels e)-h): angular correlations between the 3/2$^-$ $\rightarrow$ 3/2$^-$, 520-keV transition and the 1151-, 1482-, 2290-, 5900-keV lines depopulating the 1/2$^-$, 1/2$^-$, 1/2$^-$, and 1/2$^+$ states at 3613, 3944, 4753, and 8362 keV, respectively. Experimental fits are shown as solid red lines, while theoretical predictions are displayed as dashed blue lines (see Sec.~\ref{41Ca} and Tab.~\ref{tab:41Ca} for details).}
 	\label{fig:ang_dist_41Ca}
\end{figure*}
\noindent
The level and $\gamma$-ray decay scheme of the $^{41}$Ca nucleus is presented in Fig~\ref{fig:41Ca}, where 41 new transitions and 2 new levels obtained in this work are displayed in red (tentative levels and $\gamma$ rays are marked as dashed lines). Of particular note are the two new levels at 3564- and 6374-keV of excitation energy. In the decay scheme, the order of $\gamma$ rays was assigned on the basis of previously-known levels, as well as on the observation of parallel cascades. In the case of the 4799-1622-keV decay chain, involving the newly-found 3564-keV level, the high energy transition was tentatively assumed to depopulate directly the neutron-capture state. However, since it is not possible to firmly constrain their right order, the two transitions are displayed as dashed lines. \\
The scarce statistics collected for the $\gamma$ decays involving the new levels did not allow for the study of $\gamma$-ray angular correlations, therefore their (1/2$^\pm$,3/2$^\pm$) spin and parity is tentatively assigned on the basis of the most probable $\gamma$-ray multipolarities. On the contrary, angular correlations could be performed for a number of levels of known spin and parity, as presented in Fig.~\ref{fig:ang_dist_41Ca}. The top panels a)-d) show angular correlations between the pure E2 1943-keV ground-state decay (3/2$^-$ $\rightarrow$ 7/2$^-$), and the 520-, 727-, 1671-, and 1151-keV transitions, depopulating the 3/2$^-$, 1/2$^+$, 1/2$^-$ states at 2462 keV, 2670 keV, and 3613 keV, respectively. The solid red curve corresponds to the experimental fit, which enabled to determine the $\delta$ mixing ratios between the two most probable multipolarities, by using a $\chi^2$ minimization procedure. Theoretical predictions are also shown as dashed blue lines. In particular, a M1+E2 character, with $\delta$ = 0.13 (19), was found for the 520-keV, 3/2$^-$ $\rightarrow$ 3/2$^-$ transition. This value is in agreement, within the error, with the one reported in the literature ($\delta$ = 0.03 (12)~\cite{Nes16})  and was used to extract angular correlations for $\gamma$ rays in coincidence with the  520-keV line, as  displayed in the bottom panels e)-h) of Fig.~\ref{fig:ang_dist_41Ca}. These are the 1151-, 1482-, 2290-keV transitions depopulating the 1/2$^-$, 1/2$^-$, and 3/2$^+$ states at 3613, 3944 and 4753 keV, respectively, and the 5900-keV, 1/2$^+$ $\rightarrow$ 3/2$^-$ primary transition, for which a  E1(+M2) character was found ($\delta$ = 0.00(1)), confirming the expected dipole nature of this high-energy, primary $\gamma$ ray. \\
\noindent
The energies of levels and $\gamma$ rays, along with the $\gamma$-ray multipolarities, mixing ratios, branching ratios, and $\gamma$-ray intensities are presented in Tab.~\ref{tab:41Ca}. New results obtained in this work are marked by stars.

\begin{longtable*}{ccccccccc}
\caption{Initial and final states, $\gamma$-ray energies, multipolarities, mixing ratios, branching ratios, and $\gamma$-ray intensities normalized to the 1942.5-keV transition (100 units) of $^{41}$Ca, as observed in this work. New findings are marked by a star. Multipolarities and mixing ratios not measured in this work are also reported~\cite{Gru67,Cra70,Arn69,Nes16}.}\\
\hline 
\hline
\multicolumn{1}{c}{E$_{i}$ [keV]}& \multicolumn{1}{c}{J$^{\pi}_{i}$} & \multicolumn{1}{c}{E$_{f}$ [keV]}& \multicolumn{1}{c}{J$^{\pi}_{f}$} & \multicolumn{1}{c}{E$_\gamma$ [keV]}&\multicolumn{1}{c}{Multipolarity}&\multicolumn{1}{c}{$\delta$}&\multicolumn{1}{c}{BR$_\gamma$ }&\multicolumn{1}{c}{I$_\gamma$ }\\ 
\hline 
\endfirsthead

\multicolumn{9}{c}%
{ \tablename\ \thetable {\textit{---Continued}}}\\
\hline
\hline
\multicolumn{1}{c}{E$_{i}$ [keV]}& \multicolumn{1}{c}{J$^{\pi}_{i}$} & \multicolumn{1}{c}{E$_{f}$ [keV]}& \multicolumn{1}{c}{J$^{\pi}_{f}$} & \multicolumn{1}{c}{E$_\gamma$ [keV]}&\multicolumn{1}{c}{Multipolarity}&\multicolumn{1}{c}{$\delta$}&\multicolumn{1}{c}{BR$_\gamma$ }&\multicolumn{1}{c}{I$_\gamma$ }\\ 
\hline
\endhead
\hline
\multicolumn{7}{c}{}%
\endfoot

\hline 
\hline 
\endlastfoot

1942.5(1)  &  3/2$^-$& 0 & 7/2$^-$ & 1942.5(1)& E2 & & 1.0 & 100  \vspace{0.2cm}  \\

2009.8(1) & 3/2$^+$ & 0 &  7/2$^-$ & 2009.8(1)& M2+E3 & 0.16(2) & 1.0 &9.527(235)  \vspace{0.2cm}\\

2462.3(1) & 3/2$^-$ & 2009.8(1) & 3/2$^+$ & 452.6(2)* & & & 0.013(2)&0.148(14)\\
		 &               & 1942.5(1) & 3/2$^-$ & 519.7(1) & M1+E2 & 0.13(19)* & 0.987(2)&11.027(576)  \vspace{0.2cm}\\ 

2575.2(1)  &  5/2$^-$& 0& 7/2$^-$ & 2575.2(1)  & M1+E2 & & 1.0 &0.256(20)  \vspace{0.2cm}\\             

2605.3(1)  &  5/2$^+$& 0 & 7/2$^-$  & 2605.3(1)  & E1+M2  & -0.03(1) &  1.0&0.756(29)  \vspace{0.2cm}\\

2669.9(1) & 1/2$^+$  & 2009.8(1) & 3/2$^+$ & 660.2(1) & M1 & & 0.268(15)&0.791(44)\\
          &             &1942.5(1)& 3/2$^-$& 727.4(1) & E1(+M2) & -0.02(6)* & 0.732(15)&2.164(113) \vspace{0.2cm} \\ 

3049.2(1) & 3/2$^+$  & 2669.9(1) & 1/2$^+$ & 379.3(2) & & & 0.043(13)&0.043(14) \\
          &            & 2605.3(1) & 5/2$^+$ & 444.0(1) & & & 0.350(23)&0.355(21) \\
          &            & 2009.8(1) & 3/2$^+$ & 1039.4(1) & M1 & & 0.377(26)&0.382(27) \\          
          &            & 1942.5(1) & 3/2$^-$ & 1106.6(1) &  E1 & & 0.231(19)&0.234(19) \vspace{0.2cm}\\                  

3399.6(1)& 1/2$^+$ & 2009.8(1) & 3/2$^+$ & 1389.9(1)& M1+E2 & & 1.0&2.168(125) \vspace{0.2cm} \\

3525.3(1)  &  3/2$^+$ & 2669.9(1) &1/2$^+$ & 855.5(2) & M1+E2 & 0.22(7) & 0.090(13) & 0.070(10)    \\   
          &            & 2605.3(1) &  5/2$^+$ & 920.0(1)* & & & 0.101(10)&	0.079(8)	\\   
          &            & 2009.8(1) & 3/2$^+$ & 1515.5(1) & & &0.357(27)&0.281(27)	 \\ 
          &            & 1942.5(1) & 3/2$^-$ & 1582.8(1)& (E1+M2) & &0.453(27)&0.356(29)  \vspace{0.2cm}\\

3564.1(2)*  &  (1/2$^\pm$,3/2$^\pm$)& 1942.5(1) & 3/2$^-$ & 1621.6(2)* & & & 1.0 &0.105(17)  \vspace{0.2cm} \\       

3613.2(1)  &  1/2$^-$ & 2669.9(1) & 1/2$^+$  & 943.4(1) & (E1) & & 0.118(9) & 0.383(28)   \\ 
          &            & 2462.3(1)& 3/2$^-$& 1151.0(1) & M1+E2* & -0.49$^{+11}_{-13}$* & 0.349(21)& 	1.138(78) \\ 
          &            & 1942.5(1) & 3/2$^-$ & 1670.6(1) & M1+E2* & 0.10$^{+12}_{-10}$* & 0.533(22) & 1.736(122)  \vspace{0.2cm}\\               

3730.5(1)  &  3/2$^-$ & 2575.2(1)& 5/2$^-$ & 1155.3(2) &  && 0.249(29)& 0.102(13)  \\ 
          &            & 2462.3(1) & 3/2$^-$ & 1268.4(1) & M1+E2 && 0.428(35)&0.176(15) \\     
          &            & 1942.5(1) & 3/2$^-$& 1787.7(1) & M1+E2 && 0.181(30)&0.075(14) \\     
          &            & 0& 7/2$^-$ & 3729.8(5) &(E2)  && 0.142(39) & 	0.058(18) \vspace{0.2cm} \\                

3846.0(1) &  1/2$^+$ & 2009.8(1) & 3/2$^+$ & 1836.3(1) & & & 0.779(18)&0.949(60)    \\
          &            & 1942.5(1) & 3/2$^-$ & 1903.3(1) & & &0.221(18)& 0.269(23)  \vspace{0.2cm}\\       

3943.8(1)  &  1/2$^-$ & 3399.6(1) & 1/2$^+$ & 544.1(1)* & & & 0.003(1) &0.057(10)  \\   
          &            & 2669.9(1) & 1/2$^+$ & 1274.0(1)* &  & &0.025(2)&0.495(34) \\   
          &            & 2462.3(1) & 3/2$^-$ & 1481.6(1) & M1+E2* & 0.09(2)* & 0.067(6)&1.322(91) \\
           &            & 1942.5(1) & 3/2$^-$ & 2001.2(1) & & & 0.905(8) & 17.9(13)  \vspace{0.2cm}\\

4416.7(1)  &  3/2$^+$ & 2605.3(1) & 5/2$^+$ & 1811.7(2) && & 0.217(19) & 0.127(12)   \\  
           &            & 2009.8(1)& 3/2$^+$ & 2407.3(2) &  & &0.563(26) & 0.330(28) \\
           &            & 1942.5(1) & 3/2$^-$& 2473.6(1)* &  & &0.220(19) & 0.129(11)  \vspace{0.2cm}\\

4603.2(1)  &  3/2$^-$ & 3049.2(1) & 3/2$^+$  & 1555.1(6)* & & & 0.013(3)  & 0.037(8) \\    
           &            & 2669.9(1) & 1/2$^+$ & 1933.3(1)* &  && 0.112(11) & 0.325(30) \\ 
           &            & 2462.3(1)&3/2$^-$ & 2140.9(2)* &  & &0.022(4) & 0.064(10) \\  
           &            & 2009.8(1) & 3/2$^+$ & 2593.4(1)* &  & &0.209(32) & 0.606(114) \\
           &            & 1942.5(1) & 3/2$^-$ & 2660.8(2) &  & &0.644(31) & 1.865(127) \vspace{0.2cm} \\              

4728.0(1)  &  3/2$^+$ & 2009.8(1)& 3/2$^+$ & 2718.2(2) & & & 1.0  & 0.198(21) \vspace{0.2cm} \\       

4752.7(1)  &  1/2$^-$ & 3399.6(1) & 1/2$^+$ & 1353.1(1)* & & & 0.035(3) & 0.269(19)   \\ 
           &            & 3049.2(1) & 3/2$^+$ & 1703.4(2)* & & &0.021(2) & 0.157(16)	 \\        
           &            &2462.3(1) & 3/2$^-$ & 2290.4(1) & M1+E2* & -0.04(1)* & 0.270(17) & 2.047(141) \\  
           &            & 2009.8(1)& 3/2$^+$ & 2743.0(1)* &  & &0.040(3) & 0.307(23) \\    
           &            & 1942.5(1)& 3/2$^-$ & 2810.1(1)&  & &0.634(19) & 4.812(306) \vspace{0.2cm}\\   

4777.8(1) &  3/2$^+$ & 3525.3(1) & 3/2$^+$ & 1252.5(3)* & & & 0.049(6) & 0.109(13)   \\ 
          &            & 3399.6(1)& 1/2$^+$& 1378.2(2)* &  & &0.035(3) & 0.078(7) \\  
          &            & 3049.2(1)& 3/2$^+$ & 1728.6(1)* &  & &0.101(9) & 0.226(20) \\      
          &            & 2605.3(1)& 5/2$^+$ & 2172.6(1)* &  & &	0.104(10) & 0.232(87) \\ 
          &            & 2009.8(1)& 3/2$^+$ & 2767.9(1) &  & &0.646(20) & 1.448(104) \\ 
          &            & 1942.5(1)& 3/2$^-$ & 2835.3(2)* &  & &0.067(7) & 0.150(15)  \vspace{0.2cm}\\         

5011.8(1) &  1/2$^+$ & 3399.6(1)& 1/2$^+$ & 1612.2(2) &  & &0.284(22) & 0.115(8)   \\       
          &            & 2009.8(1)& 3/2$^+$ & 3002.0(1) &  & &0.716(22) & 0.290(23)  \vspace{0.2cm}\\   

5071.8(1)  &  1/2$^-$ & 3399.6(1)& 1/2$^+$ & 1672.2(2)* & & & 0.399(38) & 	0.096(11)    \\       
           &            & 2669.9(1)& 1/2$^+$ & 2401.9(1)* &  && 0.601(38) & 0.144(16)  \vspace{0.2cm} \\   

5449.8(3)  &  1/2$^-$ & 2462.3(1)& 3/2$^-$ & 2987.5(3)* & & & 1.0 & 0.176(16) \vspace{0.2cm}  \\      

5467.2(2) &  3/2$^-$ & 2575.2(1) & 5/2$^-$ & 2892.2(2)* & & & 0.377(42) & 0.127(19)   \\ 
         &            & 2462.3(1) & 3/2$^-$ & 3004.9(4)* & & & 0.117(22) & 0.040(8)\\
         &            & 0 & 7/2$^-$ & 5467.2(2) &  & &0.506(41) & 0.171(18) \vspace{0.2cm} \\         

5669.4(2)  &  3/2$^-$ & 2462.3(1)& 3/2$^-$  & 3207.4(2)* & & & 0.322(43) & 0.052(8)  \\  
           &            & 0& 7/2$^-$  & 5669.2(2) & & &0.678(43) & 0.110(13) \vspace{0.2cm} \\        

5703.2(2)&  1/2$^-$ & 2669.9(1) & 1/2$^+$ & 3033.3(2)* & & & 1.0   &0.041(4) \vspace{0.2cm}\\       

5888.8(1)&  1/2$^-$ & 3730.5(1) & 3/2$^-$ & 2158.3(1)* & & & 1.0  & 0.057(6)  \vspace{0.2cm} \\       

6373.8(2)*&   (1/2$^\pm$,3/2$^\pm$) & 3730.5(1)& 3/2$^-$ & 2643.1(5)* & & & 0.380(40) & 0.070(10)    \\  
         &            & 2462.3(1)& 3/2$^-$ & 3910.7(10)* & & & 0.176(33) & 0.033(7) \\
         &            & 1942.5(1) & 3/2$^-$ & 4431.2(2)* & & & 0.444(40) & 0.082(9)  \vspace{0.2cm} \\        

7190.4(2)&  1/2$^-$& 2462.3(1) & 3/2$^-$ & 4728.2(2)* & & & 1.0   & 0.027(7)   \vspace{0.2cm}   \\ 

8362.4(2)&  1/2$^+$ & 7190.4(2)& 1/2$^-$ & 1171.7(2)* &  & &    0.0003(1) & 0.029(4)  \\
         &           & 6373.8(2)*& (1/2$^\pm$,3/2$^\pm$) & 1989.5(4)* & & & 0.0018(1) & 0.198(10) \\ 
         &            & 5888.8(1) & 1/2$^-$ & 2473.6(1)* & & &  0.0005(1) & 0.050(7)\\ 
         &            & 5703.2(2) & 1/2$^-$& 2659.0(3)* & & &  0.0004(1) & 0.039(4)\\ 
         &            & 5669.4(2)&3/2$^-$ & 2693.4(2)* &  &&  0.0019(2) & 0.206(12) \\ 
         &            & 5467.2(2) & 3/2$^-$ & 2894.9(2)* &  & & 0.0035(2) & 0.382(19) \\ 
         &            & 5449.8(3) & 1/2$^-$ & 2912.7(1)* &  & & 0.0019(2) & 0.201(21)\\ 
         &            & 5071.8(1) & 1/2$^-$ & 3290.9(2)* &  &&  0.0021(2) & 0.221(21)\\
         &            & 5011.8(1) & 1/2$^+$ & 3350.8(1) &  & & 0.0043(3) & 0.458(25) \\  
         &            & 4777.8(1) & 3/2$^+$ & 3584.9(1) &  & & 0.0250(16) & 2.695(123)\\ 
         &            & 4752.7(1) & 1/2$^-$ & 3609.9(1) &  & & 0.0761(52) &8.204(441)\\ 
         &            & 4728.0(1) & 3/2$^+$ & 3634.9(1)* &  & & 0.0024(2) & 0.261(18)\\ 
         &            & 4603.2(1) & 3/2$^-$ & 3759.6(1) &  & & 0.0308(21)&3.317(166)\\ 
         &            & 4416.7(1)& 3/2$^+$ & 3945.8(1) &  & & 0.0059(4) & 0.635(30)\\ 
         &            & 3943.8(1) & 1/2$^-$ & 4418.4(1) &  & & 0.1911(131)&20.6(13) \\ 
         &            & 3846.0(1) & 1/2$^+$ & 4516.4(1) &  & & 0.0138(10) & 1.484(82)\\ 
         &            & 3730.5(1) & 3/2$^-$  & 4631.7(5)* &  & & 0.0020(2) & 0.212(13)\\ 
         &            & 3613.2(1) & 1/2$^-$  & 4749.2(1) &  & & 0.0343(24)&3.703(194)\\ 
         &            & 3564.1(2) & (1/2$^\pm$,3/2$^\pm$) & 4798.5(2)* & & & 0.0007(1) & 0.075(7) \\ 
         &            & 3525.3(1)& 3/2$^+$ & 4836.9(3) &  & &0.0078(6)& 0.840(45) \\ 
         &            & 3399.6(1) & 1/2$^+$ & 4962.9(1) &  & &0.0176(15) & 1.895(139) \\ 
         &            & 3049.2(1) &  3/2$^+$ & 5313.3(1) &  &  &0.0051(4) & 0.547(29)\\ 
         &            & 2669.9(1) & 1/2$^+$  & 5692.5(1) &  &  & 0.0174(15) & 1.872(135)\\ 
         &            & 2462.3(1) & 3/2$^-$ & 5900.0(2) & E1(+M2)* & 0.00(1)* & 0.0703(65) & 7.577(651)  \\ 
         &            & 2009.8(1)& 3/2$^+$  & 6352.6(1) &  & & 0.0074(8) & 0.800(78) \\ 
         &            & 1942.5(1) & 3/2$^-$  & 6419.7(1) &  &  & 0.4759(247)& 51.3(49)\\ 
\label{tab:41Ca}
\end{longtable*}

\subsection{$^{47}$Ca}
\label{47Ca}

\begin{figure}[htpb]
\includegraphics[width=1\columnwidth]{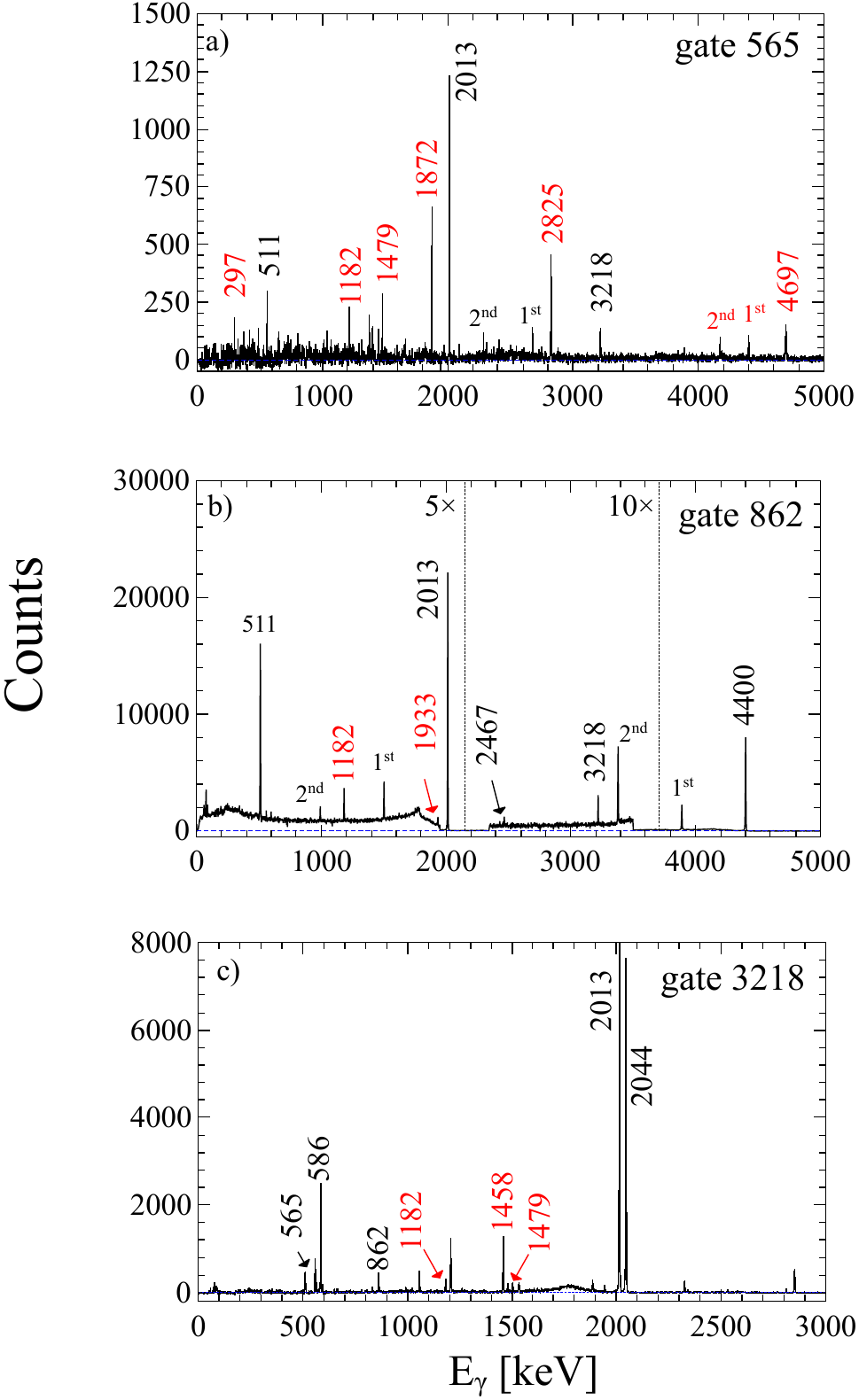}
  		\caption{(Color online) Projections of the $\gamma$-$\gamma$ coincidence matrix measured in the $^{46}$Ca(n,$\gamma$) reaction. Gates on the 565-keV a), 862-keV b), and 3218-keV c) $\gamma$ rays are presented, with new transitions, observed for the first time, displayed in red. First and second escape peaks for high-energy transitions are marked by 1$^{\text{st}}$ and 2$^{\text{nd}}$, respectively.}  
 	\label{fig:spectra_47Ca}
\end{figure}

\begin{figure*}[htpb]
 	\includegraphics[width=1.8\columnwidth]{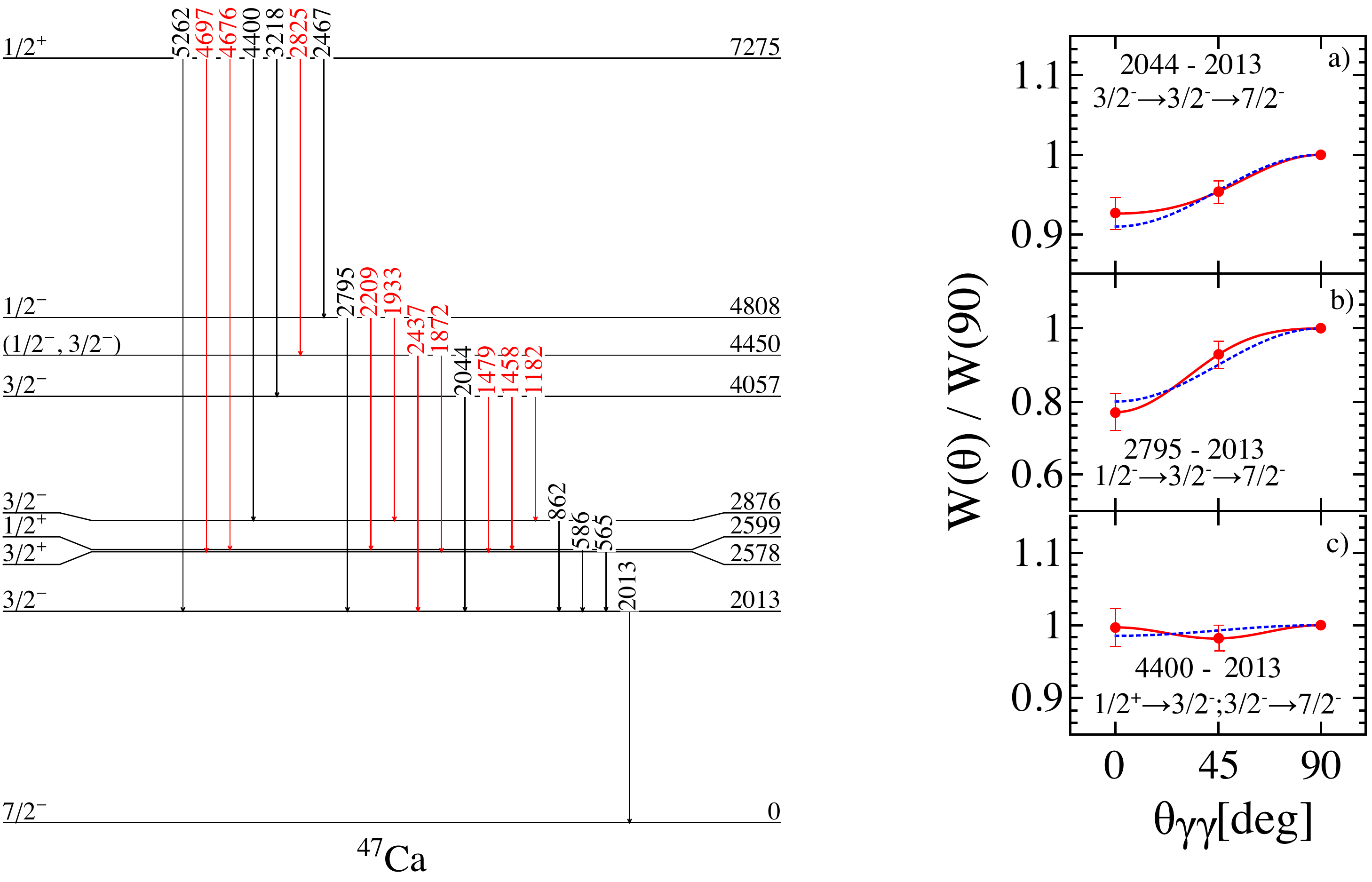}
  		\caption{(Color online) (Left) Level scheme of $^{47}$Ca as measured in the current experiment. Newly-observed $\gamma$-ray transitions are reported in red. (Right) Angular correlations in $^{47}$Ca which enabled to pin down the multipolarity of the 2044-keV a), 2795-keV b), and 4400-keV c) transitions (see Sec.~\ref{47Ca} for discussion). Experimental data are presented as dots along with experimental fits (solid red line), while theoretical predictions are shown as dashed blue line. }  
 	\label{fig:47Ca}
\end{figure*}

Projections of the $\gamma$-$\gamma$ coincidence matrix for the $^{47}$Ca nucleus are presented in Fig.~\ref{fig:spectra_47Ca}, where new $\gamma$ rays are shown in red, while transitions reported in the literature are marked in black~\citep{Cra70,Arn69,Bur07,PGAA}. The spectrum in panel a) is obtained by gating on the 565-keV transition. Of particular interest are the 4697-, 4676-, and 2825-keV $\gamma$ rays, which depopulate directly the neutron-capture level, feeding previously-known low-lying states at 2578, 2599 and 4450 keV, respectively. Panel b) shows a spectrum obtained by gating on the 862-keV line in which new $\gamma$ rays at 1182 keV and 1933 keV, populating the state at 2876 keV, can be seen. Finally, panel c) presents the spectrum measured in coincidence with the 3218-keV transition. In the picture, besides the 1182 keV transition discussed above, new $\gamma$ rays with energies 1458 keV and 1479 keV are visible. These feed the 1/2$^+$ and 3/2$^+$ states at 2599 keV and 2578 keV, respectively, from the state at 4057 keV. \\
\noindent
The level and  $\gamma$-decay scheme of the $^{47}$Ca nucleus, obtained in this work, is presented in Fig.\ref{fig:47Ca} (left), with new $\gamma$ rays shown in red. In this case, $\gamma$-ray angular correlations allowed to further characterize a number of transitions in terms of multipolarity and mixing ratios, enabling to firmly assign the spin and parity of the states involved in the decays, as presented in Fig.\ref{fig:47Ca} (right). Experimental fits are shown as solid red lines, while theoretical predictions are displayed as dashed blue lines. Angular correlations are performed against the 3/2$^-$ $\rightarrow$ 7/2$^-$ decay to the ground state, since the 2013-keV $\gamma$ ray has a pure E2 character. Panel a) shows the angular correlation for the 2044-keV line, depopulating the state at 4057 keV. The results suggest a M1+E2 character for this transition, with a mixing ratio $\delta$=0.66$^{+0.20}_{-0.11}$. The shape of the angular correlation is compatible with a 3/2$^-$ spin-parity assignment for the 4057-keV state. Panel b) presents similar results for the 2795-keV $\gamma$-ray. The experimental fit indicates a  M1+E2 character, with a mixing ratio $\delta$=0.58$^{+0.43}_{-0.14}$. In this case, the angular correlation is well reproduced assuming a 1/2$^-$ spin-parity assignment  for the 4808-keV state. Finally, panel c) shows the angular correlation for the 4400-keV, primary transition. A dominant E1 character is found, with a M2 mixing with $\delta$=-0.23(6). Moreover, the shape of the angular correlation suggests a 3/2$^-$ spin-parity assignment for the 2876-keV state. The energies of levels and $\gamma$ rays, along with the $\gamma$-ray multipolarities, mixing ratios, branching ratios, and $\gamma$-ray intensities are presented in Tab.~\ref{tab:47Ca-tab}, where new results are marked by stars.  

\subsection{$^{49}$Ca}
\label{49Ca}

\begin{figure*}[htpb]
\includegraphics[width=2\columnwidth]{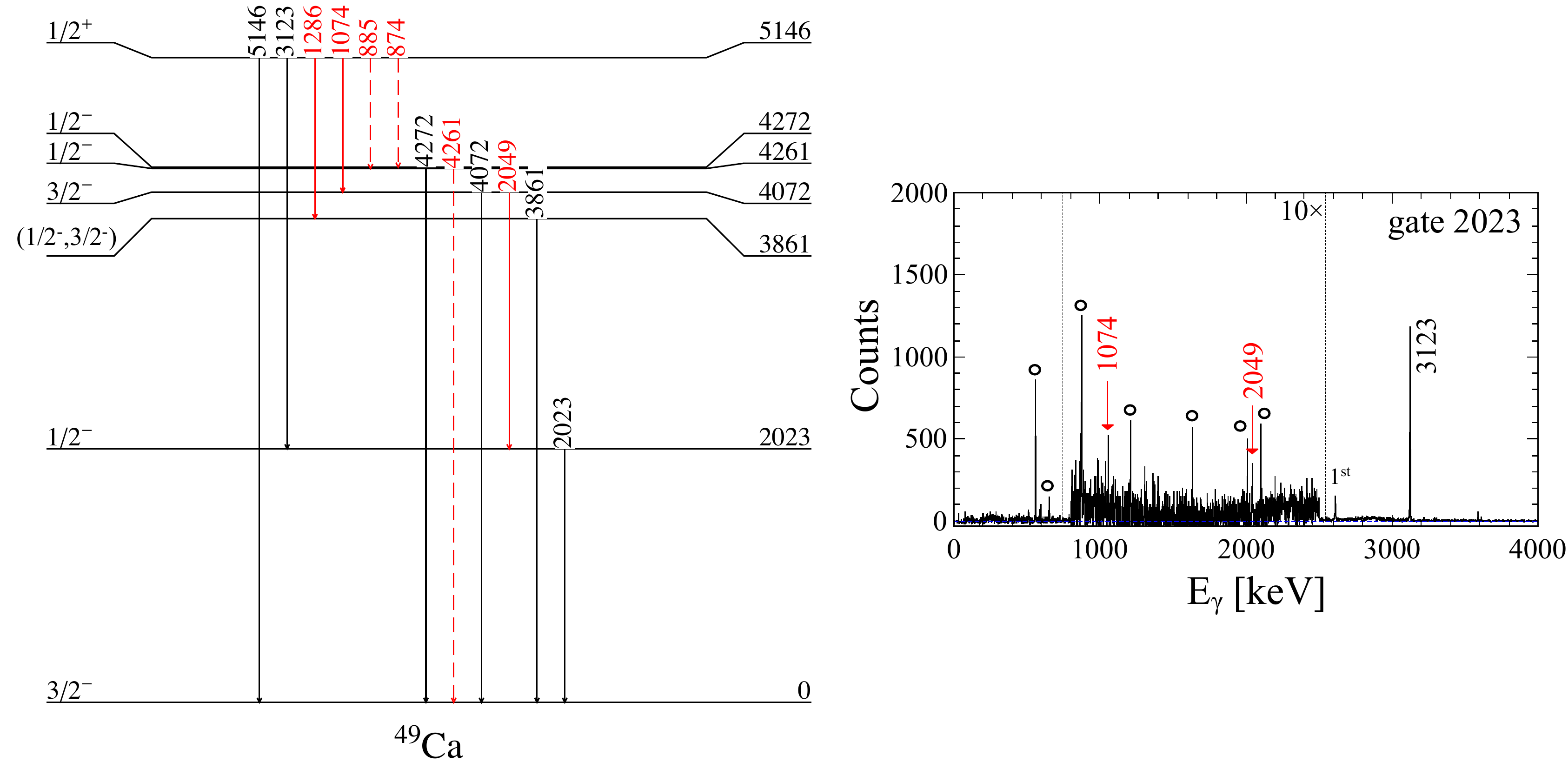}
  		\caption{(Left) Level scheme of $^{49}$Ca, as obtained in this work, with new transitions displayed in red.  (Right) Projection of the $\gamma$-$\gamma$ coincidence matrix gated on the 2023-keV transition in $^{49}$Ca, showing two new $\gamma$ rays at 1074 keV and 2049 keV. $\gamma$ rays coming from (n,$\gamma$) reactions on target contaminants are marked by circles (see Sec.~\ref{49Ca} for details). The first escape peak for the 3123-keV transition is marked by 1$^{\text{st}}$.}  
 	\label{fig:49Ca_ok}
\end{figure*}

The level scheme of the $^{49}$Ca nucleus obtained in this work is shown on the left side of Fig.~\ref{fig:49Ca_ok}. New $\gamma$ rays are marked in red, with tentative transitions displayed as dashed lines. The known $\gamma$ rays reported in the literature are shown in black~\citep{Arn69,Bur08,PGAA}. \\
\noindent
An example of $\gamma$-ray spectrum, gated on the 1/2$^-$ $\rightarrow$ 3/2$^-$, 2023-keV transition, is presented on the right side of Fig.~\ref{fig:49Ca_ok}. The most intense $\gamma$-ray, namely the 3123-keV, corresponds to the direct populations of the 1/2$^-$ state at 2023 keV, from the neutron capture level. Two new $\gamma$ lines with energies 1074 keV and 2049 keV are also present. The first corresponds to the primary transition populating the 3/2$^-$ state at 4072 keV, while the latter is the 3/2$^-$ $\rightarrow$ 1/2$^-$ decay, depopulating the 4072-keV state. Finally, the 2249-keV line depopulating the 4272-keV level and reported in~\cite{Arn69,Bur08} was not observed in the present work, suggesting a branch from this level to the 2023-keV state $\leq$0.001 \%.  It is important to note that the $^{48}$Ca sample used in the current measurement was contaminated by other nuclei with a non-negligible cross section for neutron capture. This is the case, for example, of the $^{113}$Cd nucleus, the neutron-capture cross section of which is about 20 kb. Therefore, even if present in small quantity, the $\gamma$-ray decay of the $^{114}$Cd isotope is rather strong. In Fig.~\ref{fig:49Ca_ok}, $\gamma$ rays corresponding to (n,$\gamma$) reactions on target contaminants are marked by circles. \\
\noindent
The energies of levels and $\gamma$ rays, along with branching ratios and $\gamma$-ray intensities are presented in Tab.~\ref{tab:49Ca-tab}. New results obtained in this work are marked by stars, while $\gamma$-ray multipolarities are taken from~\citep{Arn69,Bur08}. It is important to note that in the case of the 5146-keV transition, the procedure described in Sec.~\ref{Analysis} to extract branching ratios cannot be applied, since the $\gamma$ ray feeds directly the ground state. According to the literature~\citep{Arn69,Bur08}, the 5146-keV transition is 3 times larger than the 3123-

\begin{table*}[htpb]
\caption{Initial and final states, $\gamma$-ray energies, multipolarities, mixing ratios, branching ratios, and $\gamma$-ray intensities normalized to the 2013.2-keV transition (100 units) of $^{47}$Ca, as observed in this work. New findings are marked by a star. Multipolarities and mixing ratios not measured in this work are also reported~\cite{Cra70,Arn69,Bur07}.}
\begin{tabular}{ccccccccccccccccccccccccccccccccccccccccccccccccccc}
\hline
\hline
E$_i$ [keV] &J$^{\pi}_{i}$&  E$_f$ [keV] &  J$^{\pi}_{f}$ & E$_\gamma$ [keV]& Multipolarity & $\delta$ &BR$_\gamma$&I$_\gamma$  \\
\hline
2013.2(1) & 3/2$^-$  & 0 & 7/2$^-$& 2013.2(1) & E2& & 1.0 &100   \vspace{0.2cm}\\

2578.3(1) & 3/2$^+$ & 2013.2(1) & 3/2$^-$  & 565.1(1) & (E1)& & 1.0 &0.327(14) \vspace{0.2cm}\\

2599.0(1) &1/2$^+$  & 2013.2(1) & 3/2$^-$ &585.8(1) & (E1) && 1.0 &0.956(33) \vspace{0.2cm}\\

2875.6(3) & 3/2$^-$*  & 2013.2(1) & 3/2$^-$ & 862.4(1)  & & &1.0 \vspace{0.2cm}&10.922(730)\\

4057.3(2) & 3/2$^-$*  & 2875.6(3) & 3/2$^-$& 1182.4(1)* &  && 0.216(15)&1.952(116)\\
&& 2599.0(1) & 1/2$^+$&  1458.0(2)* &  && 0.060(5)&0.538(39) \\
&&2578.3(1) &3/2$^+$&1478.6(1)* &  && 0.011(2) &0.103(15)\\
&& 2013.2(1) & 3/2$^-$ & 2043.9(1)  & M1+E2* & 0.66$^{+0.20}_{-0.11}$* & 0.713(18) \vspace{0.2cm}&6.444(469)\\

4450.2(2)  &(1/2$^-$,3/2$^-$) & 2578.3(1) & 3/2$^+$& 1871.9(2)* & && 0.583(54)&0.202(14) \\ 
&&2013.2(1) &3/2$^-$&2437.3(2)* &  & &0.417(54) &0.144(31)\vspace{0.2cm}\\

4808.2(3)   &  1/2$^-$* & 2875.6(3) & 3/2$^-$& 1932.6(2)* &  & &0.052(18)&0.096(35)\\
&&2599.0(1) &1/2$^+$&2209.1(2)* &  & &0.087(9) &0.160(14)\\
&&2013.2(1) &3/2$^-$ &2794.9(1) & M1+E2* & 0.58$^{+0.43}_{-0.14}$*& 	0.861(20) \vspace{0.2cm}&1.591(106)\\

7275.4(2)  & 1/2$^+$ &4808.2(3) &1/2$^-$&2467.1(1) &  & & 0.0164(16)&1.640(111)\\
&&4450.2(2) & (1/2$^-$,3/2$^-$)&2825.1(2)* & & & 0.0040(4) &0.398(29)\\
&&4057.3(2) &3/2$^-$& 3218.3(1) &  && 0.0754(70)&7.550(502) \\ 
&& 2875.6(3) &3/2$^-$& 4400.1(1) & E1+M2* & -0.23(6)* & 0.0989(95)&9.903(716)\\ 
&&2599.0(1) &1/2$^+$&4675.7(1)* & & &  0.0021(5)&0.213(45)\\
&&2578.3(1) &3/2$^+$&4697.1(1)* &&  &  0.0009(6)&0.086(57)\\
&&2013.2(1) &3/2$^-$ &5261.8(1) &  & & 0.8024(155)\vspace{0.2cm}&80.4(70)\\   
\hline
\hline
\end{tabular}
\label{tab:47Ca-tab}
\end{table*}

\begin{table*}[htbp]
\caption{Initial and final states, $\gamma$-ray energies, branching ratios, and $\gamma$-ray intensities normalized to the 5145.9-keV transition (100 units) of $^{49}$Ca, as observed in this work. New findings are marked by a star. Multipolarities are taken from~\cite{Arn69,Bur08}.}
\begin{tabular}{ccccccccccccccccccccccccccccccccccccccccccccccccccccc}
\hline
\hline
E$_i$ [keV] &J$^{\pi}_{i}$&  E$_f$ [keV] &  J$^{\pi}_{f}$ & E$_\gamma$ [keV]& Multipolarity && $\delta$ &BR$_\gamma$&I$_\gamma$  \\
\hline
2023.2(1) &1/2$^-$  & 0 & 3/2$^-$&2023.2(1) & (M1,E2) &&& 1.0 & 33.8(22)    \vspace{0.2cm}\\

3861.2(1) &(1/2$^-$,3/2$^-$) & 0 & 3/2$^-$&3861.2(1) &  &&& 1.0 & 0.582(110) \vspace{0.2cm}\\

4072.4(2) & 3/2$^-$  &  2023.2(1) & 1/2$^-$ &2049.1(2)* & & && 0.047(18) & 0.020(15) \\
&&  0 & 3/2$^-$&4072.0(1) & (M1,E2) & &&0.953(18) & 0.406(184) \vspace{0.2cm}\\

4260.9(3) &1/2$^-$  &0 & 3/2$^-$&4260.9(3)* &  &&&1.0 & 0.181(104) \vspace{0.2cm}\\

4272.0(1) &1/2$^-$  &0 & 3/2$^-$&4272.0(1)  &  (M1,E2)&&& 1.0 & 0.090(121) \vspace{0.2cm}\\

5146.3(3)  &1/2$^+$ &4272.4(2) &1/2$^-$ &874.2(3)* & & && 0.0006(10) & 0.086(129) \\
&&4260.9(3) &1/2$^-$&885.2(3)* & & && 0.0013(9) & 0.173(118)\\
&& 4072.4(2) & 3/2$^-$&1073.8(1)* & & & &0.0022(9) & 0.300(125)\\
&&3861.2(1) &(1/2$^-$,3/2$^-$)&1286.1(2)* & & && 0.0042(9) & 0.566(118) \\ 
&&2023.2(1) &1/2$^-$&3123.4(1) & & && 0.2472(121) & 33.2(22)\\ 
&&0 & 3/2$^-$&5145.9(1) & & && 0.7444(120) & 100\vspace{0.2cm}\\     
\hline
\hline
\end{tabular}
\label{tab:49Ca-tab}
\end{table*}
\noindent
-keV transition, therefore such a value was adopted to properly normalize the branching ratios for the $\gamma$ rays depopulating the capture state.

\section{COMPARISON WITH THEORY}
\label{hybrid} 

\begin{figure*}[htbp]
\includegraphics[width=2.4 \columnwidth,angle=90]{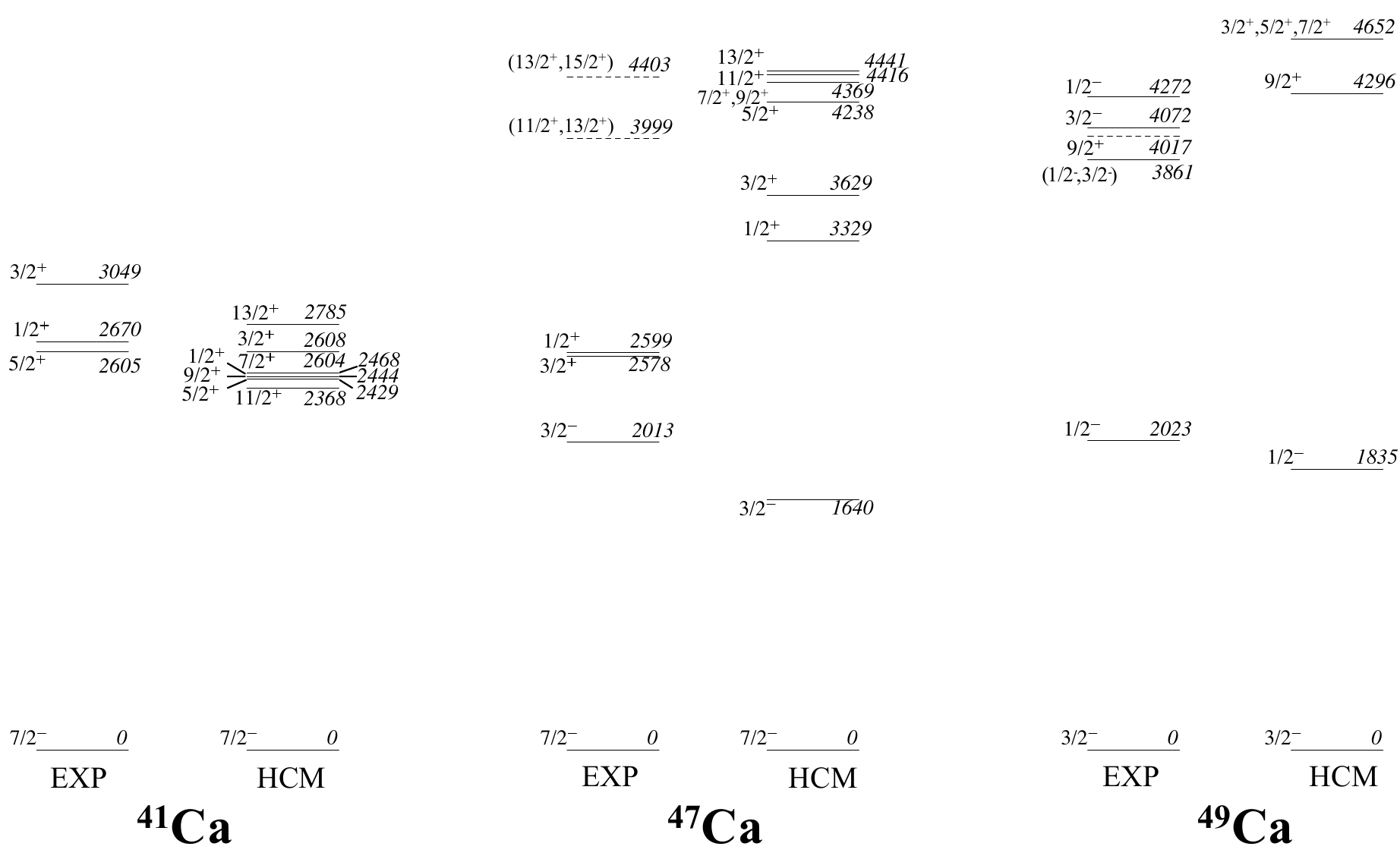}
  		\caption{Experimental low-spin states in the $^{41}$Ca, $^{47}$Ca, and $^{49}$Ca nuclei, compared with theoretical calculations performed with the HCM model, using the SkX Skyrme interaction. Higher-spin levels observed in Ref.~\cite{Mon12} are shown as dashed lines (see text for details)	.}    
 	\label{fig:theo}
\end{figure*}

The experimental excitation energy spectra of $^{41}$Ca, $^{47}$Ca and $^{49}$Ca have been partially compared with theoretical calculations performed in the framework of the Hybrid Configuration Mixing Model (HCM)~\citep{Col17,Bot19}. The model was designed to microscopically describe one-valence-particle/hole nuclei with respect to an even-even, doubly-magic ``core" with mass \textit{A} and it is based on a Hamiltonian of Skyrme type, which for the particle-core coupling case reads
\begin{eqnarray}
\label{Hamil}
&&H=H_{0}+V, \nonumber \\
&&H_{0} =\sum_{jm}\epsilon_{j}a_{jm}^{\dag}a_{jm} + \sum_{NJM}\hbar\omega_{NJ}\Gamma^{\dag}_{JM}\Gamma_{JM},  \\
&&V=\sum_{\substack{jm \\ j'm'}}\sum_{NJM}h(jm;j'm',NJM)a_{jm}[a^{\dag}_{j'm'}\otimes\Gamma^{\dag}_{JM}]_{jm}.  \nonumber
\end{eqnarray}
$H_{0}$ is the mean-field solution corresponding to Hartree-Fock (HF) particle states and Random Phase Approximation (RPA) excitations of the core calculated according to Ref.~\cite{Col13}, with $a^{\dag}$ and $\Gamma^{\dag}$ being the usual fermion-creator and boson-creator operators, respectively. \textit{V} is the coupling between single-particle states and core excitations (see Ref.~\cite{Col10} for details). A similar expression holds for the hole-core coupling case~\cite{Bot18}. Consequently, the model accounts for both single-particle/hole states and couplings with core excitations, predicting either particle/hole-phonon coupled states or 2p-1h and 2h-1p shell-model-like excitations, or hybrid mixtures, for the \textit{A+1} and \textit{A-1 }systems, respectively. It is important to note that the orthonormality and the completeness of basis states are properly taken into account by eliminating, from the model space, those spurious configurations which violate the Pauli principle. The wave functions $\ket{\Psi_n}$ for each state are then written in terms of the original basis $\ket{\alpha}$ as
\begin{equation}
\ket{\Psi_n}=\sum_{\alpha}\xi_n(\alpha) \ket{\alpha},
\end{equation}
where $\xi_n$ are the amplitudes of each component. In the case of pure single-particle/hole states, $\xi_n^2$ can be interpreted as the spectroscopic factor. 
\begin{table}[b!]
\caption{Experimental energy and B(E3; $3^- \rightarrow 0^+$) value of the ${3^-}$ phonon in the $^{40}$Ca nucleus~\cite{Kib02}, compared with RPA calculations performed with the SkX~\citep{Bro98} and SLy5~\cite{Cha98} Skyrme interactions. Calculations are done according to~\cite{Col13}. }
\begin{tabular}{ccccccccccc}
\hline
\hline
 &&& E$_{3^-}$ [MeV]   &&& B(E3; $3^- \rightarrow 0^+$) [W.u.]\\
\hline
EXP &&& 3.74 &&& 27.7(30)\vspace{0.2cm}\\
SkX &&& 2.95  &&& 17.0   \vspace{0.2cm}\\
SLy5 &&& 3.67&&& 21.5 \\
\hline
\hline
\end{tabular}
\label{RPA}
\end{table}
\noindent
Particular attention was given to the choice of the Skyrme interaction. In this work, calculations performed with the SkX parametrization~\cite{Bro98} are presented for all nuclei. This interaction was fitted on binding energies, charge radii but also single-particle energies of many doubly-magic isotopes, resulting in an effective mass m*/m~$\approx$~1.  Yet, different parametrizations were also tested. Calculations done with the SLy5 interaction (m*/m~$\approx$~0.7) better reproduce, for instance,  the properties of the 3$^-$ phonon in the $^{40}$Ca nucleus (see Tab~\ref{RPA}). As explained in Ref.~\cite{Bro98}, single-particle energies predicted by the SkX interaction specifically for this nucleus quantitatively differ from the experimental values. Such an effect can be ascribed to proton-neutron correlations, which are particularly enhanced in \textit{N=Z} systems. In $^{40}$Ca, 4p-4h and 8p-8h excitations start playing a crucial role even at low energies~\citep{Ide01}, thus affecting the  shell structure. This aspect is less pronounced in neutron-rich systems, where excitations of the neutron excess dominate over those of the symmetric core.
\begin{table}[htpb]
\caption{RPA results for $^{40}$Ca core excitations used in the HCM calculations (see text for details), showing spins, energies and main composition of the wave function, along with the squared X RPA  forward amplitudes. Only components with X$^{2}$ $\geq$ 0.1 are displayed. B(E$\lambda$; J$^{\pi}_n \rightarrow$ 0$^+_{\text{g.s.}}$) values for the $1^-$ and $3^-$ states are also reported.}
\begin{tabular}{ccccccccc}
\hline
\hline
&&J$^{\pi}_n$&& E [keV] && main w. f. && B(E$\lambda$; J$^{\pi}_n \rightarrow$ 0$^+_{\text{g.s.}}$)\\
&& && && composition && [W.u.]\\
\hline
&& && && &&  \\
{\Large \textbf{$^{40}$Ca} } \\
&&1$^-_1$ && 7290 && $\pi d_{5/2}^{-1}f_{7/2}$(0.11)&& 2.06$\cdot$10$^{-3}$  \\
&& && && $\pi d_{3/2}^{-1}p_{3/2}$(0.33) \\
&& && && $\nu d_{5/2}^{-1}f_{7/2}$(0.10) \\
&& && &&  $\nu d_{3/2}^{-1}p_{3/2}$(0.24) \vspace{0.2cm}\\
&& 2$^-_1$ &&4193 && $\pi d_{3/2}^{-1}f_{7/2}$(0.52) &&  \\
&& && && $\nu d_{3/2}^{-1}f_{7/2}$(0.45)  \vspace{0.2cm}\\
&& 2$^-_2$ &&6549 && $\nu d_{3/2}^{-1}f_{7/2}$(0.51) &&  \\
&& && && $\pi d_{3/2}^{-1}f_{7/2}$(0.44)  \vspace{0.2cm}\\
&& 3$^-_1$ &&2947&& $\pi d_{3/2}^{-1}f_{7/2}$(0.30) && 16.95  \\
&& && && $\nu d_{3/2}^{-1}f_{7/2}$(0.27)\\
&& && && $\pi s_{1/2}^{-1}f_{7/2}$(0.23)\\
&& && && $\nu s_{1/2}^{-1}f_{7/2}$(0.21)  \vspace{0.2cm}\\
&& 3$^-_2$ &&5399 && $\pi d_{3/2}^{-1}f_{7/2}$(0.62) && 0.52 \\
&& && && $\pi s_{1/2}^{-1}f_{7/2}$(0.32)  \vspace{0.2cm}\\
&& 3$^-_3$ &&5613 && $\nu d_{3/2}^{-1}f_{7/2}$(0.60) && 1.07 \\
&& && && $\nu s_{1/2}^{-1}f_{7/2}$(0.32) \vspace{0.2cm}\\
&& 3$^-_4$ &&6881 &&  $\nu s_{1/2}^{-1}f_{7/2}$(0.44) && 0.28\\
&& && &&  $\pi s_{1/2}^{-1}f_{7/2}$(0.36)   \vspace{0.2cm}\\
&& 4$^-_1$ &&7837 && $\pi d_{3/2}^{-1}f_{7/2}$(0.89) &&   \vspace{0.2cm}\\
&& 4$^-_2$ &&7837 && $\nu d_{3/2}^{-1}f_{7/2}$(0.87) &&   \vspace{0.2cm}\\
&& 4$^-_3$ &&7837 && $\pi s_{1/2}^{-1}f_{7/2}$(0.65) &&  \\
&& && && $\nu s_{1/2}^{-1}f_{7/2}$(0.34) \vspace{0.2cm}\\
&& 4$^-_4$ &&7837 && $\nu s_{1/2}^{-1}f_{7/2}$(0.63)  && \\
&& && && $\pi s_{1/2}^{-1}f_{7/2}$(0.30) \vspace{0.2cm}\\
&& 5$^-_1$ &&4998 && $\pi d_{3/2}^{-1}f_{7/2}$(0.59) && \\
&& && && $\nu d_{3/2}^{-1}f_{7/2}$(0.41) \vspace{0.2cm}\\
&& 5$^-_2$ &&6341 && $\nu d_{3/2}^{-1}f_{7/2}$(0.40)  &&  \\
&& && && $\pi d_{3/2}^{-1}f_{7/2}$ (0.40)\\
\hline
\hline
\end{tabular}
\label{table:cores1}
\end{table}
\begin{table}[htpb]
\caption{RPA results for $^{48}$Ca core excitations used in the HCM calculations (see text for details), showing spins, energies and main composition of the wave function, along with the squared X RPA  forward amplitudes. Only components with X$^{2}$ $\geq$ 0.1 are displayed. B(E$\lambda$; J$^{\pi}_n \rightarrow$ 0$^+_{\text{g.s.}}$) values for the $2^+$ and $3^-$ states are also reported.}
\begin{tabular}{ccccccccc}
\hline
\hline
&&J$^{\pi}_n$&& E [keV] && main w. f. && B(E$\lambda$; J$^{\pi}_n \rightarrow$ 0$^+_{\text{g.s.}}$)\\
&& && && composition && [W.u.]\\
\hline
&& && && &&  \\
{\Large \textbf{$^{48}$Ca} } \\
&& && && &&  \\
&& 2$^+_1$ && 2868 && $\nu f_{7/2}^{-1}p_{3/2}$(0.98) && 1.31      \vspace{0.2cm}\\
&& 3$^+_1$ && 3225 && $\nu f_{7/2}^{-1}p_{3/2}$(0.99) && \vspace{0.2cm}\\
&& 3$^+_2$ &&5016 && $\nu f_{7/2}^{-1}p_{3/2}$(0.98) && \vspace{0.2cm}\\
&& 3$^-_1$ &&4434&& $\pi s_{1/2}^{-1}f_{7/2}$(0.76) && 6.77 \\
&& && && $\pi d_{3/2}^{-1}f_{7/2}$(0.19) \vspace{0.2cm}\\
&& 3$^-_2$ &&5372&& $\pi d_{3/2}^{-1}f_{7/2}$(0.79) &&  \\
&& && && $\pi s_{1/2}^{-1}f_{7/2}$(0.20) \vspace{0.2cm}\\
&& 4$^+_1$ &&3124 && $\nu f_{7/2}^{-1}p_{3/2}$(1.00) && \vspace{0.2cm}\\
&& 4$^+_2$ &&4703&& $\nu f_{7/2}^{-1}p_{3/2}$(0.98) && \vspace{0.2cm}\\
&& 4$^-_1$ &&5111 && $\pi s_{1/2}^{-1}f_{7/2}$(0.92) && \vspace{0.2cm}\\
&& 5$^+_1$ &&3509 && $\nu f_{7/2}^{-1}p_{3/2}$(1.00) && \\
\hline
\hline
\end{tabular}
\label{table:cores2}
\end{table}
\noindent
The $^{40}$Ca and $^{48}$Ca, RPA core excitations, used in this work for HCM  calculations, are reported in Tabs.~\ref{table:cores1} and \ref{table:cores2}. Along with spins and energies, the main components of the wave functions and the B(E$\lambda$; J$^{\pi}_n \rightarrow$ 0$^+_{\text{g.s.}}$) values for low-spin states are presented for both collective phonons and non-collective excitations.
\begin{table*}[htbp]
\caption{Results of the Hybrid Configuration Mixing Model calculations for different states in $^{41}$Ca, $^{47}$Ca, and $^{49}$Ca. The main components $\ket{\alpha}$ of the wave functions are reported, along with the corresponding squared amplitudes $\xi_n^2(\alpha)$, considering only contributions with $\xi_n^2(\alpha) \geq$ 0.05 (see Sec.~\ref{hybrid} for details and discussion). }
\begin{tabular}{ccccccccccccccccccccc}
\hline
\hline
&&&&J$^{\pi}$&&&& E [keV] &&&& $\ket{\alpha}$ &&&&$\xi_n^2(\alpha)$ &&&&\\
\hline
&&&& &&&& &&&& &&&& &&&& \\
{\Large \textbf{$^{41}$Ca} } \\
&&&&7/2$^-$ &&&& 0 &&&& $\nu1f_{7/2}$  &&&&0.99 &&&&     \vspace{0.2cm}\\
&&&&11/2$^+$ &&&& 2368 &&&& $\nu1f_{7/2} \otimes 3^-_1$  &&&&0.40 &&&&     \vspace{0.2cm}\\
&&&& 5/2$^+$ &&&&2429 &&&&$\nu1f_{7/2} \otimes 3^-_1$ &&&&0.35&&&& \\
&&&&  &&&& &&&&$\nu1f_{7/2} \otimes 3^-_3$ &&&&0.08&&&& \vspace{0.2cm}\\
&&&&9/2$^+$ &&&& 2444 &&&& $\nu1f_{7/2} \otimes 3^-_1$  &&&&0.32 &&&&     \vspace{0.2cm}\\
&&&& 1/2$^+$ &&&&2468 &&&&$\nu1f_{7/2} \otimes 3^-_1$ &&&&0.33&&&&    \\
&&&&  &&&& &&&&$\nu1f_{7/2} \otimes 4^-_2$ &&&&0.10&&&& \vspace{0.2cm}\\
&&&&7/2$^+$ &&&& 2604 &&&& $\nu1f_{7/2} \otimes 3^-_1$  &&&&0.31 &&&&     \vspace{0.2cm}\\
&&&& 3/2$^+$ &&&&2608 &&&&$\nu1f_{7/2} \otimes 3^-_1$ &&&&0.28&&&&    \\
&&&&  &&&& &&&&$\nu1f_{7/2} \otimes 4^-_2$ &&&&0.13&&&& \vspace{0.2cm}\\
&&&&13/2$^+$ &&&& 2785 &&&& $\nu1f_{7/2} \otimes 3^-_1$  &&&&0.10 &&&& \\  
&&&&  &&&& &&&&$\nu1f_{7/2} \otimes 4^-_2$ &&&&0.22&&&& \vspace{0.2cm}\\
\hline
&&&& &&&& &&&& &&&& &&&& \\
{\Large \textbf{$^{47}$Ca} } \\
&&&&7/2$^-$ &&&& 0 &&&& $\nu1f_{7/2}^{-1}$  &&&&0.99 &&&&     \vspace{0.2cm}\\
&&&& 3/2$^-$ &&&&1640 &&&&$\nu1f_{7/2}^{-1} \otimes (\nu1f_{7/2}^{-1}1p_{3/2})_{2^+_1}$ &&&&0.12&&&&    \\
&&&&  &&&& &&&&$\nu1f_{7/2}^{-1} \otimes (\nu1f_{7/2}^{-1}1p_{3/2})_{4^+_1}$&&&&0.29&&&&  \vspace{0.2cm}\\
&&&& 1/2$^+$ &&&& 3329&&&&$\nu2s_{1/2}^{-1}$ &&&&0.27&&&&   \\
&&&&  &&&& &&&&$\nu1f_{7/2}^{-1} \otimes 3^-_1 $ &&&&0.68&&&&  \vspace{0.2cm}\\
&&&& 3/2$^+$ &&&& 3629&&&&$\nu2d_{3/2}^{-1}$ &&&&0.43&&&&    \\
&&&&  &&&& &&&&$\nu1f_{7/2}^{-1} \otimes 3^-_1 $ &&&&0.53&&&&  \vspace{0.2cm}\\
&&&& 5/2$^+$ &&&& 4238 &&&&$\nu1f_{7/2}^{-1} \otimes 3^-_1 $&&&&0.93&&&&    \vspace{0.2cm}\\
&&&& 7/2$^+$,9/2$^+$ &&&& 4369 &&&&$\nu1f_{7/2}^{-1} \otimes 3^-_1 $&&&&0.96&&&&    \vspace{0.2cm}\\
&&&& 11/2$^+$ &&&& 4416 &&&&$\nu1f_{7/2}^{-1} \otimes 3^-_1 $&&&&0.96&&&&    \vspace{0.2cm}\\
&&&& 13/2$^+$ &&&& 4441 &&&&$\nu1f_{7/2}^{-1} \otimes 3^-_1 $&&&&0.96&&&&    \vspace{0.2cm}\\
\hline
&&&& &&&& &&&& &&&& &&&& \\
{\Large \textbf{$^{49}$Ca} } \\
&&&&3/2$^-$ &&&& 0 &&&& $\nu1p_{3/2}$  &&&&0.92 &&&&     \vspace{0.2cm}\\
&&&& 1/2$^-$ &&&& 1835 &&&&$\nu1p_{1/2}$ &&&&0.99&&&&    \vspace{0.2cm}\\   
&&&& 9/2$^+$ &&&& 4296 &&&&$\nu1g_{9/2}$ &&&&0.11&&&&    \\
&&&&  &&&& &&&&$\nu1p_{3/2} \otimes 3^-_1 $ &&&&0.83&&&&  \vspace{0.2cm}\\
&&&& 3/2$^+$,5/2$^+$,7/2$^+$ &&&& 4652 &&&&$\nu1p_{3/2} \otimes 3^-_1 $&&&&0.94&&&&    \\
\hline
\hline
\end{tabular}
\label{table:hybrid}
\end{table*}    
\noindent
Calculations for the $^{41}$Ca nucleus were performed by assuming a $^{40}$Ca core and including neutron single-particle states of the $pfg_{9/2}$ shell and the \textit{sd} levels above the \textit{N}=50 shell gap. In the cases of the $^{47}$Ca and $^{49}$Ca isotopes, a $^{48}$Ca core was taken with the full hole space for the former and by including the $pf_{7/2}g_{9/2}$ orbitals for the latter. In the first case, $^{40}$Ca core excitations up to 8 MeV and angular momentum L=8 were considered, while for $^{48}$Ca core excitations up to 6 MeV and L=8 were taken into account. Core excitations in $^{40}$Ca are located, in general, at higher excitation energy than in $^{48}$Ca. We note that the 8 MeV, L=8 cutoffs in $^{40}$Ca select only negative-parity states, as positive-parity ones are predicted to be even higher.\\
\noindent
Fig.~\ref{fig:theo} shows a comparison between predictions from the HCM model, with the SkX Skyrme interaction, and experimental low-spin yrast states of $^{41}$Ca (left), $^{47}$Ca (middle) and $^{49}$Ca (right), obtained in this work. In the following, the comparison will be limited to the energy of the states, since very limited mixing ratios information is experimentally available for transitions depopulating such states. For detailed comparison in terms of selected B(E3) values we refer to Ref.~\cite{Mon12}.” The results of the calculations are also summarized in Tab.~\ref{table:hybrid}, for what concerns state energies and dominant wave function components. It is important to stress that similar predictions are obtained by using the SLy5 interaction: small differences are observed in the energies of the levels (within $\approx$ 300 keV on average), whereas wave function compositions are almost independent from the choice of the interaction. \\
\noindent
The $^{41}$Ca ground state is predicted to have a pure f$_{7/2}$, single-neutron nature. In the case of excited states, the comparison between experimental results and theoretical predictions is limited to positive-parity states below 4 MeV (see Fig~\ref{fig:theo}). In this region, the HCM model predicts a multiplet of states with spin 1/2$^+$, 3/2$^+$ and 5/2$^+$  at 2468 keV, 2608 keV, and 2429 keV, respectively. These states arise mainly from the coupling between a f$_{7/2}$ neutron and the octupole 3$^-$ vibration of the $^{40}$Ca core, with contributions from couplings with other phonons (see Tab.~\ref{table:hybrid}). A good correspondence in terms of level ordering and energy spacing with the lowest experimental 1/2$^+$, 3/2$^+$ and 5/2$^+$ states is found. However, the calculated multiplet is located $\approx$ 200-400 keV below the experimental one. This difference in energy might also be related to the 3$^{-}$ octupole vibration of $^{40}$Ca, which is predicted $\approx$ 800 keV below the experimental value.\\
\noindent
Calculations for the $^{47}$Ca nucleus are presented in the middle of Fig.~\ref{fig:theo}. These are the first results obtained with the HCM model for a valence-hole system in this mass region. The ground state of $^{47}$Ca is calculated as a neutron f$^{-1}_{7/2}$ configuration. Concerning the 3/2$^-$ negative-parity state, it is predicted to be the coupling between a f$^{-1}_{7/2}$ neutron hole and non-collective 1p-1h excitations of the $^{48}$Ca core, namely (f$^{-1}_{7/2}$p$_{3/2}$)$_{2^+_1}$ and (f$^{-1}_{7/2}$p$_{3/2}$)$_{4^+_1}$. It is interesting to note that the wave function composition is similar to the one obtained by shell model calculations in the full \textit{fpg} space, i.e., [$\nu f^{-2}_{7/2}p_{3/2}$]~\cite{Pov01}. The positive-parity 1/2$^+$ and 3/2$^+$ states are instead suggested to be members of the $\nu$f$^{-1}_{7/2}$ $\otimes$ 3$^-$ multiplet and are predicted about 1 MeV higher than in experiments. These results indicate that the low-spin structure of the $^{47}$Ca nucleus is more complex and probably contains more configurations than those included in the present HCM model. \\
\noindent
Finally, the level scheme of the $^{49}$Ca nucleus is presented on the right side of Fig.~\ref{fig:theo}. Its ground state is predicted to be pure, with a $\nu$p$_{3/2}$ configuration, as well as the 1/2$^-$ state which is suggested to be a p$_{1/2}$, pure single-neutron state. The latter is in good agreement with the experimental energy. The 9/2$^+$ state is instead calculated as the lowest member of the $\nu$p${3/2}$ $\otimes$ 3$^-$ multiplet.\\
\noindent
All the calculated members of the $\nu$f$_{7/2}$ $\otimes$ 3$^-$ , $\nu$f$^{-1}_{7/2}$ $\otimes$ 3$^-$ and $\nu$p${3/2}$ $\otimes$ 3$^-$ multiplets in $^{41}$Ca, $^{47}$Ca and $^{49}$Ca, respectively, are presented  in Fig.~\ref{fig:theo} and their wave function compositions are given in Tab.~\ref{table:hybrid}. These states in $^{47}$Ca and $^{49}$Ca were already investigated by this collaboration in Ref.~\cite{Mon12} and interpreted using the perturbative particle-vibration coupling approach~\cite{BM}. Similar studies were also performed in the case of the $^{65}$Cu and $^{67}$Cu nuclei, where Ni cores were considered~\cite{Boc14,NIt14}. This neutron-rich region around \textit{Z}=28 is characterized by the coexistence of different nuclear shapes~\cite{Cri16,Mor16,Mor17,Leo17,Mar20,Por20}, the emergence of which is intimately related to proton-neutron correlations and shell structure. 
In this sense, a microscopic description of the many facets of nuclear excitations is much  desirable and the Hybrid Configuration Mixing model presented in this work provides a step forward in this direction. \\
\noindent
In the case of $^{41}$Ca, low-lying positive-parity states up to 13/2$^+$ are predicted to be members of the multiplet, located around 2500 keV of excitation energy, in a range of $\approx$ 400 keV. Despite the fact that the $\nu$f$_{7/2}$ $\otimes$ 3$^-$ coupling is the dominant component in their wave function, contributions from couplings with other core excitations are significantly present. In the case of $^{47}$Ca, in addition to the 1/2$^{+}$ and 3/2$^{+}$ states discussed above, the higher spin states between 5/2$^{+}$ and 13/2$^{+}$ are displayed and compared to experimental energies obtained in previous works~\cite{Mon12} and shown by dashed lines. The HCM model also predicts the B(E3; (13/2$^+$,11/2$^+$) $\rightarrow$ 7/2$^-$) = 6.7 W.u., which is in agreement, within the error, with the 7.4(19) W.u. experimental value reported in Ref.~\cite{Mon12}. For $^{49}$Ca, the 3/2$^{+}$, 5/2$^{+}$, 7/2$^{+}$, and 9/2$^{+}$ states, members of the $\nu$p${3/2}$ $\otimes$ 3$^-$ multiplet, are also shown, and compared to the 9/2$^{+}$ state located at 4017 keV (dashed line), which is the only one known experimentally, as reported in several works~\cite{Mon11plb,Mon12,Gad16,Cra17}. On the other hand, the HCM model predicts the 3/2$^{+}$, 5/2$^{+}$, 7/2$^{+}$ to be degenerate at 4652 keV. This is due to the absence of the $d_{3/2}$, $d_{5/2}$, and $g_{7/2}$ orbitals in the configuration space. Nevertheless, the B(E3; 9/2$^+$ $\rightarrow$ 3/2$^-$) is calculated to be 5.2 W.u., which is in fair agreement with the experimental value of 7.9(20) W.u. obtained by lifetime measurements, as reported in Ref.~\cite{Mon12}\\
\noindent
The $^{47}$Ca and $^{49}$Ca isotopes were also recently investigated in neutron knockout~\cite{Cra17} and neutron pickup experiments~\cite{Gad16}, where the strength of the $\nu$1f$_{7/2}$ in $^{47}$Ca and the relative strength of the $\nu$1f$_{g/2}$ and $\nu$1f$_{5/2}$ in $^{49}$Ca were extracted. Experimental data were compared with large-scale shell-model calculations using the GXPF1 effective interaction in the \textit{sd+fp+sdg} model space, as well as NN+3N, \textit{ab initio} calculations in the \textit{pf} and \textit{pfg$_{9/2}$} model space (see ~\cite{Gad16,Cra17} and references therein). In the case of $^{47}$Ca, the strength of the $\nu$1f$_{7/2}$ orbital is found to be concentrated in the ground state, with a measured (2J+1)C$^2$S spectroscopic factor of 9.3($^{+1.1}_{-1.3})_{\text{stat}}$($\pm$1.9)$_{sys}$. This is qualitatively reproduced by shell-model calculations, which predict (2J+1)C$^2$S=7.7 and (2J+1)C$^2$S=6.7-7.0 for the GXPF1 and NN+3N interactions, respectively. Similar results are obtain in the present work with the HCM model, which estimates (2J+1)C$^2$S=7.9. \\  
\noindent
For $^{49}$Ca, results for the C$^2$S spectroscopic factor for the 9/2$^{+}$ state are summarized in Tab.~\ref{table:calc}. It can be seen that the experimental results point to a rather small value for the strength of the $\nu$1$_{g/2}$ orbital at 4296 keV, consistently with the complex octupole-coupled nature of the 9/2$^{+}$ state. This quenching is more pronounced in the case of the (d,p) measurement~\cite{Uoz94}, which is well reproduced by the HCM calculations here presented. On the other hand, shell-model calculations (see \cite{Gad16}) predict a larger spectroscopic factor, which sits in between the two experimental results. \\

\begin{table}[htbp]
\caption{Experimental and theoretical (see ~\cite{Gad16} and references therein) C$^2$S spectroscopic factors for the 9/2$^{+}$ at 4296 keV in $^{49}$Ca. }
\begin{tabular}{ccccccccccccccccc}
\hline
\hline
&&&& \textbf{Experiment} &&&& \textbf{C$^2$S} &&&& \\
\hline
&&&& (d,p)~\cite{Uoz94} &&&& 0.14 &&&&  \\
&&&& $^{12}$C+$^{48}$Ca~\cite{Gad16} &&&& 0.27(1) &&&&   \\
\hline
&&&& \textbf{Theory} &&&& \textbf{C$^2$S }&&&& \\
\hline
&&&& HCM &&&& 0.11 &&&&  \\
&&&& GXPF1~\cite{Gad16} &&&&  0.42 &&&& \\
\hline
\hline
\end{tabular}
\label{table:calc}
\end{table}

\section{CONCLUSIONS}		
\label{end}
In conclusion, the low-spin structure of the $^{41}$Ca, $^{47}$Ca, and $^{49}$Ca nuclei was investigated in the EXILL experimental campaign, following $\gamma$-ray spectroscopy of neutron-capture reactions on Ca targets. New levels, $\gamma$-ray transitions and $\gamma$-ray branching ratios and intensities were reported and a number of transitions were characterized by $\gamma$-ray angular correlations,  enabling to extract multiploraties and mixing ratios and to assign spins and parities to the states involved in the decays. \\
\noindent
Portions of the level schemes below 5 MeV were compared with theoretical calculations performed with the Hybrid Configuration Mixing Model. Despite some discrepancies, the model indicates the coexistence, at low energy, of single-particle/hole states and coupled configurations with collective and non-collective excitations of the doubly-magic core for all the nuclei studied in this work. Moreover, experimental results and beyond-mean-filed calculations obtained in this work by the HCM model for the $^{47}$Ca and $^{49}$Ca nuclei were compared with other measurements, as well as shell model and \textit{ab intio} calculations. Similar results are observed, although the HCM model better reproduces the quenching of spectroscopic factor for the 9/2$^+$ state in $^{49}$Ca, pointing to the impact of long-range correlations, such as couplings with phonons, upon the structure of nuclear excitations.  \\
\noindent
Overall, it appears
that Ca isotopes provide a fundamental play ground for
state-of-the-art theories, which, in this mass region, tend to converge to similar results, making Ca nuclei a cornerstone for a comprehensive description of nuclear structure. \\
\noindent
This 
work is also an important benchmark for the Hybrid Configuration Mixing Model, here discussed, which becomes a powerful tool to compute the complex structure of isotopes in heavier mass regions, such as the neutron-rich region around the doubly-magic $^{132}$Sn nucleus~\cite{Boc16,Bot19Zako}. Indeed, in these heavy systems, shell model calculations and \textit{ab initio} methods have severe difficulties in dealing with collective excitations of the core, due to the diverging dimension of the model space, thus resulting, up to now, in a limited description of complex excitations.

\section{ACKNOWLEDGMENTS}
\label{thanks}  

This work was supported by the Italian Istituto Nazionale di Fisica Nucleare and by the Polish National Science Centre under Contract Nos. 2014/14/M/ST2/00738 and 2013/08/M/ST2/00257. The authors would like to thank the technical services of the ILL, LPSC and GANIL for supporting the EXILL campaign, as well as the EXOGAM collaboration and the INFN Legnaro for providing HPGe detectors.

\bibliography{Ca_EXILL}

%merlin.mbs apsrev4-1.bst 2010-07-25 4.21a (PWD, AO, DPC) hacked
%Control: key (0)
%Control: author (72) initials jnrlst
%Control: editor formatted (1) identically to author
%Control: production of article title (-1) disabled
%Control: page (0) single
%Control: year (1) truncated
%Control: production of eprint (0) enabled
\begin{thebibliography}{69}%
\makeatletter
\providecommand \@ifxundefined [1]{%
 \@ifx{#1\undefined}
}%
\providecommand \@ifnum [1]{%
 \ifnum #1\expandafter \@firstoftwo
 \else \expandafter \@secondoftwo
 \fi
}%
\providecommand \@ifx [1]{%
 \ifx #1\expandafter \@firstoftwo
 \else \expandafter \@secondoftwo
 \fi
}%
\providecommand \natexlab [1]{#1}%
\providecommand \enquote  [1]{``#1''}%
\providecommand \bibnamefont  [1]{#1}%
\providecommand \bibfnamefont [1]{#1}%
\providecommand \citenamefont [1]{#1}%
\providecommand \href@noop [0]{\@secondoftwo}%
\providecommand \href [0]{\begingroup \@sanitize@url \@href}%
\providecommand \@href[1]{\@@startlink{#1}\@@href}%
\providecommand \@@href[1]{\endgroup#1\@@endlink}%
\providecommand \@sanitize@url [0]{\catcode `\\12\catcode `\$12\catcode
  `\&12\catcode `\#12\catcode `\^12\catcode `\_12\catcode `\%12\relax}%
\providecommand \@@startlink[1]{}%
\providecommand \@@endlink[0]{}%
\providecommand \url  [0]{\begingroup\@sanitize@url \@url }%
\providecommand \@url [1]{\endgroup\@href {#1}{\urlprefix }}%
\providecommand \urlprefix  [0]{URL }%
\providecommand \Eprint [0]{\href }%
\providecommand \doibase [0]{http://dx.doi.org/}%
\providecommand \selectlanguage [0]{\@gobble}%
\providecommand \bibinfo  [0]{\@secondoftwo}%
\providecommand \bibfield  [0]{\@secondoftwo}%
\providecommand \translation [1]{[#1]}%
\providecommand \BibitemOpen [0]{}%
\providecommand \bibitemStop [0]{}%
\providecommand \bibitemNoStop [0]{.\EOS\space}%
\providecommand \EOS [0]{\spacefactor3000\relax}%
\providecommand \BibitemShut  [1]{\csname bibitem#1\endcsname}%
\let\auto@bib@innerbib\@empty
%</preamble>
\bibitem [{\citenamefont {Yasuda}\ \emph {et~al.}(2010)\citenamefont {Yasuda},
  \citenamefont {Sakaguchi}, \citenamefont {Asaji}, \citenamefont {Fujita},
  \citenamefont {Hagihara}, \citenamefont {Hatanaka}, \citenamefont {Ishida},
  \citenamefont {Itoh}, \citenamefont {Kawabata}, \citenamefont {Kishi},
  \citenamefont {Noro}, \citenamefont {Sakemi}, \citenamefont {Shimizu},
  \citenamefont {Takeda}, \citenamefont {Tameshige}, \citenamefont {Terashima},
  \citenamefont {Uchida}, \citenamefont {Wakasa}, \citenamefont {Yonemura},
  \citenamefont {Yoshida}, \citenamefont {Yosoi},\ and\ \citenamefont
  {Zenihiro}}]{Yas10}%
  \BibitemOpen
  \bibfield  {author} {\bibinfo {author} {\bibfnamefont {Y.}~\bibnamefont
  {Yasuda}}, \bibinfo {author} {\bibfnamefont {H.}~\bibnamefont {Sakaguchi}},
  \bibinfo {author} {\bibfnamefont {S.}~\bibnamefont {Asaji}}, \bibinfo
  {author} {\bibfnamefont {K.}~\bibnamefont {Fujita}}, \bibinfo {author}
  {\bibfnamefont {Y.}~\bibnamefont {Hagihara}}, \bibinfo {author}
  {\bibfnamefont {K.}~\bibnamefont {Hatanaka}}, \bibinfo {author}
  {\bibfnamefont {T.}~\bibnamefont {Ishida}}, \bibinfo {author} {\bibfnamefont
  {M.}~\bibnamefont {Itoh}}, \bibinfo {author} {\bibfnamefont {T.}~\bibnamefont
  {Kawabata}}, \bibinfo {author} {\bibfnamefont {S.}~\bibnamefont {Kishi}},
  \bibinfo {author} {\bibfnamefont {T.}~\bibnamefont {Noro}}, \bibinfo {author}
  {\bibfnamefont {Y.}~\bibnamefont {Sakemi}}, \bibinfo {author} {\bibfnamefont
  {Y.}~\bibnamefont {Shimizu}}, \bibinfo {author} {\bibfnamefont
  {H.}~\bibnamefont {Takeda}}, \bibinfo {author} {\bibfnamefont
  {Y.}~\bibnamefont {Tameshige}}, \bibinfo {author} {\bibfnamefont
  {S.}~\bibnamefont {Terashima}}, \bibinfo {author} {\bibfnamefont
  {M.}~\bibnamefont {Uchida}}, \bibinfo {author} {\bibfnamefont
  {T.}~\bibnamefont {Wakasa}}, \bibinfo {author} {\bibfnamefont
  {T.}~\bibnamefont {Yonemura}}, \bibinfo {author} {\bibfnamefont {H.~P.}\
  \bibnamefont {Yoshida}}, \bibinfo {author} {\bibfnamefont {M.}~\bibnamefont
  {Yosoi}}, \ and\ \bibinfo {author} {\bibfnamefont {J.}~\bibnamefont
  {Zenihiro}},\ }\href {\doibase 10.1103/PhysRevC.81.044315} {\bibfield
  {journal} {\bibinfo  {journal} {Phys. Rev. C}\ }\textbf {\bibinfo {volume}
  {81}},\ \bibinfo {pages} {044315} (\bibinfo {year} {2010})}\BibitemShut
  {NoStop}%
\bibitem [{\citenamefont {Lui}\ \emph {et~al.}(2011)\citenamefont {Lui},
  \citenamefont {Youngblood}, \citenamefont {Shlomo}, \citenamefont {Chen},
  \citenamefont {Tokimoto}, \citenamefont {Krishichayan}, \citenamefont
  {Anders},\ and\ \citenamefont {Button}}]{Lui11}%
  \BibitemOpen
  \bibfield  {author} {\bibinfo {author} {\bibfnamefont {Y.-W.}\ \bibnamefont
  {Lui}}, \bibinfo {author} {\bibfnamefont {D.~H.}\ \bibnamefont {Youngblood}},
  \bibinfo {author} {\bibfnamefont {S.}~\bibnamefont {Shlomo}}, \bibinfo
  {author} {\bibfnamefont {X.}~\bibnamefont {Chen}}, \bibinfo {author}
  {\bibfnamefont {Y.}~\bibnamefont {Tokimoto}}, \bibinfo {author} {\bibnamefont
  {Krishichayan}}, \bibinfo {author} {\bibfnamefont {M.}~\bibnamefont
  {Anders}}, \ and\ \bibinfo {author} {\bibfnamefont {J.}~\bibnamefont
  {Button}},\ }\href {\doibase 10.1103/PhysRevC.83.044327} {\bibfield
  {journal} {\bibinfo  {journal} {Phys. Rev. C}\ }\textbf {\bibinfo {volume}
  {83}},\ \bibinfo {pages} {044327} (\bibinfo {year} {2011})}\BibitemShut
  {NoStop}%
\bibitem [{\citenamefont {Montanari}\ \emph
  {et~al.}(2011{\natexlab{a}})\citenamefont {Montanari}, \citenamefont {Leoni},
  \citenamefont {Mengoni}, \citenamefont {Benzoni}, \citenamefont {Blasi},
  \citenamefont {Bocchi}, \citenamefont {Bortignon}, \citenamefont {Bracco},
  \citenamefont {Camera}, \citenamefont {ColÃ²}, \citenamefont {Corsi},
  \citenamefont {Crespi}, \citenamefont {Million}, \citenamefont {Nicolini},
  \citenamefont {Wieland}, \citenamefont {Valiente-Dobon}, \citenamefont
  {Corradi}, \citenamefont {de~Angelis}, \citenamefont {Vedova}, \citenamefont
  {Fioretto}, \citenamefont {Gadea}, \citenamefont {Napoli}, \citenamefont
  {Orlandi}, \citenamefont {Recchia}, \citenamefont {Sahin}, \citenamefont
  {Silvestri}, \citenamefont {Stefanini}, \citenamefont {Singh}, \citenamefont
  {Szilner}, \citenamefont {Bazzacco}, \citenamefont {Farnea}, \citenamefont
  {Menegazzo}, \citenamefont {Gottardo}, \citenamefont {Lenzi}, \citenamefont
  {Lunardi}, \citenamefont {Montagnoli}, \citenamefont {Scarlassara},
  \citenamefont {Ur}, \citenamefont {Bianco}, \citenamefont {Zucchiatti},
  \citenamefont {Kmiecik}, \citenamefont {Maj}, \citenamefont {Meczynski},
  \citenamefont {Dewald}, \citenamefont {Pissulla},\ and\ \citenamefont
  {Pollarolo}}]{Mon11plb}%
  \BibitemOpen
  \bibfield  {author} {\bibinfo {author} {\bibfnamefont {D.}~\bibnamefont
  {Montanari}}, \bibinfo {author} {\bibfnamefont {S.}~\bibnamefont {Leoni}},
  \bibinfo {author} {\bibfnamefont {D.}~\bibnamefont {Mengoni}}, \bibinfo
  {author} {\bibfnamefont {G.}~\bibnamefont {Benzoni}}, \bibinfo {author}
  {\bibfnamefont {N.}~\bibnamefont {Blasi}}, \bibinfo {author} {\bibfnamefont
  {G.}~\bibnamefont {Bocchi}}, \bibinfo {author} {\bibfnamefont
  {P.}~\bibnamefont {Bortignon}}, \bibinfo {author} {\bibfnamefont
  {A.}~\bibnamefont {Bracco}}, \bibinfo {author} {\bibfnamefont
  {F.}~\bibnamefont {Camera}}, \bibinfo {author} {\bibfnamefont
  {G.}~\bibnamefont {ColÃ²}}, \bibinfo {author} {\bibfnamefont
  {A.}~\bibnamefont {Corsi}}, \bibinfo {author} {\bibfnamefont
  {F.}~\bibnamefont {Crespi}}, \bibinfo {author} {\bibfnamefont
  {B.}~\bibnamefont {Million}}, \bibinfo {author} {\bibfnamefont
  {R.}~\bibnamefont {Nicolini}}, \bibinfo {author} {\bibfnamefont
  {O.}~\bibnamefont {Wieland}}, \bibinfo {author} {\bibfnamefont
  {J.}~\bibnamefont {Valiente-Dobon}}, \bibinfo {author} {\bibfnamefont
  {L.}~\bibnamefont {Corradi}}, \bibinfo {author} {\bibfnamefont
  {G.}~\bibnamefont {de~Angelis}}, \bibinfo {author} {\bibfnamefont {F.~D.}\
  \bibnamefont {Vedova}}, \bibinfo {author} {\bibfnamefont {E.}~\bibnamefont
  {Fioretto}}, \bibinfo {author} {\bibfnamefont {A.}~\bibnamefont {Gadea}},
  \bibinfo {author} {\bibfnamefont {D.}~\bibnamefont {Napoli}}, \bibinfo
  {author} {\bibfnamefont {R.}~\bibnamefont {Orlandi}}, \bibinfo {author}
  {\bibfnamefont {F.}~\bibnamefont {Recchia}}, \bibinfo {author} {\bibfnamefont
  {E.}~\bibnamefont {Sahin}}, \bibinfo {author} {\bibfnamefont
  {R.}~\bibnamefont {Silvestri}}, \bibinfo {author} {\bibfnamefont
  {A.}~\bibnamefont {Stefanini}}, \bibinfo {author} {\bibfnamefont
  {R.}~\bibnamefont {Singh}}, \bibinfo {author} {\bibfnamefont
  {S.}~\bibnamefont {Szilner}}, \bibinfo {author} {\bibfnamefont
  {D.}~\bibnamefont {Bazzacco}}, \bibinfo {author} {\bibfnamefont
  {E.}~\bibnamefont {Farnea}}, \bibinfo {author} {\bibfnamefont
  {R.}~\bibnamefont {Menegazzo}}, \bibinfo {author} {\bibfnamefont
  {A.}~\bibnamefont {Gottardo}}, \bibinfo {author} {\bibfnamefont
  {S.}~\bibnamefont {Lenzi}}, \bibinfo {author} {\bibfnamefont
  {S.}~\bibnamefont {Lunardi}}, \bibinfo {author} {\bibfnamefont
  {G.}~\bibnamefont {Montagnoli}}, \bibinfo {author} {\bibfnamefont
  {F.}~\bibnamefont {Scarlassara}}, \bibinfo {author} {\bibfnamefont
  {C.}~\bibnamefont {Ur}}, \bibinfo {author} {\bibfnamefont {G.~L.}\
  \bibnamefont {Bianco}}, \bibinfo {author} {\bibfnamefont {A.}~\bibnamefont
  {Zucchiatti}}, \bibinfo {author} {\bibfnamefont {M.}~\bibnamefont {Kmiecik}},
  \bibinfo {author} {\bibfnamefont {A.}~\bibnamefont {Maj}}, \bibinfo {author}
  {\bibfnamefont {W.}~\bibnamefont {Meczynski}}, \bibinfo {author}
  {\bibfnamefont {A.}~\bibnamefont {Dewald}}, \bibinfo {author} {\bibfnamefont
  {T.}~\bibnamefont {Pissulla}}, \ and\ \bibinfo {author} {\bibfnamefont
  {G.}~\bibnamefont {Pollarolo}},\ }\href {\doibase
  http://doi.org/10.1016/j.physletb.2011.01.046} {\bibfield  {journal}
  {\bibinfo  {journal} {Phys. Lett. B}\ }\textbf {\bibinfo {volume} {697}},\
  \bibinfo {pages} {288 } (\bibinfo {year} {2011}{\natexlab{a}})}\BibitemShut
  {NoStop}%
\bibitem [{\citenamefont {Montanari}\ \emph
  {et~al.}(2011{\natexlab{b}})\citenamefont {Montanari}, \citenamefont {Leoni},
  \citenamefont {Corradi}, \citenamefont {Pollarolo}, \citenamefont {Benzoni},
  \citenamefont {Blasi}, \citenamefont {Bottoni}, \citenamefont {Bracco},
  \citenamefont {Camera}, \citenamefont {Corsi}, \citenamefont {Crespi},
  \citenamefont {Million}, \citenamefont {Nicolini}, \citenamefont {Wieland},
  \citenamefont {de~Angelis}, \citenamefont {Della~Vedova}, \citenamefont
  {Fioretto}, \citenamefont {Gadea}, \citenamefont {Guiot}, \citenamefont
  {Mengoni}, \citenamefont {Napoli}, \citenamefont {Orlandi}, \citenamefont
  {Recchia}, \citenamefont {Stefanini}, \citenamefont {Singh}, \citenamefont
  {Valiente-Dobon}, \citenamefont {Bazzacco}, \citenamefont {Farnea},
  \citenamefont {Lenzi}, \citenamefont {Lunardi}, \citenamefont {Montagnoli},
  \citenamefont {Scarlassara}, \citenamefont {Ur}, \citenamefont {Lo~Bianco},
  \citenamefont {Zucchiatti}, \citenamefont {Szilner}, \citenamefont {Kmiecik},
  \citenamefont {Maj},\ and\ \citenamefont {Meczynski}}]{Mon11}%
  \BibitemOpen
  \bibfield  {author} {\bibinfo {author} {\bibfnamefont {D.}~\bibnamefont
  {Montanari}}, \bibinfo {author} {\bibfnamefont {S.}~\bibnamefont {Leoni}},
  \bibinfo {author} {\bibfnamefont {L.}~\bibnamefont {Corradi}}, \bibinfo
  {author} {\bibfnamefont {G.}~\bibnamefont {Pollarolo}}, \bibinfo {author}
  {\bibfnamefont {G.}~\bibnamefont {Benzoni}}, \bibinfo {author} {\bibfnamefont
  {N.}~\bibnamefont {Blasi}}, \bibinfo {author} {\bibfnamefont
  {S.}~\bibnamefont {Bottoni}}, \bibinfo {author} {\bibfnamefont
  {A.}~\bibnamefont {Bracco}}, \bibinfo {author} {\bibfnamefont
  {F.}~\bibnamefont {Camera}}, \bibinfo {author} {\bibfnamefont
  {A.}~\bibnamefont {Corsi}}, \bibinfo {author} {\bibfnamefont {F.~C.~L.}\
  \bibnamefont {Crespi}}, \bibinfo {author} {\bibfnamefont {B.}~\bibnamefont
  {Million}}, \bibinfo {author} {\bibfnamefont {R.}~\bibnamefont {Nicolini}},
  \bibinfo {author} {\bibfnamefont {O.}~\bibnamefont {Wieland}}, \bibinfo
  {author} {\bibfnamefont {G.}~\bibnamefont {de~Angelis}}, \bibinfo {author}
  {\bibfnamefont {F.}~\bibnamefont {Della~Vedova}}, \bibinfo {author}
  {\bibfnamefont {E.}~\bibnamefont {Fioretto}}, \bibinfo {author}
  {\bibfnamefont {A.}~\bibnamefont {Gadea}}, \bibinfo {author} {\bibfnamefont
  {B.}~\bibnamefont {Guiot}}, \bibinfo {author} {\bibfnamefont
  {D.}~\bibnamefont {Mengoni}}, \bibinfo {author} {\bibfnamefont {D.~R.}\
  \bibnamefont {Napoli}}, \bibinfo {author} {\bibfnamefont {R.}~\bibnamefont
  {Orlandi}}, \bibinfo {author} {\bibfnamefont {F.}~\bibnamefont {Recchia}},
  \bibinfo {author} {\bibfnamefont {A.~M.}\ \bibnamefont {Stefanini}}, \bibinfo
  {author} {\bibfnamefont {R.~P.}\ \bibnamefont {Singh}}, \bibinfo {author}
  {\bibfnamefont {J.~J.}\ \bibnamefont {Valiente-Dobon}}, \bibinfo {author}
  {\bibfnamefont {D.}~\bibnamefont {Bazzacco}}, \bibinfo {author}
  {\bibfnamefont {E.}~\bibnamefont {Farnea}}, \bibinfo {author} {\bibfnamefont
  {S.~M.}\ \bibnamefont {Lenzi}}, \bibinfo {author} {\bibfnamefont
  {S.}~\bibnamefont {Lunardi}}, \bibinfo {author} {\bibfnamefont
  {G.}~\bibnamefont {Montagnoli}}, \bibinfo {author} {\bibfnamefont
  {F.}~\bibnamefont {Scarlassara}}, \bibinfo {author} {\bibfnamefont
  {C.}~\bibnamefont {Ur}}, \bibinfo {author} {\bibfnamefont {G.}~\bibnamefont
  {Lo~Bianco}}, \bibinfo {author} {\bibfnamefont {A.}~\bibnamefont
  {Zucchiatti}}, \bibinfo {author} {\bibfnamefont {S.}~\bibnamefont {Szilner}},
  \bibinfo {author} {\bibfnamefont {M.}~\bibnamefont {Kmiecik}}, \bibinfo
  {author} {\bibfnamefont {A.}~\bibnamefont {Maj}}, \ and\ \bibinfo {author}
  {\bibfnamefont {W.}~\bibnamefont {Meczynski}},\ }\href {\doibase
  10.1103/PhysRevC.84.054613} {\bibfield  {journal} {\bibinfo  {journal} {Phys.
  Rev. C}\ }\textbf {\bibinfo {volume} {84}},\ \bibinfo {pages} {054613}
  (\bibinfo {year} {2011}{\natexlab{b}})}\BibitemShut {NoStop}%
\bibitem [{\citenamefont {Montanari}\ \emph {et~al.}(2012)\citenamefont
  {Montanari}, \citenamefont {Leoni}, \citenamefont {Mengoni}, \citenamefont
  {Valiente-Dobon}, \citenamefont {Benzoni}, \citenamefont {Blasi},
  \citenamefont {Bocchi}, \citenamefont {Bortignon}, \citenamefont {Bottoni},
  \citenamefont {Bracco}, \citenamefont {Camera}, \citenamefont {Casati},
  \citenamefont {Col\`o}, \citenamefont {Corsi}, \citenamefont {Crespi},
  \citenamefont {Million}, \citenamefont {Nicolini}, \citenamefont {Wieland},
  \citenamefont {Bazzacco}, \citenamefont {Farnea}, \citenamefont {Germogli},
  \citenamefont {Gottardo}, \citenamefont {Lenzi}, \citenamefont {Lunardi},
  \citenamefont {Menegazzo}, \citenamefont {Montagnoli}, \citenamefont
  {Recchia}, \citenamefont {Scarlassara}, \citenamefont {Ur}, \citenamefont
  {Corradi}, \citenamefont {de~Angelis}, \citenamefont {Fioretto},
  \citenamefont {Napoli}, \citenamefont {Orlandi}, \citenamefont {Sahin},
  \citenamefont {Stefanini}, \citenamefont {Singh}, \citenamefont {Gadea},
  \citenamefont {Szilner}, \citenamefont {Kmiecik}, \citenamefont {Maj},
  \citenamefont {Meczynski}, \citenamefont {Dewald}, \citenamefont {Pissulla},\
  and\ \citenamefont {Pollarolo}}]{Mon12}%
  \BibitemOpen
  \bibfield  {author} {\bibinfo {author} {\bibfnamefont {D.}~\bibnamefont
  {Montanari}}, \bibinfo {author} {\bibfnamefont {S.}~\bibnamefont {Leoni}},
  \bibinfo {author} {\bibfnamefont {D.}~\bibnamefont {Mengoni}}, \bibinfo
  {author} {\bibfnamefont {J.~J.}\ \bibnamefont {Valiente-Dobon}}, \bibinfo
  {author} {\bibfnamefont {G.}~\bibnamefont {Benzoni}}, \bibinfo {author}
  {\bibfnamefont {N.}~\bibnamefont {Blasi}}, \bibinfo {author} {\bibfnamefont
  {G.}~\bibnamefont {Bocchi}}, \bibinfo {author} {\bibfnamefont {P.~F.}\
  \bibnamefont {Bortignon}}, \bibinfo {author} {\bibfnamefont {S.}~\bibnamefont
  {Bottoni}}, \bibinfo {author} {\bibfnamefont {A.}~\bibnamefont {Bracco}},
  \bibinfo {author} {\bibfnamefont {F.}~\bibnamefont {Camera}}, \bibinfo
  {author} {\bibfnamefont {P.}~\bibnamefont {Casati}}, \bibinfo {author}
  {\bibfnamefont {G.}~\bibnamefont {Col\`o}}, \bibinfo {author} {\bibfnamefont
  {A.}~\bibnamefont {Corsi}}, \bibinfo {author} {\bibfnamefont {F.~C.~L.}\
  \bibnamefont {Crespi}}, \bibinfo {author} {\bibfnamefont {B.}~\bibnamefont
  {Million}}, \bibinfo {author} {\bibfnamefont {R.}~\bibnamefont {Nicolini}},
  \bibinfo {author} {\bibfnamefont {O.}~\bibnamefont {Wieland}}, \bibinfo
  {author} {\bibfnamefont {D.}~\bibnamefont {Bazzacco}}, \bibinfo {author}
  {\bibfnamefont {E.}~\bibnamefont {Farnea}}, \bibinfo {author} {\bibfnamefont
  {G.}~\bibnamefont {Germogli}}, \bibinfo {author} {\bibfnamefont
  {A.}~\bibnamefont {Gottardo}}, \bibinfo {author} {\bibfnamefont {S.~M.}\
  \bibnamefont {Lenzi}}, \bibinfo {author} {\bibfnamefont {S.}~\bibnamefont
  {Lunardi}}, \bibinfo {author} {\bibfnamefont {R.}~\bibnamefont {Menegazzo}},
  \bibinfo {author} {\bibfnamefont {G.}~\bibnamefont {Montagnoli}}, \bibinfo
  {author} {\bibfnamefont {F.}~\bibnamefont {Recchia}}, \bibinfo {author}
  {\bibfnamefont {F.}~\bibnamefont {Scarlassara}}, \bibinfo {author}
  {\bibfnamefont {C.}~\bibnamefont {Ur}}, \bibinfo {author} {\bibfnamefont
  {L.}~\bibnamefont {Corradi}}, \bibinfo {author} {\bibfnamefont
  {G.}~\bibnamefont {de~Angelis}}, \bibinfo {author} {\bibfnamefont
  {E.}~\bibnamefont {Fioretto}}, \bibinfo {author} {\bibfnamefont {D.~R.}\
  \bibnamefont {Napoli}}, \bibinfo {author} {\bibfnamefont {R.}~\bibnamefont
  {Orlandi}}, \bibinfo {author} {\bibfnamefont {E.}~\bibnamefont {Sahin}},
  \bibinfo {author} {\bibfnamefont {A.~M.}\ \bibnamefont {Stefanini}}, \bibinfo
  {author} {\bibfnamefont {R.~P.}\ \bibnamefont {Singh}}, \bibinfo {author}
  {\bibfnamefont {A.}~\bibnamefont {Gadea}}, \bibinfo {author} {\bibfnamefont
  {S.}~\bibnamefont {Szilner}}, \bibinfo {author} {\bibfnamefont
  {M.}~\bibnamefont {Kmiecik}}, \bibinfo {author} {\bibfnamefont
  {A.}~\bibnamefont {Maj}}, \bibinfo {author} {\bibfnamefont {W.}~\bibnamefont
  {Meczynski}}, \bibinfo {author} {\bibfnamefont {A.}~\bibnamefont {Dewald}},
  \bibinfo {author} {\bibfnamefont {T.}~\bibnamefont {Pissulla}}, \ and\
  \bibinfo {author} {\bibfnamefont {G.}~\bibnamefont {Pollarolo}},\ }\href
  {\doibase 10.1103/PhysRevC.85.044301} {\bibfield  {journal} {\bibinfo
  {journal} {Phys. Rev. C}\ }\textbf {\bibinfo {volume} {85}},\ \bibinfo
  {pages} {044301} (\bibinfo {year} {2012})}\BibitemShut {NoStop}%
\bibitem [{\citenamefont {Garcia~Ruiz}\ \emph {et~al.}(2015)\citenamefont
  {Garcia~Ruiz}, \citenamefont {Bissell}, \citenamefont {Blaum}, \citenamefont
  {Fr\"ommgen}, \citenamefont {Hammen}, \citenamefont {Holt}, \citenamefont
  {Kowalska}, \citenamefont {Kreim}, \citenamefont {Men\'endez}, \citenamefont
  {Neugart}, \citenamefont {Neyens}, \citenamefont {N\"ortersh\"auser},
  \citenamefont {Nowacki}, \citenamefont {Papuga}, \citenamefont {Poves},
  \citenamefont {Schwenk}, \citenamefont {Simonis},\ and\ \citenamefont
  {Yordanov}}]{Gar15}%
  \BibitemOpen
  \bibfield  {author} {\bibinfo {author} {\bibfnamefont {R.~F.}\ \bibnamefont
  {Garcia~Ruiz}}, \bibinfo {author} {\bibfnamefont {M.~L.}\ \bibnamefont
  {Bissell}}, \bibinfo {author} {\bibfnamefont {K.}~\bibnamefont {Blaum}},
  \bibinfo {author} {\bibfnamefont {N.}~\bibnamefont {Fr\"ommgen}}, \bibinfo
  {author} {\bibfnamefont {M.}~\bibnamefont {Hammen}}, \bibinfo {author}
  {\bibfnamefont {J.~D.}\ \bibnamefont {Holt}}, \bibinfo {author}
  {\bibfnamefont {M.}~\bibnamefont {Kowalska}}, \bibinfo {author}
  {\bibfnamefont {K.}~\bibnamefont {Kreim}}, \bibinfo {author} {\bibfnamefont
  {J.}~\bibnamefont {Men\'endez}}, \bibinfo {author} {\bibfnamefont
  {R.}~\bibnamefont {Neugart}}, \bibinfo {author} {\bibfnamefont
  {G.}~\bibnamefont {Neyens}}, \bibinfo {author} {\bibfnamefont
  {W.}~\bibnamefont {N\"ortersh\"auser}}, \bibinfo {author} {\bibfnamefont
  {F.}~\bibnamefont {Nowacki}}, \bibinfo {author} {\bibfnamefont
  {J.}~\bibnamefont {Papuga}}, \bibinfo {author} {\bibfnamefont
  {A.}~\bibnamefont {Poves}}, \bibinfo {author} {\bibfnamefont
  {A.}~\bibnamefont {Schwenk}}, \bibinfo {author} {\bibfnamefont
  {J.}~\bibnamefont {Simonis}}, \ and\ \bibinfo {author} {\bibfnamefont
  {D.~T.}\ \bibnamefont {Yordanov}},\ }\href {\doibase
  10.1103/PhysRevC.91.041304} {\bibfield  {journal} {\bibinfo  {journal} {Phys.
  Rev. C}\ }\textbf {\bibinfo {volume} {91}},\ \bibinfo {pages} {041304}
  (\bibinfo {year} {2015})}\BibitemShut {NoStop}%
\bibitem [{\citenamefont {Noji}\ \emph {et~al.}(2015)\citenamefont {Noji},
  \citenamefont {Zegers}, \citenamefont {Austin}, \citenamefont {Baugher},
  \citenamefont {Bazin}, \citenamefont {Brown}, \citenamefont {Campbell},
  \citenamefont {Cole}, \citenamefont {Doster}, \citenamefont {Gade},
  \citenamefont {Guess}, \citenamefont {Gupta}, \citenamefont {Hitt},
  \citenamefont {Langer}, \citenamefont {Lipschutz}, \citenamefont
  {Lunderberg}, \citenamefont {Meharchand}, \citenamefont {Meisel},
  \citenamefont {Perdikakis}, \citenamefont {Pereira}, \citenamefont {Recchia},
  \citenamefont {Schatz}, \citenamefont {Scott}, \citenamefont {Stroberg},
  \citenamefont {Sullivan}, \citenamefont {Valdez}, \citenamefont {Walz},
  \citenamefont {Weisshaar}, \citenamefont {Williams},\ and\ \citenamefont
  {Wimmer}}]{Noj15}%
  \BibitemOpen
  \bibfield  {author} {\bibinfo {author} {\bibfnamefont {S.}~\bibnamefont
  {Noji}}, \bibinfo {author} {\bibfnamefont {R.~G.~T.}\ \bibnamefont {Zegers}},
  \bibinfo {author} {\bibfnamefont {S.~M.}\ \bibnamefont {Austin}}, \bibinfo
  {author} {\bibfnamefont {T.}~\bibnamefont {Baugher}}, \bibinfo {author}
  {\bibfnamefont {D.}~\bibnamefont {Bazin}}, \bibinfo {author} {\bibfnamefont
  {B.~A.}\ \bibnamefont {Brown}}, \bibinfo {author} {\bibfnamefont {C.~M.}\
  \bibnamefont {Campbell}}, \bibinfo {author} {\bibfnamefont {A.~L.}\
  \bibnamefont {Cole}}, \bibinfo {author} {\bibfnamefont {H.~J.}\ \bibnamefont
  {Doster}}, \bibinfo {author} {\bibfnamefont {A.}~\bibnamefont {Gade}},
  \bibinfo {author} {\bibfnamefont {C.~J.}\ \bibnamefont {Guess}}, \bibinfo
  {author} {\bibfnamefont {S.}~\bibnamefont {Gupta}}, \bibinfo {author}
  {\bibfnamefont {G.~W.}\ \bibnamefont {Hitt}}, \bibinfo {author}
  {\bibfnamefont {C.}~\bibnamefont {Langer}}, \bibinfo {author} {\bibfnamefont
  {S.}~\bibnamefont {Lipschutz}}, \bibinfo {author} {\bibfnamefont
  {E.}~\bibnamefont {Lunderberg}}, \bibinfo {author} {\bibfnamefont
  {R.}~\bibnamefont {Meharchand}}, \bibinfo {author} {\bibfnamefont
  {Z.}~\bibnamefont {Meisel}}, \bibinfo {author} {\bibfnamefont
  {G.}~\bibnamefont {Perdikakis}}, \bibinfo {author} {\bibfnamefont
  {J.}~\bibnamefont {Pereira}}, \bibinfo {author} {\bibfnamefont
  {F.}~\bibnamefont {Recchia}}, \bibinfo {author} {\bibfnamefont
  {H.}~\bibnamefont {Schatz}}, \bibinfo {author} {\bibfnamefont
  {M.}~\bibnamefont {Scott}}, \bibinfo {author} {\bibfnamefont {S.~R.}\
  \bibnamefont {Stroberg}}, \bibinfo {author} {\bibfnamefont {C.}~\bibnamefont
  {Sullivan}}, \bibinfo {author} {\bibfnamefont {L.}~\bibnamefont {Valdez}},
  \bibinfo {author} {\bibfnamefont {C.}~\bibnamefont {Walz}}, \bibinfo {author}
  {\bibfnamefont {D.}~\bibnamefont {Weisshaar}}, \bibinfo {author}
  {\bibfnamefont {S.~J.}\ \bibnamefont {Williams}}, \ and\ \bibinfo {author}
  {\bibfnamefont {K.}~\bibnamefont {Wimmer}},\ }\href {\doibase
  10.1103/PhysRevC.92.024312} {\bibfield  {journal} {\bibinfo  {journal} {Phys.
  Rev. C}\ }\textbf {\bibinfo {volume} {92}},\ \bibinfo {pages} {024312}
  (\bibinfo {year} {2015})}\BibitemShut {NoStop}%
\bibitem [{\citenamefont {Gade}\ \emph {et~al.}(2016)\citenamefont {Gade},
  \citenamefont {Tostevin}, \citenamefont {Bader}, \citenamefont {Baugher},
  \citenamefont {Bazin}, \citenamefont {Berryman}, \citenamefont {Brown},
  \citenamefont {Hartley}, \citenamefont {Lunderberg}, \citenamefont {Recchia},
  \citenamefont {Stroberg}, \citenamefont {Utsuno}, \citenamefont {Weisshaar},\
  and\ \citenamefont {Wimmer}}]{Gad16}%
  \BibitemOpen
  \bibfield  {author} {\bibinfo {author} {\bibfnamefont {A.}~\bibnamefont
  {Gade}}, \bibinfo {author} {\bibfnamefont {J.~A.}\ \bibnamefont {Tostevin}},
  \bibinfo {author} {\bibfnamefont {V.}~\bibnamefont {Bader}}, \bibinfo
  {author} {\bibfnamefont {T.}~\bibnamefont {Baugher}}, \bibinfo {author}
  {\bibfnamefont {D.}~\bibnamefont {Bazin}}, \bibinfo {author} {\bibfnamefont
  {J.~S.}\ \bibnamefont {Berryman}}, \bibinfo {author} {\bibfnamefont {B.~A.}\
  \bibnamefont {Brown}}, \bibinfo {author} {\bibfnamefont {D.~J.}\ \bibnamefont
  {Hartley}}, \bibinfo {author} {\bibfnamefont {E.}~\bibnamefont {Lunderberg}},
  \bibinfo {author} {\bibfnamefont {F.}~\bibnamefont {Recchia}}, \bibinfo
  {author} {\bibfnamefont {S.~R.}\ \bibnamefont {Stroberg}}, \bibinfo {author}
  {\bibfnamefont {Y.}~\bibnamefont {Utsuno}}, \bibinfo {author} {\bibfnamefont
  {D.}~\bibnamefont {Weisshaar}}, \ and\ \bibinfo {author} {\bibfnamefont
  {K.}~\bibnamefont {Wimmer}},\ }\href {\doibase 10.1103/PhysRevC.93.031601}
  {\bibfield  {journal} {\bibinfo  {journal} {Phys. Rev. C}\ }\textbf {\bibinfo
  {volume} {93}},\ \bibinfo {pages} {031601(R)} (\bibinfo {year}
  {2016})}\BibitemShut {NoStop}%
\bibitem [{\citenamefont {Hady\ifmmode \acute{n}\else
  \'{n}\fi{}ska-Kl\ifmmode~\mbox{\c{e}}\else \c{e}\fi{}k}\ \emph
  {et~al.}(2016)\citenamefont {Hady\ifmmode \acute{n}\else
  \'{n}\fi{}ska-Kl\ifmmode~\mbox{\c{e}}\else \c{e}\fi{}k}, \citenamefont
  {Napiorkowski}, \citenamefont {Zieli\ifmmode~\acute{n}\else \'{n}\fi{}ska},
  \citenamefont {Srebrny}, \citenamefont {Maj}, \citenamefont {Azaiez},
  \citenamefont {Valiente~Dob\'on}, \citenamefont {Kici\ifmmode
  \acute{n}\else~\'{n}\fi{}ska Habior}, \citenamefont {Nowacki}, \citenamefont
  {Na\"{\i}dja}, \citenamefont {Bounthong}, \citenamefont {Rodr\'{\i}guez},
  \citenamefont {de~Angelis}, \citenamefont {Abraham}, \citenamefont
  {Anil~Kumar}, \citenamefont {Bazzacco}, \citenamefont {Bellato},
  \citenamefont {Bortolato}, \citenamefont {Bednarczyk}, \citenamefont
  {Benzoni}, \citenamefont {Berti}, \citenamefont {Birkenbach}, \citenamefont
  {Bruyneel}, \citenamefont {Brambilla}, \citenamefont {Camera}, \citenamefont
  {Chavas}, \citenamefont {Cederwall}, \citenamefont {Charles}, \citenamefont
  {Ciema\l{}a}, \citenamefont {Cocconi}, \citenamefont {Coleman-Smith},
  \citenamefont {Colombo}, \citenamefont {Corsi}, \citenamefont {Crespi},
  \citenamefont {Cullen}, \citenamefont {Czermak}, \citenamefont
  {D\'esesquelles}, \citenamefont {Doherty}, \citenamefont {Dulny},
  \citenamefont {Eberth}, \citenamefont {Farnea}, \citenamefont {Fornal},
  \citenamefont {Franchoo}, \citenamefont {Gadea}, \citenamefont {Giaz},
  \citenamefont {Gottardo}, \citenamefont {Grave}, \citenamefont
  {Gr\ifmmode~\mbox{\c{e}}\else \c{e}\fi{}bosz}, \citenamefont {G\"orgen},
  \citenamefont {Gulmini}, \citenamefont {Habermann}, \citenamefont {Hess},
  \citenamefont {Isocrate}, \citenamefont {Iwanicki}, \citenamefont {Jaworski},
  \citenamefont {Judson}, \citenamefont {Jungclaus}, \citenamefont {Karkour},
  \citenamefont {Kmiecik}, \citenamefont {Karpi\ifmmode~\acute{n}\else
  \'{n}\fi{}ski}, \citenamefont {Kisieli\ifmmode~\acute{n}\else \'{n}\fi{}ski},
  \citenamefont {Kondratyev}, \citenamefont {Korichi}, \citenamefont
  {Komorowska}, \citenamefont {Kowalczyk}, \citenamefont {Korten},
  \citenamefont {Krzysiek}, \citenamefont {Lehaut}, \citenamefont {Leoni},
  \citenamefont {Ljungvall}, \citenamefont {Lopez-Martens}, \citenamefont
  {Lunardi}, \citenamefont {Maron}, \citenamefont {Mazurek}, \citenamefont
  {Menegazzo}, \citenamefont {Mengoni}, \citenamefont {Merch\'an},
  \citenamefont {M\ifmmode \mbox{\c{e}}\else
  \c{e}\fi{}czy\ifmmode~\acute{n}\else \'{n}\fi{}ski}, \citenamefont
  {Michelagnoli}, \citenamefont {Mierzejewski}, \citenamefont {Million},
  \citenamefont {Myalski}, \citenamefont {Napoli}, \citenamefont {Nicolini},
  \citenamefont {Niikura}, \citenamefont {Obertelli}, \citenamefont {\"Ozmen},
  \citenamefont {Palacz}, \citenamefont {Pr\'ochniak}, \citenamefont {Pullia},
  \citenamefont {Quintana}, \citenamefont {Rampazzo}, \citenamefont {Recchia},
  \citenamefont {Redon}, \citenamefont {Reiter}, \citenamefont {Rosso},
  \citenamefont {Rusek}, \citenamefont {Sahin}, \citenamefont {Salsac},
  \citenamefont {S\"oderstr\"om}, \citenamefont {Stefan}, \citenamefont
  {St\'ezowski}, \citenamefont {Stycze\ifmmode~\acute{n}\else \'{n}\fi{}},
  \citenamefont {Theisen}, \citenamefont {Toniolo}, \citenamefont {Ur},
  \citenamefont {Vandone}, \citenamefont {Wadsworth}, \citenamefont
  {Wasilewska}, \citenamefont {Wiens}, \citenamefont {Wood}, \citenamefont
  {Wrzosek-Lipska},\ and\ \citenamefont {Zi\ifmmode \mbox{\c{e}}\else
  \c{e}\fi{}bli\ifmmode~\acute{n}\else \'{n}\fi{}ski}}]{Had16}%
  \BibitemOpen
  \bibfield  {author} {\bibinfo {author} {\bibfnamefont {K.}~\bibnamefont
  {Hady\ifmmode \acute{n}\else \'{n}\fi{}ska-Kl\ifmmode~\mbox{\c{e}}\else
  \c{e}\fi{}k}}, \bibinfo {author} {\bibfnamefont {P.~J.}\ \bibnamefont
  {Napiorkowski}}, \bibinfo {author} {\bibfnamefont {M.}~\bibnamefont
  {Zieli\ifmmode~\acute{n}\else \'{n}\fi{}ska}}, \bibinfo {author}
  {\bibfnamefont {J.}~\bibnamefont {Srebrny}}, \bibinfo {author} {\bibfnamefont
  {A.}~\bibnamefont {Maj}}, \bibinfo {author} {\bibfnamefont {F.}~\bibnamefont
  {Azaiez}}, \bibinfo {author} {\bibfnamefont {J.~J.}\ \bibnamefont
  {Valiente~Dob\'on}}, \bibinfo {author} {\bibfnamefont {M.}~\bibnamefont
  {Kici\ifmmode \acute{n}\else~\'{n}\fi{}ska Habior}}, \bibinfo {author}
  {\bibfnamefont {F.}~\bibnamefont {Nowacki}}, \bibinfo {author} {\bibfnamefont
  {H.}~\bibnamefont {Na\"{\i}dja}}, \bibinfo {author} {\bibfnamefont
  {B.}~\bibnamefont {Bounthong}}, \bibinfo {author} {\bibfnamefont {T.~R.}\
  \bibnamefont {Rodr\'{\i}guez}}, \bibinfo {author} {\bibfnamefont
  {G.}~\bibnamefont {de~Angelis}}, \bibinfo {author} {\bibfnamefont
  {T.}~\bibnamefont {Abraham}}, \bibinfo {author} {\bibfnamefont
  {G.}~\bibnamefont {Anil~Kumar}}, \bibinfo {author} {\bibfnamefont
  {D.}~\bibnamefont {Bazzacco}}, \bibinfo {author} {\bibfnamefont
  {M.}~\bibnamefont {Bellato}}, \bibinfo {author} {\bibfnamefont
  {D.}~\bibnamefont {Bortolato}}, \bibinfo {author} {\bibfnamefont
  {P.}~\bibnamefont {Bednarczyk}}, \bibinfo {author} {\bibfnamefont
  {G.}~\bibnamefont {Benzoni}}, \bibinfo {author} {\bibfnamefont
  {L.}~\bibnamefont {Berti}}, \bibinfo {author} {\bibfnamefont
  {B.}~\bibnamefont {Birkenbach}}, \bibinfo {author} {\bibfnamefont
  {B.}~\bibnamefont {Bruyneel}}, \bibinfo {author} {\bibfnamefont
  {S.}~\bibnamefont {Brambilla}}, \bibinfo {author} {\bibfnamefont
  {F.}~\bibnamefont {Camera}}, \bibinfo {author} {\bibfnamefont
  {J.}~\bibnamefont {Chavas}}, \bibinfo {author} {\bibfnamefont
  {B.}~\bibnamefont {Cederwall}}, \bibinfo {author} {\bibfnamefont
  {L.}~\bibnamefont {Charles}}, \bibinfo {author} {\bibfnamefont
  {M.}~\bibnamefont {Ciema\l{}a}}, \bibinfo {author} {\bibfnamefont
  {P.}~\bibnamefont {Cocconi}}, \bibinfo {author} {\bibfnamefont
  {P.}~\bibnamefont {Coleman-Smith}}, \bibinfo {author} {\bibfnamefont
  {A.}~\bibnamefont {Colombo}}, \bibinfo {author} {\bibfnamefont
  {A.}~\bibnamefont {Corsi}}, \bibinfo {author} {\bibfnamefont {F.~C.~L.}\
  \bibnamefont {Crespi}}, \bibinfo {author} {\bibfnamefont {D.~M.}\
  \bibnamefont {Cullen}}, \bibinfo {author} {\bibfnamefont {A.}~\bibnamefont
  {Czermak}}, \bibinfo {author} {\bibfnamefont {P.}~\bibnamefont
  {D\'esesquelles}}, \bibinfo {author} {\bibfnamefont {D.~T.}\ \bibnamefont
  {Doherty}}, \bibinfo {author} {\bibfnamefont {B.}~\bibnamefont {Dulny}},
  \bibinfo {author} {\bibfnamefont {J.}~\bibnamefont {Eberth}}, \bibinfo
  {author} {\bibfnamefont {E.}~\bibnamefont {Farnea}}, \bibinfo {author}
  {\bibfnamefont {B.}~\bibnamefont {Fornal}}, \bibinfo {author} {\bibfnamefont
  {S.}~\bibnamefont {Franchoo}}, \bibinfo {author} {\bibfnamefont
  {A.}~\bibnamefont {Gadea}}, \bibinfo {author} {\bibfnamefont
  {A.}~\bibnamefont {Giaz}}, \bibinfo {author} {\bibfnamefont {A.}~\bibnamefont
  {Gottardo}}, \bibinfo {author} {\bibfnamefont {X.}~\bibnamefont {Grave}},
  \bibinfo {author} {\bibfnamefont {J.}~\bibnamefont
  {Gr\ifmmode~\mbox{\c{e}}\else \c{e}\fi{}bosz}}, \bibinfo {author}
  {\bibfnamefont {A.}~\bibnamefont {G\"orgen}}, \bibinfo {author}
  {\bibfnamefont {M.}~\bibnamefont {Gulmini}}, \bibinfo {author} {\bibfnamefont
  {T.}~\bibnamefont {Habermann}}, \bibinfo {author} {\bibfnamefont
  {H.}~\bibnamefont {Hess}}, \bibinfo {author} {\bibfnamefont {R.}~\bibnamefont
  {Isocrate}}, \bibinfo {author} {\bibfnamefont {J.}~\bibnamefont {Iwanicki}},
  \bibinfo {author} {\bibfnamefont {G.}~\bibnamefont {Jaworski}}, \bibinfo
  {author} {\bibfnamefont {D.~S.}\ \bibnamefont {Judson}}, \bibinfo {author}
  {\bibfnamefont {A.}~\bibnamefont {Jungclaus}}, \bibinfo {author}
  {\bibfnamefont {N.}~\bibnamefont {Karkour}}, \bibinfo {author} {\bibfnamefont
  {M.}~\bibnamefont {Kmiecik}}, \bibinfo {author} {\bibfnamefont
  {D.}~\bibnamefont {Karpi\ifmmode~\acute{n}\else \'{n}\fi{}ski}}, \bibinfo
  {author} {\bibfnamefont {M.}~\bibnamefont {Kisieli\ifmmode~\acute{n}\else
  \'{n}\fi{}ski}}, \bibinfo {author} {\bibfnamefont {N.}~\bibnamefont
  {Kondratyev}}, \bibinfo {author} {\bibfnamefont {A.}~\bibnamefont {Korichi}},
  \bibinfo {author} {\bibfnamefont {M.}~\bibnamefont {Komorowska}}, \bibinfo
  {author} {\bibfnamefont {M.}~\bibnamefont {Kowalczyk}}, \bibinfo {author}
  {\bibfnamefont {W.}~\bibnamefont {Korten}}, \bibinfo {author} {\bibfnamefont
  {M.}~\bibnamefont {Krzysiek}}, \bibinfo {author} {\bibfnamefont
  {G.}~\bibnamefont {Lehaut}}, \bibinfo {author} {\bibfnamefont
  {S.}~\bibnamefont {Leoni}}, \bibinfo {author} {\bibfnamefont
  {J.}~\bibnamefont {Ljungvall}}, \bibinfo {author} {\bibfnamefont
  {A.}~\bibnamefont {Lopez-Martens}}, \bibinfo {author} {\bibfnamefont
  {S.}~\bibnamefont {Lunardi}}, \bibinfo {author} {\bibfnamefont
  {G.}~\bibnamefont {Maron}}, \bibinfo {author} {\bibfnamefont
  {K.}~\bibnamefont {Mazurek}}, \bibinfo {author} {\bibfnamefont
  {R.}~\bibnamefont {Menegazzo}}, \bibinfo {author} {\bibfnamefont
  {D.}~\bibnamefont {Mengoni}}, \bibinfo {author} {\bibfnamefont
  {E.}~\bibnamefont {Merch\'an}}, \bibinfo {author} {\bibfnamefont
  {W.}~\bibnamefont {M\ifmmode \mbox{\c{e}}\else
  \c{e}\fi{}czy\ifmmode~\acute{n}\else \'{n}\fi{}ski}}, \bibinfo {author}
  {\bibfnamefont {C.}~\bibnamefont {Michelagnoli}}, \bibinfo {author}
  {\bibfnamefont {J.}~\bibnamefont {Mierzejewski}}, \bibinfo {author}
  {\bibfnamefont {B.}~\bibnamefont {Million}}, \bibinfo {author} {\bibfnamefont
  {S.}~\bibnamefont {Myalski}}, \bibinfo {author} {\bibfnamefont {D.~R.}\
  \bibnamefont {Napoli}}, \bibinfo {author} {\bibfnamefont {R.}~\bibnamefont
  {Nicolini}}, \bibinfo {author} {\bibfnamefont {M.}~\bibnamefont {Niikura}},
  \bibinfo {author} {\bibfnamefont {A.}~\bibnamefont {Obertelli}}, \bibinfo
  {author} {\bibfnamefont {S.~F.}\ \bibnamefont {\"Ozmen}}, \bibinfo {author}
  {\bibfnamefont {M.}~\bibnamefont {Palacz}}, \bibinfo {author} {\bibfnamefont
  {L.}~\bibnamefont {Pr\'ochniak}}, \bibinfo {author} {\bibfnamefont
  {A.}~\bibnamefont {Pullia}}, \bibinfo {author} {\bibfnamefont
  {B.}~\bibnamefont {Quintana}}, \bibinfo {author} {\bibfnamefont
  {G.}~\bibnamefont {Rampazzo}}, \bibinfo {author} {\bibfnamefont
  {F.}~\bibnamefont {Recchia}}, \bibinfo {author} {\bibfnamefont
  {N.}~\bibnamefont {Redon}}, \bibinfo {author} {\bibfnamefont
  {P.}~\bibnamefont {Reiter}}, \bibinfo {author} {\bibfnamefont
  {D.}~\bibnamefont {Rosso}}, \bibinfo {author} {\bibfnamefont
  {K.}~\bibnamefont {Rusek}}, \bibinfo {author} {\bibfnamefont
  {E.}~\bibnamefont {Sahin}}, \bibinfo {author} {\bibfnamefont {M.-D.}\
  \bibnamefont {Salsac}}, \bibinfo {author} {\bibfnamefont {P.-A.}\
  \bibnamefont {S\"oderstr\"om}}, \bibinfo {author} {\bibfnamefont
  {I.}~\bibnamefont {Stefan}}, \bibinfo {author} {\bibfnamefont
  {O.}~\bibnamefont {St\'ezowski}}, \bibinfo {author} {\bibfnamefont
  {J.}~\bibnamefont {Stycze\ifmmode~\acute{n}\else \'{n}\fi{}}}, \bibinfo
  {author} {\bibfnamefont {C.}~\bibnamefont {Theisen}}, \bibinfo {author}
  {\bibfnamefont {N.}~\bibnamefont {Toniolo}}, \bibinfo {author} {\bibfnamefont
  {C.~A.}\ \bibnamefont {Ur}}, \bibinfo {author} {\bibfnamefont
  {V.}~\bibnamefont {Vandone}}, \bibinfo {author} {\bibfnamefont
  {R.}~\bibnamefont {Wadsworth}}, \bibinfo {author} {\bibfnamefont
  {B.}~\bibnamefont {Wasilewska}}, \bibinfo {author} {\bibfnamefont
  {A.}~\bibnamefont {Wiens}}, \bibinfo {author} {\bibfnamefont {J.~L.}\
  \bibnamefont {Wood}}, \bibinfo {author} {\bibfnamefont {K.}~\bibnamefont
  {Wrzosek-Lipska}}, \ and\ \bibinfo {author} {\bibfnamefont {M.}~\bibnamefont
  {Zi\ifmmode \mbox{\c{e}}\else \c{e}\fi{}bli\ifmmode~\acute{n}\else
  \'{n}\fi{}ski}},\ }\href {\doibase 10.1103/PhysRevLett.117.062501} {\bibfield
   {journal} {\bibinfo  {journal} {Phys. Rev. Lett.}\ }\textbf {\bibinfo
  {volume} {117}},\ \bibinfo {pages} {062501} (\bibinfo {year}
  {2016})}\BibitemShut {NoStop}%
\bibitem [{\citenamefont {Riley}\ \emph {et~al.}(2016)\citenamefont {Riley},
  \citenamefont {McPherson}, \citenamefont {Agiorgousis}, \citenamefont
  {Baugher}, \citenamefont {Bazin}, \citenamefont {Bowry}, \citenamefont
  {Cottle}, \citenamefont {DeVone}, \citenamefont {Gade}, \citenamefont
  {Glowacki}, \citenamefont {Gregory}, \citenamefont {Haldeman}, \citenamefont
  {Kemper}, \citenamefont {Lunderberg}, \citenamefont {Noji}, \citenamefont
  {Recchia}, \citenamefont {Sadler}, \citenamefont {Scott}, \citenamefont
  {Weisshaar},\ and\ \citenamefont {Zegers}}]{Ril16}%
  \BibitemOpen
  \bibfield  {author} {\bibinfo {author} {\bibfnamefont {L.~A.}\ \bibnamefont
  {Riley}}, \bibinfo {author} {\bibfnamefont {D.~M.}\ \bibnamefont
  {McPherson}}, \bibinfo {author} {\bibfnamefont {M.~L.}\ \bibnamefont
  {Agiorgousis}}, \bibinfo {author} {\bibfnamefont {T.~R.}\ \bibnamefont
  {Baugher}}, \bibinfo {author} {\bibfnamefont {D.}~\bibnamefont {Bazin}},
  \bibinfo {author} {\bibfnamefont {M.}~\bibnamefont {Bowry}}, \bibinfo
  {author} {\bibfnamefont {P.~D.}\ \bibnamefont {Cottle}}, \bibinfo {author}
  {\bibfnamefont {F.~G.}\ \bibnamefont {DeVone}}, \bibinfo {author}
  {\bibfnamefont {A.}~\bibnamefont {Gade}}, \bibinfo {author} {\bibfnamefont
  {M.~T.}\ \bibnamefont {Glowacki}}, \bibinfo {author} {\bibfnamefont {S.~D.}\
  \bibnamefont {Gregory}}, \bibinfo {author} {\bibfnamefont {E.~B.}\
  \bibnamefont {Haldeman}}, \bibinfo {author} {\bibfnamefont {K.~W.}\
  \bibnamefont {Kemper}}, \bibinfo {author} {\bibfnamefont {E.}~\bibnamefont
  {Lunderberg}}, \bibinfo {author} {\bibfnamefont {S.}~\bibnamefont {Noji}},
  \bibinfo {author} {\bibfnamefont {F.}~\bibnamefont {Recchia}}, \bibinfo
  {author} {\bibfnamefont {B.~V.}\ \bibnamefont {Sadler}}, \bibinfo {author}
  {\bibfnamefont {M.}~\bibnamefont {Scott}}, \bibinfo {author} {\bibfnamefont
  {D.}~\bibnamefont {Weisshaar}}, \ and\ \bibinfo {author} {\bibfnamefont
  {R.~G.~T.}\ \bibnamefont {Zegers}},\ }\href {\doibase
  10.1103/PhysRevC.93.044327} {\bibfield  {journal} {\bibinfo  {journal} {Phys.
  Rev. C}\ }\textbf {\bibinfo {volume} {93}},\ \bibinfo {pages} {044327}
  (\bibinfo {year} {2016})}\BibitemShut {NoStop}%
\bibitem [{\citenamefont {Crawford}\ \emph {et~al.}(2017)\citenamefont
  {Crawford}, \citenamefont {Macchiavelli}, \citenamefont {Fallon},
  \citenamefont {Albers}, \citenamefont {Bader}, \citenamefont {Bazin},
  \citenamefont {Campbell}, \citenamefont {Clark}, \citenamefont {Cromaz},
  \citenamefont {Dilling}, \citenamefont {Gade}, \citenamefont {Gallant},
  \citenamefont {Holt}, \citenamefont {Janssens}, \citenamefont {Kr\"ucken},
  \citenamefont {Langer}, \citenamefont {Lauritsen}, \citenamefont {Lee},
  \citenamefont {Men\'endez}, \citenamefont {Noji}, \citenamefont {Paschalis},
  \citenamefont {Recchia}, \citenamefont {Rissanen}, \citenamefont {Schwenk},
  \citenamefont {Scott}, \citenamefont {Simonis}, \citenamefont {Stroberg},
  \citenamefont {Tostevin}, \citenamefont {Walz}, \citenamefont {Weisshaar},
  \citenamefont {Wiens}, \citenamefont {Wimmer},\ and\ \citenamefont
  {Zhu}}]{Cra17}%
  \BibitemOpen
  \bibfield  {author} {\bibinfo {author} {\bibfnamefont {H.~L.}\ \bibnamefont
  {Crawford}}, \bibinfo {author} {\bibfnamefont {A.~O.}\ \bibnamefont
  {Macchiavelli}}, \bibinfo {author} {\bibfnamefont {P.}~\bibnamefont
  {Fallon}}, \bibinfo {author} {\bibfnamefont {M.}~\bibnamefont {Albers}},
  \bibinfo {author} {\bibfnamefont {V.~M.}\ \bibnamefont {Bader}}, \bibinfo
  {author} {\bibfnamefont {D.}~\bibnamefont {Bazin}}, \bibinfo {author}
  {\bibfnamefont {C.~M.}\ \bibnamefont {Campbell}}, \bibinfo {author}
  {\bibfnamefont {R.~M.}\ \bibnamefont {Clark}}, \bibinfo {author}
  {\bibfnamefont {M.}~\bibnamefont {Cromaz}}, \bibinfo {author} {\bibfnamefont
  {J.}~\bibnamefont {Dilling}}, \bibinfo {author} {\bibfnamefont
  {A.}~\bibnamefont {Gade}}, \bibinfo {author} {\bibfnamefont {A.}~\bibnamefont
  {Gallant}}, \bibinfo {author} {\bibfnamefont {J.~D.}\ \bibnamefont {Holt}},
  \bibinfo {author} {\bibfnamefont {R.~V.~F.}\ \bibnamefont {Janssens}},
  \bibinfo {author} {\bibfnamefont {R.}~\bibnamefont {Kr\"ucken}}, \bibinfo
  {author} {\bibfnamefont {C.}~\bibnamefont {Langer}}, \bibinfo {author}
  {\bibfnamefont {T.}~\bibnamefont {Lauritsen}}, \bibinfo {author}
  {\bibfnamefont {I.~Y.}\ \bibnamefont {Lee}}, \bibinfo {author} {\bibfnamefont
  {J.}~\bibnamefont {Men\'endez}}, \bibinfo {author} {\bibfnamefont
  {S.}~\bibnamefont {Noji}}, \bibinfo {author} {\bibfnamefont {S.}~\bibnamefont
  {Paschalis}}, \bibinfo {author} {\bibfnamefont {F.}~\bibnamefont {Recchia}},
  \bibinfo {author} {\bibfnamefont {J.}~\bibnamefont {Rissanen}}, \bibinfo
  {author} {\bibfnamefont {A.}~\bibnamefont {Schwenk}}, \bibinfo {author}
  {\bibfnamefont {M.}~\bibnamefont {Scott}}, \bibinfo {author} {\bibfnamefont
  {J.}~\bibnamefont {Simonis}}, \bibinfo {author} {\bibfnamefont {S.~R.}\
  \bibnamefont {Stroberg}}, \bibinfo {author} {\bibfnamefont {J.~A.}\
  \bibnamefont {Tostevin}}, \bibinfo {author} {\bibfnamefont {C.}~\bibnamefont
  {Walz}}, \bibinfo {author} {\bibfnamefont {D.}~\bibnamefont {Weisshaar}},
  \bibinfo {author} {\bibfnamefont {A.}~\bibnamefont {Wiens}}, \bibinfo
  {author} {\bibfnamefont {K.}~\bibnamefont {Wimmer}}, \ and\ \bibinfo {author}
  {\bibfnamefont {S.}~\bibnamefont {Zhu}},\ }\href {\doibase
  10.1103/PhysRevC.95.064317} {\bibfield  {journal} {\bibinfo  {journal} {Phys.
  Rev. C}\ }\textbf {\bibinfo {volume} {95}},\ \bibinfo {pages} {064317}
  (\bibinfo {year} {2017})}\BibitemShut {NoStop}%
\bibitem [{\citenamefont {Talwar}\ \emph {et~al.}(2018)\citenamefont {Talwar},
  \citenamefont {Bojazi}, \citenamefont {Mohr}, \citenamefont {Auranen},
  \citenamefont {Avila}, \citenamefont {Ayangeakaa}, \citenamefont {Harker},
  \citenamefont {Hoffman}, \citenamefont {Jiang}, \citenamefont {Kuvin},
  \citenamefont {Meyer}, \citenamefont {Rehm}, \citenamefont
  {Santiago-Gonzalez}, \citenamefont {Sethi}, \citenamefont {Ugalde},\ and\
  \citenamefont {Winkelbauer}}]{Tal18}%
  \BibitemOpen
  \bibfield  {author} {\bibinfo {author} {\bibfnamefont {R.}~\bibnamefont
  {Talwar}}, \bibinfo {author} {\bibfnamefont {M.~J.}\ \bibnamefont {Bojazi}},
  \bibinfo {author} {\bibfnamefont {P.}~\bibnamefont {Mohr}}, \bibinfo {author}
  {\bibfnamefont {K.}~\bibnamefont {Auranen}}, \bibinfo {author} {\bibfnamefont
  {M.~L.}\ \bibnamefont {Avila}}, \bibinfo {author} {\bibfnamefont {A.~D.}\
  \bibnamefont {Ayangeakaa}}, \bibinfo {author} {\bibfnamefont
  {J.}~\bibnamefont {Harker}}, \bibinfo {author} {\bibfnamefont {C.~R.}\
  \bibnamefont {Hoffman}}, \bibinfo {author} {\bibfnamefont {C.~L.}\
  \bibnamefont {Jiang}}, \bibinfo {author} {\bibfnamefont {S.~A.}\ \bibnamefont
  {Kuvin}}, \bibinfo {author} {\bibfnamefont {B.~S.}\ \bibnamefont {Meyer}},
  \bibinfo {author} {\bibfnamefont {K.~E.}\ \bibnamefont {Rehm}}, \bibinfo
  {author} {\bibfnamefont {D.}~\bibnamefont {Santiago-Gonzalez}}, \bibinfo
  {author} {\bibfnamefont {J.}~\bibnamefont {Sethi}}, \bibinfo {author}
  {\bibfnamefont {C.}~\bibnamefont {Ugalde}}, \ and\ \bibinfo {author}
  {\bibfnamefont {J.~R.}\ \bibnamefont {Winkelbauer}},\ }\href {\doibase
  10.1103/PhysRevC.97.055801} {\bibfield  {journal} {\bibinfo  {journal} {Phys.
  Rev. C}\ }\textbf {\bibinfo {volume} {97}},\ \bibinfo {pages} {055801}
  (\bibinfo {year} {2018})}\BibitemShut {NoStop}%
\bibitem [{\citenamefont {Hady\ifmmode \acute{n}\else
  \'{n}\fi{}ska-Kl\ifmmode~\mbox{\c{e}}\else \c{e}\fi{}k}\ \emph
  {et~al.}(2018)\citenamefont {Hady\ifmmode \acute{n}\else
  \'{n}\fi{}ska-Kl\ifmmode~\mbox{\c{e}}\else \c{e}\fi{}k}, \citenamefont
  {Napiorkowski}, \citenamefont {Zieli\ifmmode~\acute{n}\else \'{n}\fi{}ska},
  \citenamefont {Srebrny}, \citenamefont {Maj}, \citenamefont {Azaiez},
  \citenamefont {Valiente~Dob\'on}, \citenamefont {Kici\ifmmode
  \acute{n}\else~\'{n}\fi{}ska Habior}, \citenamefont {Nowacki}, \citenamefont
  {Na\"{\i}dja}, \citenamefont {Bounthong}, \citenamefont {Rodr\'{\i}guez},
  \citenamefont {de~Angelis}, \citenamefont {Abraham}, \citenamefont
  {Anil~Kumar}, \citenamefont {Bazzacco}, \citenamefont {Bellato},
  \citenamefont {Bortolato}, \citenamefont {Bednarczyk}, \citenamefont
  {Benzoni}, \citenamefont {Berti}, \citenamefont {Birkenbach}, \citenamefont
  {Bruyneel}, \citenamefont {Brambilla}, \citenamefont {Camera}, \citenamefont
  {Chavas}, \citenamefont {Cederwall}, \citenamefont {Charles}, \citenamefont
  {Ciema\l{}a}, \citenamefont {Cocconi}, \citenamefont {Coleman-Smith},
  \citenamefont {Colombo}, \citenamefont {Corsi}, \citenamefont {Crespi},
  \citenamefont {Cullen}, \citenamefont {Czermak}, \citenamefont
  {D\'esesquelles}, \citenamefont {Doherty}, \citenamefont {Dulny},
  \citenamefont {Eberth}, \citenamefont {Farnea}, \citenamefont {Fornal},
  \citenamefont {Franchoo}, \citenamefont {Gadea}, \citenamefont {Giaz},
  \citenamefont {Gottardo}, \citenamefont {Grave}, \citenamefont
  {Gr\ifmmode~\mbox{\c{e}}\else \c{e}\fi{}bosz}, \citenamefont {G\"orgen},
  \citenamefont {Gulmini}, \citenamefont {Habermann}, \citenamefont {Hess},
  \citenamefont {Isocrate}, \citenamefont {Iwanicki}, \citenamefont {Jaworski},
  \citenamefont {Judson}, \citenamefont {Jungclaus}, \citenamefont {Karkour},
  \citenamefont {Kmiecik}, \citenamefont {Karpi\ifmmode~\acute{n}\else
  \'{n}\fi{}ski}, \citenamefont {Kisieli\ifmmode~\acute{n}\else \'{n}\fi{}ski},
  \citenamefont {Kondratyev}, \citenamefont {Korichi}, \citenamefont
  {Komorowska}, \citenamefont {Kowalczyk}, \citenamefont {Korten},
  \citenamefont {Krzysiek}, \citenamefont {Lehaut}, \citenamefont {Leoni},
  \citenamefont {Ljungvall}, \citenamefont {Lopez-Martens}, \citenamefont
  {Lunardi}, \citenamefont {Maron}, \citenamefont {Mazurek}, \citenamefont
  {Menegazzo}, \citenamefont {Mengoni}, \citenamefont {Merch\'an},
  \citenamefont {M\ifmmode \mbox{\c{e}}\else
  \c{e}\fi{}czy\ifmmode~\acute{n}\else \'{n}\fi{}ski}, \citenamefont
  {Michelagnoli}, \citenamefont {Million}, \citenamefont {Myalski},
  \citenamefont {Napoli}, \citenamefont {Niikura}, \citenamefont {Obertelli},
  \citenamefont {\"Ozmen}, \citenamefont {Palacz}, \citenamefont {Pr\'ochniak},
  \citenamefont {Pullia}, \citenamefont {Quintana}, \citenamefont {Rampazzo},
  \citenamefont {Recchia}, \citenamefont {Redon}, \citenamefont {Reiter},
  \citenamefont {Rosso}, \citenamefont {Rusek}, \citenamefont {Sahin},
  \citenamefont {Salsac}, \citenamefont {S\"oderstr\"om}, \citenamefont
  {Stefan}, \citenamefont {St\'ezowski}, \citenamefont
  {Stycze\ifmmode~\acute{n}\else \'{n}\fi{}}, \citenamefont {Theisen},
  \citenamefont {Toniolo}, \citenamefont {Ur}, \citenamefont {Wadsworth},
  \citenamefont {Wasilewska}, \citenamefont {Wiens}, \citenamefont {Wood},
  \citenamefont {Wrzosek-Lipska},\ and\ \citenamefont {Zi\ifmmode
  \mbox{\c{e}}\else \c{e}\fi{}bli\ifmmode~\acute{n}\else
  \'{n}\fi{}ski}}]{Had18}%
  \BibitemOpen
  \bibfield  {author} {\bibinfo {author} {\bibfnamefont {K.}~\bibnamefont
  {Hady\ifmmode \acute{n}\else \'{n}\fi{}ska-Kl\ifmmode~\mbox{\c{e}}\else
  \c{e}\fi{}k}}, \bibinfo {author} {\bibfnamefont {P.~J.}\ \bibnamefont
  {Napiorkowski}}, \bibinfo {author} {\bibfnamefont {M.}~\bibnamefont
  {Zieli\ifmmode~\acute{n}\else \'{n}\fi{}ska}}, \bibinfo {author}
  {\bibfnamefont {J.}~\bibnamefont {Srebrny}}, \bibinfo {author} {\bibfnamefont
  {A.}~\bibnamefont {Maj}}, \bibinfo {author} {\bibfnamefont {F.}~\bibnamefont
  {Azaiez}}, \bibinfo {author} {\bibfnamefont {J.~J.}\ \bibnamefont
  {Valiente~Dob\'on}}, \bibinfo {author} {\bibfnamefont {M.}~\bibnamefont
  {Kici\ifmmode \acute{n}\else~\'{n}\fi{}ska Habior}}, \bibinfo {author}
  {\bibfnamefont {F.}~\bibnamefont {Nowacki}}, \bibinfo {author} {\bibfnamefont
  {H.}~\bibnamefont {Na\"{\i}dja}}, \bibinfo {author} {\bibfnamefont
  {B.}~\bibnamefont {Bounthong}}, \bibinfo {author} {\bibfnamefont {T.~R.}\
  \bibnamefont {Rodr\'{\i}guez}}, \bibinfo {author} {\bibfnamefont
  {G.}~\bibnamefont {de~Angelis}}, \bibinfo {author} {\bibfnamefont
  {T.}~\bibnamefont {Abraham}}, \bibinfo {author} {\bibfnamefont
  {G.}~\bibnamefont {Anil~Kumar}}, \bibinfo {author} {\bibfnamefont
  {D.}~\bibnamefont {Bazzacco}}, \bibinfo {author} {\bibfnamefont
  {M.}~\bibnamefont {Bellato}}, \bibinfo {author} {\bibfnamefont
  {D.}~\bibnamefont {Bortolato}}, \bibinfo {author} {\bibfnamefont
  {P.}~\bibnamefont {Bednarczyk}}, \bibinfo {author} {\bibfnamefont
  {G.}~\bibnamefont {Benzoni}}, \bibinfo {author} {\bibfnamefont
  {L.}~\bibnamefont {Berti}}, \bibinfo {author} {\bibfnamefont
  {B.}~\bibnamefont {Birkenbach}}, \bibinfo {author} {\bibfnamefont
  {B.}~\bibnamefont {Bruyneel}}, \bibinfo {author} {\bibfnamefont
  {S.}~\bibnamefont {Brambilla}}, \bibinfo {author} {\bibfnamefont
  {F.}~\bibnamefont {Camera}}, \bibinfo {author} {\bibfnamefont
  {J.}~\bibnamefont {Chavas}}, \bibinfo {author} {\bibfnamefont
  {B.}~\bibnamefont {Cederwall}}, \bibinfo {author} {\bibfnamefont
  {L.}~\bibnamefont {Charles}}, \bibinfo {author} {\bibfnamefont
  {M.}~\bibnamefont {Ciema\l{}a}}, \bibinfo {author} {\bibfnamefont
  {P.}~\bibnamefont {Cocconi}}, \bibinfo {author} {\bibfnamefont
  {P.}~\bibnamefont {Coleman-Smith}}, \bibinfo {author} {\bibfnamefont
  {A.}~\bibnamefont {Colombo}}, \bibinfo {author} {\bibfnamefont
  {A.}~\bibnamefont {Corsi}}, \bibinfo {author} {\bibfnamefont {F.~C.~L.}\
  \bibnamefont {Crespi}}, \bibinfo {author} {\bibfnamefont {D.~M.}\
  \bibnamefont {Cullen}}, \bibinfo {author} {\bibfnamefont {A.}~\bibnamefont
  {Czermak}}, \bibinfo {author} {\bibfnamefont {P.}~\bibnamefont
  {D\'esesquelles}}, \bibinfo {author} {\bibfnamefont {D.~T.}\ \bibnamefont
  {Doherty}}, \bibinfo {author} {\bibfnamefont {B.}~\bibnamefont {Dulny}},
  \bibinfo {author} {\bibfnamefont {J.}~\bibnamefont {Eberth}}, \bibinfo
  {author} {\bibfnamefont {E.}~\bibnamefont {Farnea}}, \bibinfo {author}
  {\bibfnamefont {B.}~\bibnamefont {Fornal}}, \bibinfo {author} {\bibfnamefont
  {S.}~\bibnamefont {Franchoo}}, \bibinfo {author} {\bibfnamefont
  {A.}~\bibnamefont {Gadea}}, \bibinfo {author} {\bibfnamefont
  {A.}~\bibnamefont {Giaz}}, \bibinfo {author} {\bibfnamefont {A.}~\bibnamefont
  {Gottardo}}, \bibinfo {author} {\bibfnamefont {X.}~\bibnamefont {Grave}},
  \bibinfo {author} {\bibfnamefont {J.}~\bibnamefont
  {Gr\ifmmode~\mbox{\c{e}}\else \c{e}\fi{}bosz}}, \bibinfo {author}
  {\bibfnamefont {A.}~\bibnamefont {G\"orgen}}, \bibinfo {author}
  {\bibfnamefont {M.}~\bibnamefont {Gulmini}}, \bibinfo {author} {\bibfnamefont
  {T.}~\bibnamefont {Habermann}}, \bibinfo {author} {\bibfnamefont
  {H.}~\bibnamefont {Hess}}, \bibinfo {author} {\bibfnamefont {R.}~\bibnamefont
  {Isocrate}}, \bibinfo {author} {\bibfnamefont {J.}~\bibnamefont {Iwanicki}},
  \bibinfo {author} {\bibfnamefont {G.}~\bibnamefont {Jaworski}}, \bibinfo
  {author} {\bibfnamefont {D.~S.}\ \bibnamefont {Judson}}, \bibinfo {author}
  {\bibfnamefont {A.}~\bibnamefont {Jungclaus}}, \bibinfo {author}
  {\bibfnamefont {N.}~\bibnamefont {Karkour}}, \bibinfo {author} {\bibfnamefont
  {M.}~\bibnamefont {Kmiecik}}, \bibinfo {author} {\bibfnamefont
  {D.}~\bibnamefont {Karpi\ifmmode~\acute{n}\else \'{n}\fi{}ski}}, \bibinfo
  {author} {\bibfnamefont {M.}~\bibnamefont {Kisieli\ifmmode~\acute{n}\else
  \'{n}\fi{}ski}}, \bibinfo {author} {\bibfnamefont {N.}~\bibnamefont
  {Kondratyev}}, \bibinfo {author} {\bibfnamefont {A.}~\bibnamefont {Korichi}},
  \bibinfo {author} {\bibfnamefont {M.}~\bibnamefont {Komorowska}}, \bibinfo
  {author} {\bibfnamefont {M.}~\bibnamefont {Kowalczyk}}, \bibinfo {author}
  {\bibfnamefont {W.}~\bibnamefont {Korten}}, \bibinfo {author} {\bibfnamefont
  {M.}~\bibnamefont {Krzysiek}}, \bibinfo {author} {\bibfnamefont
  {G.}~\bibnamefont {Lehaut}}, \bibinfo {author} {\bibfnamefont
  {S.}~\bibnamefont {Leoni}}, \bibinfo {author} {\bibfnamefont
  {J.}~\bibnamefont {Ljungvall}}, \bibinfo {author} {\bibfnamefont
  {A.}~\bibnamefont {Lopez-Martens}}, \bibinfo {author} {\bibfnamefont
  {S.}~\bibnamefont {Lunardi}}, \bibinfo {author} {\bibfnamefont
  {G.}~\bibnamefont {Maron}}, \bibinfo {author} {\bibfnamefont
  {K.}~\bibnamefont {Mazurek}}, \bibinfo {author} {\bibfnamefont
  {R.}~\bibnamefont {Menegazzo}}, \bibinfo {author} {\bibfnamefont
  {D.}~\bibnamefont {Mengoni}}, \bibinfo {author} {\bibfnamefont
  {E.}~\bibnamefont {Merch\'an}}, \bibinfo {author} {\bibfnamefont
  {W.}~\bibnamefont {M\ifmmode \mbox{\c{e}}\else
  \c{e}\fi{}czy\ifmmode~\acute{n}\else \'{n}\fi{}ski}}, \bibinfo {author}
  {\bibfnamefont {C.}~\bibnamefont {Michelagnoli}}, \bibinfo {author}
  {\bibfnamefont {B.}~\bibnamefont {Million}}, \bibinfo {author} {\bibfnamefont
  {S.}~\bibnamefont {Myalski}}, \bibinfo {author} {\bibfnamefont {D.~R.}\
  \bibnamefont {Napoli}}, \bibinfo {author} {\bibfnamefont {M.}~\bibnamefont
  {Niikura}}, \bibinfo {author} {\bibfnamefont {A.}~\bibnamefont {Obertelli}},
  \bibinfo {author} {\bibfnamefont {S.~F.}\ \bibnamefont {\"Ozmen}}, \bibinfo
  {author} {\bibfnamefont {M.}~\bibnamefont {Palacz}}, \bibinfo {author}
  {\bibfnamefont {L.}~\bibnamefont {Pr\'ochniak}}, \bibinfo {author}
  {\bibfnamefont {A.}~\bibnamefont {Pullia}}, \bibinfo {author} {\bibfnamefont
  {B.}~\bibnamefont {Quintana}}, \bibinfo {author} {\bibfnamefont
  {G.}~\bibnamefont {Rampazzo}}, \bibinfo {author} {\bibfnamefont
  {F.}~\bibnamefont {Recchia}}, \bibinfo {author} {\bibfnamefont
  {N.}~\bibnamefont {Redon}}, \bibinfo {author} {\bibfnamefont
  {P.}~\bibnamefont {Reiter}}, \bibinfo {author} {\bibfnamefont
  {D.}~\bibnamefont {Rosso}}, \bibinfo {author} {\bibfnamefont
  {K.}~\bibnamefont {Rusek}}, \bibinfo {author} {\bibfnamefont
  {E.}~\bibnamefont {Sahin}}, \bibinfo {author} {\bibfnamefont {M.-D.}\
  \bibnamefont {Salsac}}, \bibinfo {author} {\bibfnamefont {P.-A.}\
  \bibnamefont {S\"oderstr\"om}}, \bibinfo {author} {\bibfnamefont
  {I.}~\bibnamefont {Stefan}}, \bibinfo {author} {\bibfnamefont
  {O.}~\bibnamefont {St\'ezowski}}, \bibinfo {author} {\bibfnamefont
  {J.}~\bibnamefont {Stycze\ifmmode~\acute{n}\else \'{n}\fi{}}}, \bibinfo
  {author} {\bibfnamefont {C.}~\bibnamefont {Theisen}}, \bibinfo {author}
  {\bibfnamefont {N.}~\bibnamefont {Toniolo}}, \bibinfo {author} {\bibfnamefont
  {C.~A.}\ \bibnamefont {Ur}}, \bibinfo {author} {\bibfnamefont
  {R.}~\bibnamefont {Wadsworth}}, \bibinfo {author} {\bibfnamefont
  {B.}~\bibnamefont {Wasilewska}}, \bibinfo {author} {\bibfnamefont
  {A.}~\bibnamefont {Wiens}}, \bibinfo {author} {\bibfnamefont {J.~L.}\
  \bibnamefont {Wood}}, \bibinfo {author} {\bibfnamefont {K.}~\bibnamefont
  {Wrzosek-Lipska}}, \ and\ \bibinfo {author} {\bibfnamefont {M.}~\bibnamefont
  {Zi\ifmmode \mbox{\c{e}}\else \c{e}\fi{}bli\ifmmode~\acute{n}\else
  \'{n}\fi{}ski}},\ }\href {\doibase 10.1103/PhysRevC.97.024326} {\bibfield
  {journal} {\bibinfo  {journal} {Phys. Rev. C}\ }\textbf {\bibinfo {volume}
  {97}},\ \bibinfo {pages} {024326} (\bibinfo {year} {2018})}\BibitemShut
  {NoStop}%
\bibitem [{\citenamefont {D.~Bodansky}\ and\ \citenamefont
  {Fowler}(1968)}]{Bod68}%
  \BibitemOpen
  \bibfield  {author} {\bibinfo {author} {\bibfnamefont {F.}~\bibnamefont
  {D.~Bodansky}, \bibfnamefont {D.~Clayton}}\ and\ \bibinfo {author}
  {\bibfnamefont {A.}~\bibnamefont {Fowler}},\ }\href@noop {} {\bibfield
  {journal} {\bibinfo  {journal} {Astrophys. J. Suppl. S.}\ }\textbf {\bibinfo
  {volume} {16}},\ \bibinfo {pages} {299} (\bibinfo {year} {1968})}\BibitemShut
  {NoStop}%
\bibitem [{\citenamefont {Truran}(1972)}]{Tru72}%
  \BibitemOpen
  \bibfield  {author} {\bibinfo {author} {\bibfnamefont {J.~W.}\ \bibnamefont
  {Truran}},\ }\href@noop {} {\bibfield  {journal} {\bibinfo  {journal} {Ap.
  Sapce. Sci.}\ }\textbf {\bibinfo {volume} {18}},\ \bibinfo {pages} {306}
  (\bibinfo {year} {1972})}\BibitemShut {NoStop}%
\bibitem [{\citenamefont {K\"appeler}\ \emph {et~al.}(1985)\citenamefont
  {K\"appeler}, \citenamefont {Walter},\ and\ \citenamefont {Mathews}}]{Kap84}%
  \BibitemOpen
  \bibfield  {author} {\bibinfo {author} {\bibfnamefont {F.}~\bibnamefont
  {K\"appeler}}, \bibinfo {author} {\bibfnamefont {G.}~\bibnamefont {Walter}},
  \ and\ \bibinfo {author} {\bibfnamefont {J.}~\bibnamefont {Mathews}},\
  }\href@noop {} {\bibfield  {journal} {\bibinfo  {journal} {Astrophys. J.}\
  }\textbf {\bibinfo {volume} {291}},\ \bibinfo {pages} {319} (\bibinfo {year}
  {1985})}\BibitemShut {NoStop}%
\bibitem [{\citenamefont {S.~E.~Woosley}\ and\ \citenamefont
  {Clayton}(1973)}]{Woo73}%
  \BibitemOpen
  \bibfield  {author} {\bibinfo {author} {\bibfnamefont {D.}~\bibnamefont
  {S.~E.~Woosley}, \bibfnamefont {W.~David~Arnett}}\ and\ \bibinfo {author}
  {\bibfnamefont {D.}~\bibnamefont {Clayton}},\ }\href@noop {} {\bibfield
  {journal} {\bibinfo  {journal} {Astrophys. J. Suppl. S.}\ }\textbf {\bibinfo
  {volume} {26}},\ \bibinfo {pages} {231} (\bibinfo {year} {1973})}\BibitemShut
  {NoStop}%
\bibitem [{\citenamefont {K\"appeler}\ \emph {et~al.}(1982)\citenamefont
  {K\"appeler}, \citenamefont {Beer}, \citenamefont {Wisshak},\ and\
  \citenamefont {Clayton}}]{Kap82}%
  \BibitemOpen
  \bibfield  {author} {\bibinfo {author} {\bibfnamefont {F.}~\bibnamefont
  {K\"appeler}}, \bibinfo {author} {\bibfnamefont {H.}~\bibnamefont {Beer}},
  \bibinfo {author} {\bibfnamefont {K.}~\bibnamefont {Wisshak}}, \ and\
  \bibinfo {author} {\bibfnamefont {D.~D.}\ \bibnamefont {Clayton}},\
  }\href@noop {} {\bibfield  {journal} {\bibinfo  {journal} {Astrophys. J.
  Suppl. S.}\ }\textbf {\bibinfo {volume} {257}},\ \bibinfo {pages} {821}
  (\bibinfo {year} {1982})}\BibitemShut {NoStop}%
\bibitem [{\citenamefont {Cameron}(1979)}]{Cam79}%
  \BibitemOpen
  \bibfield  {author} {\bibinfo {author} {\bibfnamefont {A.~G.~W.}\
  \bibnamefont {Cameron}},\ }\href@noop {} {\bibfield  {journal} {\bibinfo
  {journal} {Astrophys. J.}\ }\textbf {\bibinfo {volume} {230}},\ \bibinfo
  {pages} {53} (\bibinfo {year} {1979})}\BibitemShut {NoStop}%
\bibitem [{\citenamefont {Kubik}\ \emph {et~al.}(1986)\citenamefont {Kubik},
  \citenamefont {Elmore}, \citenamefont {Conard}, \citenamefont {Nishiizumi},\
  and\ \citenamefont {Arnold}}]{Kub86}%
  \BibitemOpen
  \bibfield  {author} {\bibinfo {author} {\bibfnamefont {P.~W.}\ \bibnamefont
  {Kubik}}, \bibinfo {author} {\bibfnamefont {D.}~\bibnamefont {Elmore}},
  \bibinfo {author} {\bibfnamefont {N.~J.}\ \bibnamefont {Conard}}, \bibinfo
  {author} {\bibfnamefont {K.}~\bibnamefont {Nishiizumi}}, \ and\ \bibinfo
  {author} {\bibfnamefont {J.~R.}\ \bibnamefont {Arnold}},\ }\href {\doibase
  10.1038/319568a0} {\bibfield  {journal} {\bibinfo  {journal} {Nature}\
  }\textbf {\bibinfo {volume} {319}},\ \bibinfo {pages} {568} (\bibinfo {year}
  {1986})}\BibitemShut {NoStop}%
\bibitem [{\citenamefont {Honma}\ \emph {et~al.}(2002)\citenamefont {Honma},
  \citenamefont {Otsuka}, \citenamefont {Brown},\ and\ \citenamefont
  {Mizusaki}}]{Hon02}%
  \BibitemOpen
  \bibfield  {author} {\bibinfo {author} {\bibfnamefont {M.}~\bibnamefont
  {Honma}}, \bibinfo {author} {\bibfnamefont {T.}~\bibnamefont {Otsuka}},
  \bibinfo {author} {\bibfnamefont {B.~A.}\ \bibnamefont {Brown}}, \ and\
  \bibinfo {author} {\bibfnamefont {T.}~\bibnamefont {Mizusaki}},\ }\href
  {\doibase 10.1103/PhysRevC.65.061301} {\bibfield  {journal} {\bibinfo
  {journal} {Phys. Rev. C}\ }\textbf {\bibinfo {volume} {65}},\ \bibinfo
  {pages} {061301} (\bibinfo {year} {2002})}\BibitemShut {NoStop}%
\bibitem [{\citenamefont {Honma}\ \emph {et~al.}(2005)\citenamefont {Honma},
  \citenamefont {Otsuka}, \citenamefont {Brown},\ and\ \citenamefont
  {Mizusaki}}]{Hon05}%
  \BibitemOpen
  \bibfield  {author} {\bibinfo {author} {\bibfnamefont {M.}~\bibnamefont
  {Honma}}, \bibinfo {author} {\bibfnamefont {T.}~\bibnamefont {Otsuka}},
  \bibinfo {author} {\bibfnamefont {B.~A.}\ \bibnamefont {Brown}}, \ and\
  \bibinfo {author} {\bibfnamefont {T.}~\bibnamefont {Mizusaki}},\ }\href
  {\doibase 10.1140/epjad/i2005-06-032-2} {\bibfield  {journal} {\bibinfo
  {journal} {Eur. Phys. Journ. A.}\ }\textbf {\bibinfo {volume} {25}},\
  \bibinfo {pages} {499} (\bibinfo {year} {2005})}\BibitemShut {NoStop}%
\bibitem [{\citenamefont {Utsuno}\ \emph {et~al.}(2012)\citenamefont {Utsuno},
  \citenamefont {Otsuka}, \citenamefont {Brown}, \citenamefont {Honma},
  \citenamefont {Mizusaki},\ and\ \citenamefont {Shimizu}}]{Uts12}%
  \BibitemOpen
  \bibfield  {author} {\bibinfo {author} {\bibfnamefont {Y.}~\bibnamefont
  {Utsuno}}, \bibinfo {author} {\bibfnamefont {T.}~\bibnamefont {Otsuka}},
  \bibinfo {author} {\bibfnamefont {B.~A.}\ \bibnamefont {Brown}}, \bibinfo
  {author} {\bibfnamefont {M.}~\bibnamefont {Honma}}, \bibinfo {author}
  {\bibfnamefont {T.}~\bibnamefont {Mizusaki}}, \ and\ \bibinfo {author}
  {\bibfnamefont {N.}~\bibnamefont {Shimizu}},\ }\href {\doibase
  10.1143/PTPS.196.304} {\bibfield  {journal} {\bibinfo  {journal} {Progr.
  Theor. Phys. Suppl.}\ }\textbf {\bibinfo {volume} {196}},\ \bibinfo {pages}
  {304} (\bibinfo {year} {2012})}\BibitemShut {NoStop}%
\bibitem [{\citenamefont {Tsunoda}\ \emph {et~al.}(2014)\citenamefont
  {Tsunoda}, \citenamefont {Takayanagi}, \citenamefont {Hjorth-Jensen},\ and\
  \citenamefont {Otsuka}}]{Tsu13}%
  \BibitemOpen
  \bibfield  {author} {\bibinfo {author} {\bibfnamefont {N.}~\bibnamefont
  {Tsunoda}}, \bibinfo {author} {\bibfnamefont {K.}~\bibnamefont {Takayanagi}},
  \bibinfo {author} {\bibfnamefont {M.}~\bibnamefont {Hjorth-Jensen}}, \ and\
  \bibinfo {author} {\bibfnamefont {T.}~\bibnamefont {Otsuka}},\ }\href
  {\doibase 10.1103/PhysRevC.89.024313} {\bibfield  {journal} {\bibinfo
  {journal} {Phys. Rev. C}\ }\textbf {\bibinfo {volume} {89}},\ \bibinfo
  {pages} {024313} (\bibinfo {year} {2014})}\BibitemShut {NoStop}%
\bibitem [{\citenamefont {Hebeler}\ \emph {et~al.}(2011)\citenamefont
  {Hebeler}, \citenamefont {Bogner}, \citenamefont {Furnstahl}, \citenamefont
  {Nogga},\ and\ \citenamefont {Schwenk}}]{Heb11}%
  \BibitemOpen
  \bibfield  {author} {\bibinfo {author} {\bibfnamefont {K.}~\bibnamefont
  {Hebeler}}, \bibinfo {author} {\bibfnamefont {S.~K.}\ \bibnamefont {Bogner}},
  \bibinfo {author} {\bibfnamefont {R.~J.}\ \bibnamefont {Furnstahl}}, \bibinfo
  {author} {\bibfnamefont {A.}~\bibnamefont {Nogga}}, \ and\ \bibinfo {author}
  {\bibfnamefont {A.}~\bibnamefont {Schwenk}},\ }\href {\doibase
  10.1103/PhysRevC.83.031301} {\bibfield  {journal} {\bibinfo  {journal} {Phys.
  Rev. C}\ }\textbf {\bibinfo {volume} {83}},\ \bibinfo {pages} {031301}
  (\bibinfo {year} {2011})}\BibitemShut {NoStop}%
\bibitem [{\citenamefont {Holt}\ \emph {et~al.}(2012)\citenamefont {Holt},
  \citenamefont {Otsuka}, \citenamefont {Schwenk},\ and\ \citenamefont
  {Suzuki}}]{Hol12}%
  \BibitemOpen
  \bibfield  {author} {\bibinfo {author} {\bibfnamefont {J.~D.}\ \bibnamefont
  {Holt}}, \bibinfo {author} {\bibfnamefont {T.}~\bibnamefont {Otsuka}},
  \bibinfo {author} {\bibfnamefont {A.}~\bibnamefont {Schwenk}}, \ and\
  \bibinfo {author} {\bibfnamefont {T.}~\bibnamefont {Suzuki}},\ }\href
  {\doibase 10.1088/0954-3899/39/8/085111} {\bibfield  {journal} {\bibinfo
  {journal} {Journ. Phys. G: Nucl. and Part. Phys.}\ }\textbf {\bibinfo
  {volume} {39}},\ \bibinfo {pages} {085111} (\bibinfo {year}
  {2012})}\BibitemShut {NoStop}%
\bibitem [{\citenamefont {Holt}\ \emph {et~al.}(2014)\citenamefont {Holt},
  \citenamefont {Men\'endez}, \citenamefont {Simonis},\ and\ \citenamefont
  {Schwenk}}]{Hol14}%
  \BibitemOpen
  \bibfield  {author} {\bibinfo {author} {\bibfnamefont {J.~D.}\ \bibnamefont
  {Holt}}, \bibinfo {author} {\bibfnamefont {J.}~\bibnamefont {Men\'endez}},
  \bibinfo {author} {\bibfnamefont {J.}~\bibnamefont {Simonis}}, \ and\
  \bibinfo {author} {\bibfnamefont {A.}~\bibnamefont {Schwenk}},\ }\href
  {\doibase 10.1103/PhysRevC.90.024312} {\bibfield  {journal} {\bibinfo
  {journal} {Phys. Rev. C}\ }\textbf {\bibinfo {volume} {90}},\ \bibinfo
  {pages} {024312} (\bibinfo {year} {2014})}\BibitemShut {NoStop}%
\bibitem [{\citenamefont {Simonis}\ \emph {et~al.}(2016)\citenamefont
  {Simonis}, \citenamefont {Hebeler}, \citenamefont {Holt}, \citenamefont
  {Men\'endez},\ and\ \citenamefont {Schwenk}}]{Sim16}%
  \BibitemOpen
  \bibfield  {author} {\bibinfo {author} {\bibfnamefont {J.}~\bibnamefont
  {Simonis}}, \bibinfo {author} {\bibfnamefont {K.}~\bibnamefont {Hebeler}},
  \bibinfo {author} {\bibfnamefont {J.~D.}\ \bibnamefont {Holt}}, \bibinfo
  {author} {\bibfnamefont {J.}~\bibnamefont {Men\'endez}}, \ and\ \bibinfo
  {author} {\bibfnamefont {A.}~\bibnamefont {Schwenk}},\ }\href {\doibase
  10.1103/PhysRevC.93.011302} {\bibfield  {journal} {\bibinfo  {journal} {Phys.
  Rev. C}\ }\textbf {\bibinfo {volume} {93}},\ \bibinfo {pages} {011302}
  (\bibinfo {year} {2016})}\BibitemShut {NoStop}%
\bibitem [{\citenamefont {Kibedi}\ and\ \citenamefont
  {R.H.Spear}(2002)}]{Kib02}%
  \BibitemOpen
  \bibfield  {author} {\bibinfo {author} {\bibfnamefont {T.}~\bibnamefont
  {Kibedi}}\ and\ \bibinfo {author} {\bibnamefont {R.H.Spear}},\ }\href
  {\doibase https://doi.org/10.1006/adnd.2001.0871} {\bibfield  {journal}
  {\bibinfo  {journal} {Atom. Nucl. Data Tables}\ }\textbf {\bibinfo {volume}
  {80}},\ \bibinfo {pages} {35} (\bibinfo {year} {2002})}\BibitemShut {NoStop}%
\bibitem [{\citenamefont {Ideguchi}\ \emph {et~al.}(2001)\citenamefont
  {Ideguchi}, \citenamefont {Sarantites}, \citenamefont {Reviol}, \citenamefont
  {Afanasjev}, \citenamefont {Devlin}, \citenamefont {Baktash}, \citenamefont
  {Janssens}, \citenamefont {Rudolph}, \citenamefont {Axelsson}, \citenamefont
  {Carpenter}, \citenamefont {Galindo-Uribarri}, \citenamefont {LaFosse},
  \citenamefont {Lauritsen}, \citenamefont {Lerma}, \citenamefont {Lister},
  \citenamefont {Reiter}, \citenamefont {Seweryniak}, \citenamefont
  {Weiszflog},\ and\ \citenamefont {Wilson}}]{Ide01}%
  \BibitemOpen
  \bibfield  {author} {\bibinfo {author} {\bibfnamefont {E.}~\bibnamefont
  {Ideguchi}}, \bibinfo {author} {\bibfnamefont {D.~G.}\ \bibnamefont
  {Sarantites}}, \bibinfo {author} {\bibfnamefont {W.}~\bibnamefont {Reviol}},
  \bibinfo {author} {\bibfnamefont {A.~V.}\ \bibnamefont {Afanasjev}}, \bibinfo
  {author} {\bibfnamefont {M.}~\bibnamefont {Devlin}}, \bibinfo {author}
  {\bibfnamefont {C.}~\bibnamefont {Baktash}}, \bibinfo {author} {\bibfnamefont
  {R.~V.~F.}\ \bibnamefont {Janssens}}, \bibinfo {author} {\bibfnamefont
  {D.}~\bibnamefont {Rudolph}}, \bibinfo {author} {\bibfnamefont
  {A.}~\bibnamefont {Axelsson}}, \bibinfo {author} {\bibfnamefont {M.~P.}\
  \bibnamefont {Carpenter}}, \bibinfo {author} {\bibfnamefont {A.}~\bibnamefont
  {Galindo-Uribarri}}, \bibinfo {author} {\bibfnamefont {D.~R.}\ \bibnamefont
  {LaFosse}}, \bibinfo {author} {\bibfnamefont {T.}~\bibnamefont {Lauritsen}},
  \bibinfo {author} {\bibfnamefont {F.}~\bibnamefont {Lerma}}, \bibinfo
  {author} {\bibfnamefont {C.~J.}\ \bibnamefont {Lister}}, \bibinfo {author}
  {\bibfnamefont {P.}~\bibnamefont {Reiter}}, \bibinfo {author} {\bibfnamefont
  {D.}~\bibnamefont {Seweryniak}}, \bibinfo {author} {\bibfnamefont
  {M.}~\bibnamefont {Weiszflog}}, \ and\ \bibinfo {author} {\bibfnamefont
  {J.~N.}\ \bibnamefont {Wilson}},\ }\href {\doibase
  10.1103/PhysRevLett.87.222501} {\bibfield  {journal} {\bibinfo  {journal}
  {Phys. Rev. Lett.}\ }\textbf {\bibinfo {volume} {87}},\ \bibinfo {pages}
  {222501} (\bibinfo {year} {2001})}\BibitemShut {NoStop}%
\bibitem [{\citenamefont {Burrows}(2006)}]{Bur06}%
  \BibitemOpen
  \bibfield  {author} {\bibinfo {author} {\bibfnamefont {T.~W.}\ \bibnamefont
  {Burrows}},\ }\href
  {https://www.sciencedirect.com/science/article/pii/S0090375206000482}
  {\bibfield  {journal} {\bibinfo  {journal} {Nucl. Data Sheet}\ }\textbf
  {\bibinfo {volume} {107}},\ \bibinfo {pages} {1747} (\bibinfo {year}
  {2006})}\BibitemShut {NoStop}%
\bibitem [{\citenamefont {Bohr}\ and\ \citenamefont {Mottelson}()}]{BM}%
  \BibitemOpen
  \bibfield  {author} {\bibinfo {author} {\bibfnamefont {A.}~\bibnamefont
  {Bohr}}\ and\ \bibinfo {author} {\bibfnamefont {B.~M.}\ \bibnamefont
  {Mottelson}},\ }\href@noop {} {\bibinfo  {journal} {Nuclear Structure. Volume
  II: Nuclear Deformations (W. A. Benjamin, New York, 1980)}\ }\BibitemShut
  {NoStop}%
\bibitem [{\citenamefont {Hamamoto}(1969)}]{Ham69}%
  \BibitemOpen
\bibfield  {journal} {  }\bibfield  {author} {\bibinfo {author} {\bibfnamefont
  {I.}~\bibnamefont {Hamamoto}},\ }\href {\doibase
  https://doi.org/10.1016/0375-9474(69)90846-X} {\bibfield  {journal} {\bibinfo
   {journal} {Nucl. Phys. A}\ }\textbf {\bibinfo {volume} {126}},\ \bibinfo
  {pages} {545 } (\bibinfo {year} {1969})}\BibitemShut {NoStop}%
\bibitem [{\citenamefont {Hamamoto}(1974)}]{Ham74}%
  \BibitemOpen
  \bibfield  {author} {\bibinfo {author} {\bibfnamefont {I.}~\bibnamefont
  {Hamamoto}},\ }\href {\doibase https://doi.org/10.1016/0370-1573(74)90019-2}
  {\bibfield  {journal} {\bibinfo  {journal} {Phys. Rep.}\ }\textbf {\bibinfo
  {volume} {10}},\ \bibinfo {pages} {63 } (\bibinfo {year} {1974})}\BibitemShut
  {NoStop}%
\bibitem [{\citenamefont {Col\`o}\ \emph {et~al.}(2017)\citenamefont {Col\`o},
  \citenamefont {Bortignon},\ and\ \citenamefont {Bocchi}}]{Col17}%
  \BibitemOpen
  \bibfield  {author} {\bibinfo {author} {\bibfnamefont {G.}~\bibnamefont
  {Col\`o}}, \bibinfo {author} {\bibfnamefont {P.~F.}\ \bibnamefont
  {Bortignon}}, \ and\ \bibinfo {author} {\bibfnamefont {G.}~\bibnamefont
  {Bocchi}},\ }\href {\doibase 10.1103/PhysRevC.95.034303} {\bibfield
  {journal} {\bibinfo  {journal} {Phys. Rev. C}\ }\textbf {\bibinfo {volume}
  {95}},\ \bibinfo {pages} {034303} (\bibinfo {year} {2017})}\BibitemShut
  {NoStop}%
\bibitem [{\citenamefont {Bottoni}\ \emph {et~al.}()\citenamefont {Bottoni},
  \citenamefont {Col\`o}, \citenamefont {Niu},\ and\ \citenamefont
  {Bortignon}}]{Bot19}%
  \BibitemOpen
  \bibfield  {author} {\bibinfo {author} {\bibfnamefont {S.}~\bibnamefont
  {Bottoni}}, \bibinfo {author} {\bibfnamefont {G.}~\bibnamefont {Col\`o}},
  \bibinfo {author} {\bibfnamefont {Y.}~\bibnamefont {Niu}}, \ and\ \bibinfo
  {author} {\bibfnamefont {P.~F.}\ \bibnamefont {Bortignon}},\ }\href@noop {}
  {\bibinfo  {journal} {to be published}\ }\BibitemShut {NoStop}%
\bibitem [{\citenamefont {Jentschel}\ \emph {et~al.}(2017)\citenamefont
  {Jentschel}, \citenamefont {Blanc}, \citenamefont {de~France}, \citenamefont
  {K\"oster}, \citenamefont {Leoni}, \citenamefont {Mutti}, \citenamefont
  {Simpson}, \citenamefont {Soldner}, \citenamefont {Ur}, \citenamefont
  {Urban}, \citenamefont {Ahmed}, \citenamefont {Astier}, \citenamefont
  {Augey}, \citenamefont {Back}, \citenamefont {Baczyk}, \citenamefont
  {Bajoga}, \citenamefont {Balabanski}, \citenamefont {Belgya}, \citenamefont
  {Benzoni}, \citenamefont {Bernards}, \citenamefont {Biswas}, \citenamefont
  {Bocchi}, \citenamefont {Bottoni}, \citenamefont {Britton}, \citenamefont
  {Bruyneel}, \citenamefont {Burnett}, \citenamefont {Cakirli}, \citenamefont
  {Carroll}, \citenamefont {Catford}, \citenamefont {Cederwall}, \citenamefont
  {Celikovic}, \citenamefont {Cieplicka-Ory{\'{n}}czak}, \citenamefont
  {Clement}, \citenamefont {Cooper}, \citenamefont {Crespi}, \citenamefont
  {Csatlos}, \citenamefont {Curien}, \citenamefont {Czerwi{\'{n}}ski},
  \citenamefont {Danu}, \citenamefont {Davies}, \citenamefont {Didierjean},
  \citenamefont {Drouet}, \citenamefont {Duch{\^{e}}ne}, \citenamefont
  {Ducoin}, \citenamefont {Eberhardt}, \citenamefont {Erturk}, \citenamefont
  {Fraile}, \citenamefont {Gottardo}, \citenamefont {Grente}, \citenamefont
  {Grocutt}, \citenamefont {Guerrero}, \citenamefont {Guinet}, \citenamefont
  {Hartig}, \citenamefont {Henrich}, \citenamefont {Ignatov}, \citenamefont
  {Ilieva}, \citenamefont {Ivanova}, \citenamefont {John}, \citenamefont
  {John}, \citenamefont {Jolie}, \citenamefont {Kisyov}, \citenamefont
  {Krticka}, \citenamefont {Konstantinopoulos}, \citenamefont {Korgul},
  \citenamefont {Krasznahorkay}, \citenamefont {Kr\"oll}, \citenamefont
  {Kurpeta}, \citenamefont {Kuti}, \citenamefont {Lalkovski}, \citenamefont
  {Larijani}, \citenamefont {Leguillon}, \citenamefont {Lica}, \citenamefont
  {Litaize}, \citenamefont {Lozeva}, \citenamefont {Magron}, \citenamefont
  {Mancuso}, \citenamefont {Martinez}, \citenamefont {Massarczyk},
  \citenamefont {Mazzocchi}, \citenamefont {Melon}, \citenamefont {Mengoni},
  \citenamefont {Michelagnoli}, \citenamefont {Million}, \citenamefont {Mokry},
  \citenamefont {Mukhopadhyay}, \citenamefont {Mulholland}, \citenamefont
  {Nannini}, \citenamefont {Napoli}, \citenamefont {Olaizola}, \citenamefont
  {Orlandi}, \citenamefont {Patel}, \citenamefont {Paziy}, \citenamefont
  {Petrache}, \citenamefont {Pfeiffer}, \citenamefont {Pietralla},
  \citenamefont {Podolyak}, \citenamefont {Ramdhane}, \citenamefont {Redon},
  \citenamefont {Regan}, \citenamefont {Regis}, \citenamefont {Regnier},
  \citenamefont {Oliver}, \citenamefont {Rudigier}, \citenamefont {Runke},
  \citenamefont {Rzaca-Urban}, \citenamefont {Saed-Samii}, \citenamefont
  {Salsac}, \citenamefont {Scheck}, \citenamefont {Schwengner}, \citenamefont
  {Sengele}, \citenamefont {Singh}, \citenamefont {Smith}, \citenamefont
  {Stezowski}, \citenamefont {Szpak}, \citenamefont {Thomas}, \citenamefont
  {Th\"urauf}, \citenamefont {Timar}, \citenamefont {Tom}, \citenamefont
  {Tomandl}, \citenamefont {Tornyi}, \citenamefont {Townsley}, \citenamefont
  {Tuerler}, \citenamefont {Valenta}, \citenamefont {Vancraeyenest},
  \citenamefont {Vandone}, \citenamefont {Vanhoy}, \citenamefont {Vedia},
  \citenamefont {Warr}, \citenamefont {Werner}, \citenamefont {Wilmsen},
  \citenamefont {Wilson}, \citenamefont {Zerrouki},\ and\ \citenamefont
  {Zielinska}}]{Jen17}%
  \BibitemOpen
\bibfield  {journal} {  }\bibfield  {author} {\bibinfo {author} {\bibfnamefont
  {M.}~\bibnamefont {Jentschel}}, \bibinfo {author} {\bibfnamefont
  {A.}~\bibnamefont {Blanc}}, \bibinfo {author} {\bibfnamefont
  {G.}~\bibnamefont {de~France}}, \bibinfo {author} {\bibfnamefont
  {U.}~\bibnamefont {K\"oster}}, \bibinfo {author} {\bibfnamefont
  {S.}~\bibnamefont {Leoni}}, \bibinfo {author} {\bibfnamefont
  {P.}~\bibnamefont {Mutti}}, \bibinfo {author} {\bibfnamefont
  {G.}~\bibnamefont {Simpson}}, \bibinfo {author} {\bibfnamefont
  {T.}~\bibnamefont {Soldner}}, \bibinfo {author} {\bibfnamefont
  {C.}~\bibnamefont {Ur}}, \bibinfo {author} {\bibfnamefont {W.}~\bibnamefont
  {Urban}}, \bibinfo {author} {\bibfnamefont {S.}~\bibnamefont {Ahmed}},
  \bibinfo {author} {\bibfnamefont {A.}~\bibnamefont {Astier}}, \bibinfo
  {author} {\bibfnamefont {L.}~\bibnamefont {Augey}}, \bibinfo {author}
  {\bibfnamefont {T.}~\bibnamefont {Back}}, \bibinfo {author} {\bibfnamefont
  {P.}~\bibnamefont {Baczyk}}, \bibinfo {author} {\bibfnamefont
  {A.}~\bibnamefont {Bajoga}}, \bibinfo {author} {\bibfnamefont
  {D.}~\bibnamefont {Balabanski}}, \bibinfo {author} {\bibfnamefont
  {T.}~\bibnamefont {Belgya}}, \bibinfo {author} {\bibfnamefont
  {G.}~\bibnamefont {Benzoni}}, \bibinfo {author} {\bibfnamefont
  {C.}~\bibnamefont {Bernards}}, \bibinfo {author} {\bibfnamefont
  {D.}~\bibnamefont {Biswas}}, \bibinfo {author} {\bibfnamefont
  {G.}~\bibnamefont {Bocchi}}, \bibinfo {author} {\bibfnamefont
  {S.}~\bibnamefont {Bottoni}}, \bibinfo {author} {\bibfnamefont
  {R.}~\bibnamefont {Britton}}, \bibinfo {author} {\bibfnamefont
  {B.}~\bibnamefont {Bruyneel}}, \bibinfo {author} {\bibfnamefont
  {J.}~\bibnamefont {Burnett}}, \bibinfo {author} {\bibfnamefont
  {R.}~\bibnamefont {Cakirli}}, \bibinfo {author} {\bibfnamefont
  {R.}~\bibnamefont {Carroll}}, \bibinfo {author} {\bibfnamefont
  {W.}~\bibnamefont {Catford}}, \bibinfo {author} {\bibfnamefont
  {B.}~\bibnamefont {Cederwall}}, \bibinfo {author} {\bibfnamefont
  {I.}~\bibnamefont {Celikovic}}, \bibinfo {author} {\bibfnamefont
  {N.}~\bibnamefont {Cieplicka-Ory{\'{n}}czak}}, \bibinfo {author}
  {\bibfnamefont {E.}~\bibnamefont {Clement}}, \bibinfo {author} {\bibfnamefont
  {N.}~\bibnamefont {Cooper}}, \bibinfo {author} {\bibfnamefont
  {F.}~\bibnamefont {Crespi}}, \bibinfo {author} {\bibfnamefont
  {M.}~\bibnamefont {Csatlos}}, \bibinfo {author} {\bibfnamefont
  {D.}~\bibnamefont {Curien}}, \bibinfo {author} {\bibfnamefont
  {M.}~\bibnamefont {Czerwi{\'{n}}ski}}, \bibinfo {author} {\bibfnamefont
  {L.}~\bibnamefont {Danu}}, \bibinfo {author} {\bibfnamefont {A.}~\bibnamefont
  {Davies}}, \bibinfo {author} {\bibfnamefont {F.}~\bibnamefont {Didierjean}},
  \bibinfo {author} {\bibfnamefont {F.}~\bibnamefont {Drouet}}, \bibinfo
  {author} {\bibfnamefont {G.}~\bibnamefont {Duch{\^{e}}ne}}, \bibinfo {author}
  {\bibfnamefont {C.}~\bibnamefont {Ducoin}}, \bibinfo {author} {\bibfnamefont
  {K.}~\bibnamefont {Eberhardt}}, \bibinfo {author} {\bibfnamefont
  {S.}~\bibnamefont {Erturk}}, \bibinfo {author} {\bibfnamefont
  {L.}~\bibnamefont {Fraile}}, \bibinfo {author} {\bibfnamefont
  {A.}~\bibnamefont {Gottardo}}, \bibinfo {author} {\bibfnamefont
  {L.}~\bibnamefont {Grente}}, \bibinfo {author} {\bibfnamefont
  {L.}~\bibnamefont {Grocutt}}, \bibinfo {author} {\bibfnamefont
  {C.}~\bibnamefont {Guerrero}}, \bibinfo {author} {\bibfnamefont
  {D.}~\bibnamefont {Guinet}}, \bibinfo {author} {\bibfnamefont {A.-L.}\
  \bibnamefont {Hartig}}, \bibinfo {author} {\bibfnamefont {C.}~\bibnamefont
  {Henrich}}, \bibinfo {author} {\bibfnamefont {A.}~\bibnamefont {Ignatov}},
  \bibinfo {author} {\bibfnamefont {S.}~\bibnamefont {Ilieva}}, \bibinfo
  {author} {\bibfnamefont {D.}~\bibnamefont {Ivanova}}, \bibinfo {author}
  {\bibfnamefont {B.}~\bibnamefont {John}}, \bibinfo {author} {\bibfnamefont
  {R.}~\bibnamefont {John}}, \bibinfo {author} {\bibfnamefont {J.}~\bibnamefont
  {Jolie}}, \bibinfo {author} {\bibfnamefont {S.}~\bibnamefont {Kisyov}},
  \bibinfo {author} {\bibfnamefont {M.}~\bibnamefont {Krticka}}, \bibinfo
  {author} {\bibfnamefont {T.}~\bibnamefont {Konstantinopoulos}}, \bibinfo
  {author} {\bibfnamefont {A.}~\bibnamefont {Korgul}}, \bibinfo {author}
  {\bibfnamefont {A.}~\bibnamefont {Krasznahorkay}}, \bibinfo {author}
  {\bibfnamefont {T.}~\bibnamefont {Kr\"oll}}, \bibinfo {author} {\bibfnamefont
  {J.}~\bibnamefont {Kurpeta}}, \bibinfo {author} {\bibfnamefont
  {I.}~\bibnamefont {Kuti}}, \bibinfo {author} {\bibfnamefont {S.}~\bibnamefont
  {Lalkovski}}, \bibinfo {author} {\bibfnamefont {C.}~\bibnamefont {Larijani}},
  \bibinfo {author} {\bibfnamefont {R.}~\bibnamefont {Leguillon}}, \bibinfo
  {author} {\bibfnamefont {R.}~\bibnamefont {Lica}}, \bibinfo {author}
  {\bibfnamefont {O.}~\bibnamefont {Litaize}}, \bibinfo {author} {\bibfnamefont
  {R.}~\bibnamefont {Lozeva}}, \bibinfo {author} {\bibfnamefont
  {C.}~\bibnamefont {Magron}}, \bibinfo {author} {\bibfnamefont
  {C.}~\bibnamefont {Mancuso}}, \bibinfo {author} {\bibfnamefont {E.~R.}\
  \bibnamefont {Martinez}}, \bibinfo {author} {\bibfnamefont {R.}~\bibnamefont
  {Massarczyk}}, \bibinfo {author} {\bibfnamefont {C.}~\bibnamefont
  {Mazzocchi}}, \bibinfo {author} {\bibfnamefont {B.}~\bibnamefont {Melon}},
  \bibinfo {author} {\bibfnamefont {D.}~\bibnamefont {Mengoni}}, \bibinfo
  {author} {\bibfnamefont {C.}~\bibnamefont {Michelagnoli}}, \bibinfo {author}
  {\bibfnamefont {B.}~\bibnamefont {Million}}, \bibinfo {author} {\bibfnamefont
  {C.}~\bibnamefont {Mokry}}, \bibinfo {author} {\bibfnamefont
  {S.}~\bibnamefont {Mukhopadhyay}}, \bibinfo {author} {\bibfnamefont
  {K.}~\bibnamefont {Mulholland}}, \bibinfo {author} {\bibfnamefont
  {A.}~\bibnamefont {Nannini}}, \bibinfo {author} {\bibfnamefont
  {D.}~\bibnamefont {Napoli}}, \bibinfo {author} {\bibfnamefont
  {B.}~\bibnamefont {Olaizola}}, \bibinfo {author} {\bibfnamefont
  {R.}~\bibnamefont {Orlandi}}, \bibinfo {author} {\bibfnamefont
  {Z.}~\bibnamefont {Patel}}, \bibinfo {author} {\bibfnamefont
  {V.}~\bibnamefont {Paziy}}, \bibinfo {author} {\bibfnamefont
  {C.}~\bibnamefont {Petrache}}, \bibinfo {author} {\bibfnamefont
  {M.}~\bibnamefont {Pfeiffer}}, \bibinfo {author} {\bibfnamefont
  {N.}~\bibnamefont {Pietralla}}, \bibinfo {author} {\bibfnamefont
  {Z.}~\bibnamefont {Podolyak}}, \bibinfo {author} {\bibfnamefont
  {M.}~\bibnamefont {Ramdhane}}, \bibinfo {author} {\bibfnamefont
  {N.}~\bibnamefont {Redon}}, \bibinfo {author} {\bibfnamefont
  {P.}~\bibnamefont {Regan}}, \bibinfo {author} {\bibfnamefont
  {J.}~\bibnamefont {Regis}}, \bibinfo {author} {\bibfnamefont
  {D.}~\bibnamefont {Regnier}}, \bibinfo {author} {\bibfnamefont {R.~J.}\
  \bibnamefont {Oliver}}, \bibinfo {author} {\bibfnamefont {M.}~\bibnamefont
  {Rudigier}}, \bibinfo {author} {\bibfnamefont {J.}~\bibnamefont {Runke}},
  \bibinfo {author} {\bibfnamefont {T.}~\bibnamefont {Rzaca-Urban}}, \bibinfo
  {author} {\bibfnamefont {N.}~\bibnamefont {Saed-Samii}}, \bibinfo {author}
  {\bibfnamefont {M.}~\bibnamefont {Salsac}}, \bibinfo {author} {\bibfnamefont
  {M.}~\bibnamefont {Scheck}}, \bibinfo {author} {\bibfnamefont
  {R.}~\bibnamefont {Schwengner}}, \bibinfo {author} {\bibfnamefont
  {L.}~\bibnamefont {Sengele}}, \bibinfo {author} {\bibfnamefont
  {P.}~\bibnamefont {Singh}}, \bibinfo {author} {\bibfnamefont
  {J.}~\bibnamefont {Smith}}, \bibinfo {author} {\bibfnamefont
  {O.}~\bibnamefont {Stezowski}}, \bibinfo {author} {\bibfnamefont
  {B.}~\bibnamefont {Szpak}}, \bibinfo {author} {\bibfnamefont
  {T.}~\bibnamefont {Thomas}}, \bibinfo {author} {\bibfnamefont
  {M.}~\bibnamefont {Th\"urauf}}, \bibinfo {author} {\bibfnamefont
  {J.}~\bibnamefont {Timar}}, \bibinfo {author} {\bibfnamefont
  {A.}~\bibnamefont {Tom}}, \bibinfo {author} {\bibfnamefont {I.}~\bibnamefont
  {Tomandl}}, \bibinfo {author} {\bibfnamefont {T.}~\bibnamefont {Tornyi}},
  \bibinfo {author} {\bibfnamefont {C.}~\bibnamefont {Townsley}}, \bibinfo
  {author} {\bibfnamefont {A.}~\bibnamefont {Tuerler}}, \bibinfo {author}
  {\bibfnamefont {S.}~\bibnamefont {Valenta}}, \bibinfo {author} {\bibfnamefont
  {A.}~\bibnamefont {Vancraeyenest}}, \bibinfo {author} {\bibfnamefont
  {V.}~\bibnamefont {Vandone}}, \bibinfo {author} {\bibfnamefont
  {J.}~\bibnamefont {Vanhoy}}, \bibinfo {author} {\bibfnamefont
  {V.}~\bibnamefont {Vedia}}, \bibinfo {author} {\bibfnamefont
  {N.}~\bibnamefont {Warr}}, \bibinfo {author} {\bibfnamefont {V.}~\bibnamefont
  {Werner}}, \bibinfo {author} {\bibfnamefont {D.}~\bibnamefont {Wilmsen}},
  \bibinfo {author} {\bibfnamefont {E.}~\bibnamefont {Wilson}}, \bibinfo
  {author} {\bibfnamefont {T.}~\bibnamefont {Zerrouki}}, \ and\ \bibinfo
  {author} {\bibfnamefont {M.}~\bibnamefont {Zielinska}},\ }\href {\doibase
  10.1088/1748-0221/12/11/p11003} {\bibfield  {journal} {\bibinfo  {journal}
  {J. Instrum.}\ }\textbf {\bibinfo {volume} {12}},\ \bibinfo {pages} {11003}
  (\bibinfo {year} {2017})}\BibitemShut {NoStop}%
\bibitem [{\citenamefont {Moll}(1974)}]{Mol74}%
  \BibitemOpen
  \bibfield  {author} {\bibinfo {author} {\bibfnamefont {E.}~\bibnamefont
  {Moll}},\ }\href@noop {} {\bibfield  {journal} {\bibinfo  {journal} {The
  Franco-German High Flux Reactor and its Facilities for Nuclear Research,
  Nuclear Structure Study with Neutrons (Springer, Boston, MA)}\ } (\bibinfo
  {year} {1974})}\BibitemShut {NoStop}%
\bibitem [{\citenamefont {Abele}\ \emph {et~al.}(2006)\citenamefont {Abele},
  \citenamefont {Dubbers}, \citenamefont {HÃ€se}, \citenamefont {Klein},
  \citenamefont {Knapfler}, \citenamefont {Kreuz}, \citenamefont {Lauer},
  \citenamefont {MÃ€rkisch}, \citenamefont {Mund}, \citenamefont
  {Nesvizhevsky}, \citenamefont {Petoukhov}, \citenamefont {Schmidt},
  \citenamefont {Schumann},\ and\ \citenamefont {Soldner}}]{Abe06}%
  \BibitemOpen
  \bibfield  {author} {\bibinfo {author} {\bibfnamefont {H.}~\bibnamefont
  {Abele}}, \bibinfo {author} {\bibfnamefont {D.}~\bibnamefont {Dubbers}},
  \bibinfo {author} {\bibfnamefont {H.}~\bibnamefont {HÃ€se}}, \bibinfo
  {author} {\bibfnamefont {M.}~\bibnamefont {Klein}}, \bibinfo {author}
  {\bibfnamefont {A.}~\bibnamefont {Knapfler}}, \bibinfo {author}
  {\bibfnamefont {M.}~\bibnamefont {Kreuz}}, \bibinfo {author} {\bibfnamefont
  {T.}~\bibnamefont {Lauer}}, \bibinfo {author} {\bibfnamefont
  {B.}~\bibnamefont {MÃ€rkisch}}, \bibinfo {author} {\bibfnamefont
  {D.}~\bibnamefont {Mund}}, \bibinfo {author} {\bibfnamefont {V.}~\bibnamefont
  {Nesvizhevsky}}, \bibinfo {author} {\bibfnamefont {A.}~\bibnamefont
  {Petoukhov}}, \bibinfo {author} {\bibfnamefont {C.}~\bibnamefont {Schmidt}},
  \bibinfo {author} {\bibfnamefont {M.}~\bibnamefont {Schumann}}, \ and\
  \bibinfo {author} {\bibfnamefont {T.}~\bibnamefont {Soldner}},\ }\href
  {\doibase https://doi.org/10.1016/j.nima.2006.03.020} {\bibfield  {journal}
  {\bibinfo  {journal} {Nucl. Instrum. Meth. A}\ }\textbf {\bibinfo {volume}
  {562}},\ \bibinfo {pages} {407 } (\bibinfo {year} {2006})}\BibitemShut
  {NoStop}%
\bibitem [{\citenamefont {Simpson}\ \emph {et~al.}(2000)\citenamefont
  {Simpson}, \citenamefont {Azaiez}, \citenamefont {DeFrance}, \citenamefont
  {Fouan}, \citenamefont {Gerl}, \citenamefont {Julin}, \citenamefont {Korten},
  \citenamefont {Nolan}, \citenamefont {NyakÃ¶}, \citenamefont {Sletten},\
  and\ \citenamefont {Walker}}]{Sim00}%
  \BibitemOpen
  \bibfield  {author} {\bibinfo {author} {\bibfnamefont {J.}~\bibnamefont
  {Simpson}}, \bibinfo {author} {\bibfnamefont {F.}~\bibnamefont {Azaiez}},
  \bibinfo {author} {\bibfnamefont {G.}~\bibnamefont {DeFrance}}, \bibinfo
  {author} {\bibfnamefont {J.}~\bibnamefont {Fouan}}, \bibinfo {author}
  {\bibfnamefont {J.}~\bibnamefont {Gerl}}, \bibinfo {author} {\bibfnamefont
  {R.}~\bibnamefont {Julin}}, \bibinfo {author} {\bibfnamefont
  {W.}~\bibnamefont {Korten}}, \bibinfo {author} {\bibfnamefont
  {P.}~\bibnamefont {Nolan}}, \bibinfo {author} {\bibfnamefont
  {B.}~\bibnamefont {NyakÃ¶}}, \bibinfo {author} {\bibfnamefont
  {G.}~\bibnamefont {Sletten}}, \ and\ \bibinfo {author} {\bibfnamefont
  {P.}~\bibnamefont {Walker}},\ }\href
  {https://www.scopus.com/inward/record.uri?eid=2-s2.0-0043141469&partnerID=40&md5=15b88e5cc2ae06c3c8e05c7677d8a8d6}
  {\bibfield  {journal} {\bibinfo  {journal} {Acta Phys. Hung., New Ser. Heavy
  Ion Phys.}\ }\textbf {\bibinfo {volume} {11}},\ \bibinfo {pages} {159}
  (\bibinfo {year} {2000})}\BibitemShut {NoStop}%
\bibitem [{\citenamefont {Alvarez}(1993)}]{Alv93}%
  \BibitemOpen
  \bibfield  {author} {\bibinfo {author} {\bibfnamefont {C.~R.}\ \bibnamefont
  {Alvarez}},\ }\href@noop {} {\bibfield  {journal} {\bibinfo  {journal} {Nucl.
  Phys. News}\ ,\ \bibinfo {pages} {3}} (\bibinfo {year} {1993})}\BibitemShut
  {NoStop}%
\bibitem [{\citenamefont {Roberts}\ \emph {et~al.}(2014)\citenamefont
  {Roberts}, \citenamefont {Bruce}, \citenamefont {Regan}, \citenamefont
  {Podolyak}, \citenamefont {Townsley}, \citenamefont {Smith}, \citenamefont
  {Mulholland},\ and\ \citenamefont {Smith}}]{Rob14}%
  \BibitemOpen
  \bibfield  {author} {\bibinfo {author} {\bibfnamefont {O.~J.}\ \bibnamefont
  {Roberts}}, \bibinfo {author} {\bibfnamefont {A.~M.}\ \bibnamefont {Bruce}},
  \bibinfo {author} {\bibfnamefont {P.~H.}\ \bibnamefont {Regan}}, \bibinfo
  {author} {\bibfnamefont {Z.}~\bibnamefont {Podolyak}}, \bibinfo {author}
  {\bibfnamefont {C.~M.}\ \bibnamefont {Townsley}}, \bibinfo {author}
  {\bibfnamefont {J.~F.}\ \bibnamefont {Smith}}, \bibinfo {author}
  {\bibfnamefont {K.~F.}\ \bibnamefont {Mulholland}}, \ and\ \bibinfo {author}
  {\bibfnamefont {A.}~\bibnamefont {Smith}},\ }\href {\doibase
  https://doi.org/10.1016/j.nima.2014.02.037} {\bibfield  {journal} {\bibinfo
  {journal} {Nucl. Instrum. Meth. A}\ }\textbf {\bibinfo {volume} {748}},\
  \bibinfo {pages} {91} (\bibinfo {year} {2014})}\BibitemShut {NoStop}%
\bibitem [{\citenamefont {Regis}\ \emph {et~al.}(2013)\citenamefont {Regis},
  \citenamefont {Mach}, \citenamefont {Simpson}, \citenamefont {Jolie},
  \citenamefont {Pascovici}, \citenamefont {Saed-Samii}, \citenamefont {Warr},
  \citenamefont {Bruce}, \citenamefont {Degenkolb}, \citenamefont {Fraile},
  \citenamefont {Fransen}, \citenamefont {Ghita}, \citenamefont {Kisyov},
  \citenamefont {Koester}, \citenamefont {Korgul}, \citenamefont {Lalkovski},
  \citenamefont {Marginean}, \citenamefont {Mutti}, \citenamefont {Olaizola},
  \citenamefont {Podolyak}, \citenamefont {Regan}, \citenamefont {Roberts},
  \citenamefont {Rudigier}, \citenamefont {Stroe}, \citenamefont {Urban},\ and\
  \citenamefont {Wilmsen}}]{Reg13}%
  \BibitemOpen
  \bibfield  {author} {\bibinfo {author} {\bibfnamefont {J.-M.}\ \bibnamefont
  {Regis}}, \bibinfo {author} {\bibfnamefont {H.}~\bibnamefont {Mach}},
  \bibinfo {author} {\bibfnamefont {G.}~\bibnamefont {Simpson}}, \bibinfo
  {author} {\bibfnamefont {J.}~\bibnamefont {Jolie}}, \bibinfo {author}
  {\bibfnamefont {G.}~\bibnamefont {Pascovici}}, \bibinfo {author}
  {\bibfnamefont {N.}~\bibnamefont {Saed-Samii}}, \bibinfo {author}
  {\bibfnamefont {N.}~\bibnamefont {Warr}}, \bibinfo {author} {\bibfnamefont
  {A.}~\bibnamefont {Bruce}}, \bibinfo {author} {\bibfnamefont
  {J.}~\bibnamefont {Degenkolb}}, \bibinfo {author} {\bibfnamefont
  {L.}~\bibnamefont {Fraile}}, \bibinfo {author} {\bibfnamefont
  {C.}~\bibnamefont {Fransen}}, \bibinfo {author} {\bibfnamefont
  {D.}~\bibnamefont {Ghita}}, \bibinfo {author} {\bibfnamefont
  {S.}~\bibnamefont {Kisyov}}, \bibinfo {author} {\bibfnamefont
  {U.}~\bibnamefont {Koester}}, \bibinfo {author} {\bibfnamefont
  {A.}~\bibnamefont {Korgul}}, \bibinfo {author} {\bibfnamefont
  {S.}~\bibnamefont {Lalkovski}}, \bibinfo {author} {\bibfnamefont
  {N.}~\bibnamefont {Marginean}}, \bibinfo {author} {\bibfnamefont
  {P.}~\bibnamefont {Mutti}}, \bibinfo {author} {\bibfnamefont
  {B.}~\bibnamefont {Olaizola}}, \bibinfo {author} {\bibfnamefont
  {Z.}~\bibnamefont {Podolyak}}, \bibinfo {author} {\bibfnamefont
  {P.}~\bibnamefont {Regan}}, \bibinfo {author} {\bibfnamefont
  {O.}~\bibnamefont {Roberts}}, \bibinfo {author} {\bibfnamefont
  {M.}~\bibnamefont {Rudigier}}, \bibinfo {author} {\bibfnamefont
  {L.}~\bibnamefont {Stroe}}, \bibinfo {author} {\bibfnamefont
  {W.}~\bibnamefont {Urban}}, \ and\ \bibinfo {author} {\bibfnamefont
  {D.}~\bibnamefont {Wilmsen}},\ }\href {\doibase
  https://doi.org/10.1016/j.nima.2013.05.126} {\bibfield  {journal} {\bibinfo
  {journal} {Nucl. Instrum. Meth. A}\ }\textbf {\bibinfo {volume} {726}},\
  \bibinfo {pages} {191 } (\bibinfo {year} {2013})}\BibitemShut {NoStop}%
\bibitem [{\citenamefont {Morinaga}(1976)}]{Mor76}%
  \BibitemOpen
  \bibfield  {author} {\bibinfo {author} {\bibfnamefont {H.}~\bibnamefont
  {Morinaga}},\ }\href@noop {} {\bibfield  {journal} {\bibinfo  {journal}
  {In-beam gamma-ray spectroscopy (North-Holland; Amsterdam, Netherlands)}\ }
  (\bibinfo {year} {1976})}\BibitemShut {NoStop}%
\bibitem [{\citenamefont {Mughabghab}()}]{Mug18}%
  \BibitemOpen
  \bibfield  {author} {\bibinfo {author} {\bibfnamefont {S.~F.}\ \bibnamefont
  {Mughabghab}},\ }\href
  {https://www.sciencedirect.com/book/9780444637864/atlas-of-neutron-resonances}
  {\bibinfo  {journal} {Atlas of Neutron Resonances 6$^{th}$ edn. (Elsevier
  Upton, 2018)}\ }\BibitemShut {NoStop}%
\bibitem [{\citenamefont {Gruppelaar}\ and\ \citenamefont
  {Spilling}(1967)}]{Gru67}%
  \BibitemOpen
\bibfield  {journal} {  }\bibfield  {author} {\bibinfo {author} {\bibfnamefont
  {H.}~\bibnamefont {Gruppelaar}}\ and\ \bibinfo {author} {\bibfnamefont
  {P.}~\bibnamefont {Spilling}},\ }\href {\doibase
  https://doi.org/10.1016/0375-9474(67)90333-8} {\bibfield  {journal} {\bibinfo
   {journal} {Nucl. Phys. A}\ }\textbf {\bibinfo {volume} {102}},\ \bibinfo
  {pages} {226} (\bibinfo {year} {1967})}\BibitemShut {NoStop}%
\bibitem [{\citenamefont {Cranston}\ \emph {et~al.}(1970)\citenamefont
  {Cranston}, \citenamefont {Birkett}, \citenamefont {White},\ and\
  \citenamefont {Hughes}}]{Cra70}%
  \BibitemOpen
  \bibfield  {author} {\bibinfo {author} {\bibfnamefont {F.}~\bibnamefont
  {Cranston}}, \bibinfo {author} {\bibfnamefont {R.}~\bibnamefont {Birkett}},
  \bibinfo {author} {\bibfnamefont {D.}~\bibnamefont {White}}, \ and\ \bibinfo
  {author} {\bibfnamefont {J.}~\bibnamefont {Hughes}},\ }\href {\doibase
  https://doi.org/10.1016/0375-9474(70)90780-3} {\bibfield  {journal} {\bibinfo
   {journal} {Nucl. Phys. A}\ }\textbf {\bibinfo {volume} {153}},\ \bibinfo
  {pages} {413} (\bibinfo {year} {1970})}\BibitemShut {NoStop}%
\bibitem [{\citenamefont {{Arnell}}\ \emph {et~al.}(1969)\citenamefont
  {{Arnell}}, \citenamefont {{Hardell}}, \citenamefont {{Skeppstedt}},\ and\
  \citenamefont {{Wallander}}}]{Arn69}%
  \BibitemOpen
  \bibfield  {author} {\bibinfo {author} {\bibfnamefont {S.~E.}\ \bibnamefont
  {{Arnell}}}, \bibinfo {author} {\bibfnamefont {R.}~\bibnamefont {{Hardell}}},
  \bibinfo {author} {\bibfnamefont {O.}~\bibnamefont {{Skeppstedt}}}, \ and\
  \bibinfo {author} {\bibfnamefont {E.}~\bibnamefont {{Wallander}}},\
  }\bibfield  {booktitle} {\emph {\bibinfo {booktitle}
  {Proc.Intern.Symp.Neutron Capture Gamma-Ray Spectroscopy}},\ }\href@noop {}
  {\  (\bibinfo {year} {1969})}\BibitemShut {NoStop}%
\bibitem [{\citenamefont {Nesaraja}\ and\ \citenamefont
  {McCutchan}(2016)}]{Nes16}%
  \BibitemOpen
  \bibfield  {author} {\bibinfo {author} {\bibfnamefont {C.}~\bibnamefont
  {Nesaraja}}\ and\ \bibinfo {author} {\bibfnamefont {E.}~\bibnamefont
  {McCutchan}},\ }\href {\doibase https://doi.org/10.1016/j.nds.2016.02.001}
  {\bibfield  {journal} {\bibinfo  {journal} {Nucl. Data Sheets}\ }\textbf
  {\bibinfo {volume} {133}},\ \bibinfo {pages} {1 } (\bibinfo {year}
  {2016})}\BibitemShut {NoStop}%
\bibitem [{\citenamefont {Burrows}(2007)}]{Bur07}%
  \BibitemOpen
  \bibfield  {author} {\bibinfo {author} {\bibfnamefont {T.}~\bibnamefont
  {Burrows}},\ }\href {\doibase https://doi.org/10.1016/j.nds.2007.04.002}
  {\bibfield  {journal} {\bibinfo  {journal} {Nucl. Data Sheets}\ }\textbf
  {\bibinfo {volume} {108}},\ \bibinfo {pages} {923 } (\bibinfo {year}
  {2007})}\BibitemShut {NoStop}%
\bibitem [{\citenamefont {Burrows}(2008)}]{Bur08}%
  \BibitemOpen
  \bibfield  {author} {\bibinfo {author} {\bibfnamefont {T.}~\bibnamefont
  {Burrows}},\ }\href {\doibase https://doi.org/10.1016/j.nds.2008.07.001}
  {\bibfield  {journal} {\bibinfo  {journal} {Nucl. Data Sheets}\ }\textbf
  {\bibinfo {volume} {109}},\ \bibinfo {pages} {1879 } (\bibinfo {year}
  {2008})}\BibitemShut {NoStop}%
\bibitem [{PGA()}]{PGAA}%
  \BibitemOpen
  \href {https://www-nds.iaea.org/pgaa/} {\bibinfo  {journal} {Nucler Data
  Services, IAEA, Database for Prompt Gamma-ray Neutron Activation Analysis
  (2003) https://www-nds.iaea.org/pgaa/}\ }\BibitemShut {NoStop}%
\bibitem [{\citenamefont {Col\'o}\ \emph {et~al.}(2013)\citenamefont {Col\'o},
  \citenamefont {Cao}, \citenamefont {Giai},\ and\ \citenamefont
  {Capelli}}]{Col13}%
  \BibitemOpen
\bibfield  {journal} {  }\bibfield  {author} {\bibinfo {author} {\bibfnamefont
  {G.}~\bibnamefont {Col\'o}}, \bibinfo {author} {\bibfnamefont
  {L.}~\bibnamefont {Cao}}, \bibinfo {author} {\bibfnamefont {N.~V.}\
  \bibnamefont {Giai}}, \ and\ \bibinfo {author} {\bibfnamefont
  {L.}~\bibnamefont {Capelli}},\ }\href {\doibase
  https://doi.org/10.1016/j.cpc.2012.07.016} {\bibfield  {journal} {\bibinfo
  {journal} {Comp. Phys. Comm.}\ }\textbf {\bibinfo {volume} {184}},\ \bibinfo
  {pages} {142 } (\bibinfo {year} {2013})}\BibitemShut {NoStop}%
\bibitem [{\citenamefont {Col\`o}\ \emph {et~al.}(2010)\citenamefont {Col\`o},
  \citenamefont {Sagawa},\ and\ \citenamefont {Bortignon}}]{Col10}%
  \BibitemOpen
  \bibfield  {author} {\bibinfo {author} {\bibfnamefont {G.}~\bibnamefont
  {Col\`o}}, \bibinfo {author} {\bibfnamefont {H.}~\bibnamefont {Sagawa}}, \
  and\ \bibinfo {author} {\bibfnamefont {P.~F.}\ \bibnamefont {Bortignon}},\
  }\href {\doibase 10.1103/PhysRevC.82.064307} {\bibfield  {journal} {\bibinfo
  {journal} {Phys. Rev. C}\ }\textbf {\bibinfo {volume} {82}},\ \bibinfo
  {pages} {064307} (\bibinfo {year} {2010})}\BibitemShut {NoStop}%
\bibitem [{\citenamefont {Bottoni}\ \emph {et~al.}(2018)\citenamefont
  {Bottoni}, \citenamefont {Cieplicka-Ory\'nczak}, \citenamefont {Bocchi},
  \citenamefont {Leoni}, \citenamefont {Fornal}, \citenamefont {Col\`o},
  \citenamefont {Bortignon}, \citenamefont {Benzoni}, \citenamefont {Blanc},
  \citenamefont {Bracco}, \citenamefont {Crespi}, \citenamefont {Jentschel},
  \citenamefont {K\"oster}, \citenamefont {Michelagnoli}, \citenamefont
  {Million}, \citenamefont {Mutti}, \citenamefont {Soldner}, \citenamefont
  {T\"urler}, \citenamefont {Ur},\ and\ \citenamefont {Urban}}]{Bot18}%
  \BibitemOpen
  \bibfield  {author} {\bibinfo {author} {\bibfnamefont {S.}~\bibnamefont
  {Bottoni}}, \bibinfo {author} {\bibfnamefont {N.}~\bibnamefont
  {Cieplicka-Ory\'nczak}}, \bibinfo {author} {\bibfnamefont {G.}~\bibnamefont
  {Bocchi}}, \bibinfo {author} {\bibfnamefont {S.}~\bibnamefont {Leoni}},
  \bibinfo {author} {\bibfnamefont {B.}~\bibnamefont {Fornal}}, \bibinfo
  {author} {\bibfnamefont {G.}~\bibnamefont {Col\`o}}, \bibinfo {author}
  {\bibfnamefont {P.}~\bibnamefont {Bortignon}}, \bibinfo {author}
  {\bibfnamefont {G.}~\bibnamefont {Benzoni}}, \bibinfo {author} {\bibfnamefont
  {A.}~\bibnamefont {Blanc}}, \bibinfo {author} {\bibfnamefont
  {A.}~\bibnamefont {Bracco}}, \bibinfo {author} {\bibfnamefont
  {F.}~\bibnamefont {Crespi}}, \bibinfo {author} {\bibfnamefont
  {M.}~\bibnamefont {Jentschel}}, \bibinfo {author} {\bibfnamefont
  {U.}~\bibnamefont {K\"oster}}, \bibinfo {author} {\bibfnamefont
  {C.}~\bibnamefont {Michelagnoli}}, \bibinfo {author} {\bibfnamefont
  {B.}~\bibnamefont {Million}}, \bibinfo {author} {\bibfnamefont
  {P.}~\bibnamefont {Mutti}}, \bibinfo {author} {\bibfnamefont
  {T.}~\bibnamefont {Soldner}}, \bibinfo {author} {\bibfnamefont
  {A.}~\bibnamefont {T\"urler}}, \bibinfo {author} {\bibfnamefont
  {C.}~\bibnamefont {Ur}}, \ and\ \bibinfo {author} {\bibfnamefont
  {W.}~\bibnamefont {Urban}},\ }\href {\doibase 10.1051/epjconf/201819305001}
  {\bibfield  {journal} {\bibinfo  {journal} {EPJ Web Conf.}\ }\textbf
  {\bibinfo {volume} {193}},\ \bibinfo {pages} {05001} (\bibinfo {year}
  {2018})}\BibitemShut {NoStop}%
\bibitem [{\citenamefont {Alex~Brown}(1998)}]{Bro98}%
  \BibitemOpen
  \bibfield  {author} {\bibinfo {author} {\bibfnamefont {B.}~\bibnamefont
  {Alex~Brown}},\ }\href {\doibase 10.1103/PhysRevC.58.220} {\bibfield
  {journal} {\bibinfo  {journal} {Phys. Rev. C}\ }\textbf {\bibinfo {volume}
  {58}},\ \bibinfo {pages} {220} (\bibinfo {year} {1998})}\BibitemShut
  {NoStop}%
\bibitem [{\citenamefont {Chabanat}\ \emph {et~al.}(1998)\citenamefont
  {Chabanat}, \citenamefont {Bonche}, \citenamefont {Haensel}, \citenamefont
  {Meyer},\ and\ \citenamefont {Schaeffer}}]{Cha98}%
  \BibitemOpen
  \bibfield  {author} {\bibinfo {author} {\bibfnamefont {E.}~\bibnamefont
  {Chabanat}}, \bibinfo {author} {\bibfnamefont {P.}~\bibnamefont {Bonche}},
  \bibinfo {author} {\bibfnamefont {P.}~\bibnamefont {Haensel}}, \bibinfo
  {author} {\bibfnamefont {J.}~\bibnamefont {Meyer}}, \ and\ \bibinfo {author}
  {\bibfnamefont {R.}~\bibnamefont {Schaeffer}},\ }\href {\doibase
  https://doi.org/10.1016/S0375-9474(98)00180-8} {\bibfield  {journal}
  {\bibinfo  {journal} {Nucl. Phys. A}\ }\textbf {\bibinfo {volume} {635}},\
  \bibinfo {pages} {231 } (\bibinfo {year} {1998})}\BibitemShut {NoStop}%
\bibitem [{\citenamefont {Poves}\ \emph {et~al.}(2001)\citenamefont {Poves},
  \citenamefont {Sanchez-Solano}, \citenamefont {Caurier},\ and\ \citenamefont
  {Nowacki}}]{Pov01}%
  \BibitemOpen
  \bibfield  {author} {\bibinfo {author} {\bibfnamefont {A.}~\bibnamefont
  {Poves}}, \bibinfo {author} {\bibfnamefont {J.}~\bibnamefont
  {Sanchez-Solano}}, \bibinfo {author} {\bibfnamefont {E.}~\bibnamefont
  {Caurier}}, \ and\ \bibinfo {author} {\bibfnamefont {F.}~\bibnamefont
  {Nowacki}},\ }\href {\doibase https://doi.org/10.1016/S0375-9474(01)00967-8}
  {\bibfield  {journal} {\bibinfo  {journal} {Nucl. Phys. A}\ }\textbf
  {\bibinfo {volume} {694}},\ \bibinfo {pages} {157 } (\bibinfo {year}
  {2001})}\BibitemShut {NoStop}%
\bibitem [{\citenamefont {Bocchi}\ \emph {et~al.}(2014)\citenamefont {Bocchi},
  \citenamefont {Leoni}, \citenamefont {Bottoni}, \citenamefont {Benzoni},
  \citenamefont {Bracco}, \citenamefont {Bortignon}, \citenamefont {Col\`o},
  \citenamefont {Belvito}, \citenamefont {Ni\ifmmode \mbox{\c{t}}\else
  \c{t}\fi{}\ifmmode~\u{a}\else \u{a}\fi{}}, \citenamefont {Marginean},
  \citenamefont {Filipescu}, \citenamefont {Ghita}, \citenamefont {Glodariu},
  \citenamefont {Lica}, \citenamefont {Marginean}, \citenamefont {Mihai},
  \citenamefont {Negret}, \citenamefont {Sava}, \citenamefont {Stroe},
  \citenamefont {Toma}, \citenamefont {Bucurescu}, \citenamefont {Georghe},
  \citenamefont {Suv\ifmmode \u{a}\else \u{a}\fi{}il\ifmmode~\u{a}\else
  \u{a}\fi{}}, \citenamefont {Deleanu}, \citenamefont {Ur},\ and\ \citenamefont
  {Aydin}}]{Boc14}%
  \BibitemOpen
  \bibfield  {author} {\bibinfo {author} {\bibfnamefont {G.}~\bibnamefont
  {Bocchi}}, \bibinfo {author} {\bibfnamefont {S.}~\bibnamefont {Leoni}},
  \bibinfo {author} {\bibfnamefont {S.}~\bibnamefont {Bottoni}}, \bibinfo
  {author} {\bibfnamefont {G.}~\bibnamefont {Benzoni}}, \bibinfo {author}
  {\bibfnamefont {A.}~\bibnamefont {Bracco}}, \bibinfo {author} {\bibfnamefont
  {P.~F.}\ \bibnamefont {Bortignon}}, \bibinfo {author} {\bibfnamefont
  {G.}~\bibnamefont {Col\`o}}, \bibinfo {author} {\bibfnamefont
  {B.}~\bibnamefont {Belvito}}, \bibinfo {author} {\bibfnamefont {C.~R.}\
  \bibnamefont {Ni\ifmmode \mbox{\c{t}}\else \c{t}\fi{}\ifmmode~\u{a}\else
  \u{a}\fi{}}}, \bibinfo {author} {\bibfnamefont {N.}~\bibnamefont
  {Marginean}}, \bibinfo {author} {\bibfnamefont {D.}~\bibnamefont
  {Filipescu}}, \bibinfo {author} {\bibfnamefont {D.}~\bibnamefont {Ghita}},
  \bibinfo {author} {\bibfnamefont {T.}~\bibnamefont {Glodariu}}, \bibinfo
  {author} {\bibfnamefont {R.}~\bibnamefont {Lica}}, \bibinfo {author}
  {\bibfnamefont {R.}~\bibnamefont {Marginean}}, \bibinfo {author}
  {\bibfnamefont {C.}~\bibnamefont {Mihai}}, \bibinfo {author} {\bibfnamefont
  {A.}~\bibnamefont {Negret}}, \bibinfo {author} {\bibfnamefont
  {T.}~\bibnamefont {Sava}}, \bibinfo {author} {\bibfnamefont {L.}~\bibnamefont
  {Stroe}}, \bibinfo {author} {\bibfnamefont {S.}~\bibnamefont {Toma}},
  \bibinfo {author} {\bibfnamefont {D.}~\bibnamefont {Bucurescu}}, \bibinfo
  {author} {\bibfnamefont {I.}~\bibnamefont {Georghe}}, \bibinfo {author}
  {\bibfnamefont {R.}~\bibnamefont {Suv\ifmmode \u{a}\else
  \u{a}\fi{}il\ifmmode~\u{a}\else \u{a}\fi{}}}, \bibinfo {author}
  {\bibfnamefont {D.}~\bibnamefont {Deleanu}}, \bibinfo {author} {\bibfnamefont
  {C.~A.}\ \bibnamefont {Ur}}, \ and\ \bibinfo {author} {\bibfnamefont
  {S.}~\bibnamefont {Aydin}},\ }\href {\doibase 10.1103/PhysRevC.89.054302}
  {\bibfield  {journal} {\bibinfo  {journal} {Phys. Rev. C}\ }\textbf {\bibinfo
  {volume} {89}},\ \bibinfo {pages} {054302} (\bibinfo {year}
  {2014})}\BibitemShut {NoStop}%
\bibitem [{\citenamefont {Ni\ifmmode \mbox{\c{t}}\else
  \c{t}\fi{}\ifmmode~\u{a}\else \u{a}\fi{}}\ \emph {et~al.}(2014)\citenamefont
  {Ni\ifmmode \mbox{\c{t}}\else \c{t}\fi{}\ifmmode~\u{a}\else \u{a}\fi{}},
  \citenamefont {Bucurescu}, \citenamefont {M\ifmmode~\u{a}\else
  \u{a}\fi{}rginean}, \citenamefont {Avrigeanu}, \citenamefont {Bocchi},
  \citenamefont {Bottoni}, \citenamefont {Bracco}, \citenamefont {Bruce},
  \citenamefont {C\ifmmode \u{a}\else~\u{a}\fi{}ta Danil}, \citenamefont
  {Col\'o}, \citenamefont {Deleanu}, \citenamefont {Filipescu}, \citenamefont
  {Ghi\ifmmode \mbox{\c{t}}\else \c{t}\fi{}\ifmmode~\u{a}\else \u{a}\fi{}},
  \citenamefont {Glodariu}, \citenamefont {Leoni}, \citenamefont {Mihai},
  \citenamefont {Mason}, \citenamefont {M\ifmmode~\u{a}\else
  \u{a}\fi{}rginean}, \citenamefont {Negret}, \citenamefont
  {Pantelic\ifmmode~\u{a}\else \u{a}\fi{}}, \citenamefont {Podolyak},
  \citenamefont {Regan}, \citenamefont {Sava}, \citenamefont {Stroe},
  \citenamefont {Toma}, \citenamefont {Ur},\ and\ \citenamefont
  {Wilson}}]{NIt14}%
  \BibitemOpen
  \bibfield  {author} {\bibinfo {author} {\bibfnamefont {C.~R.}\ \bibnamefont
  {Ni\ifmmode \mbox{\c{t}}\else \c{t}\fi{}\ifmmode~\u{a}\else \u{a}\fi{}}},
  \bibinfo {author} {\bibfnamefont {D.}~\bibnamefont {Bucurescu}}, \bibinfo
  {author} {\bibfnamefont {N.}~\bibnamefont {M\ifmmode~\u{a}\else
  \u{a}\fi{}rginean}}, \bibinfo {author} {\bibfnamefont {M.}~\bibnamefont
  {Avrigeanu}}, \bibinfo {author} {\bibfnamefont {G.}~\bibnamefont {Bocchi}},
  \bibinfo {author} {\bibfnamefont {S.}~\bibnamefont {Bottoni}}, \bibinfo
  {author} {\bibfnamefont {A.}~\bibnamefont {Bracco}}, \bibinfo {author}
  {\bibfnamefont {A.~M.}\ \bibnamefont {Bruce}}, \bibinfo {author}
  {\bibfnamefont {G.}~\bibnamefont {C\ifmmode \u{a}\else~\u{a}\fi{}ta Danil}},
  \bibinfo {author} {\bibfnamefont {G.}~\bibnamefont {Col\'o}}, \bibinfo
  {author} {\bibfnamefont {D.}~\bibnamefont {Deleanu}}, \bibinfo {author}
  {\bibfnamefont {D.}~\bibnamefont {Filipescu}}, \bibinfo {author}
  {\bibfnamefont {D.~G.}\ \bibnamefont {Ghi\ifmmode \mbox{\c{t}}\else
  \c{t}\fi{}\ifmmode~\u{a}\else \u{a}\fi{}}}, \bibinfo {author} {\bibfnamefont
  {T.}~\bibnamefont {Glodariu}}, \bibinfo {author} {\bibfnamefont
  {S.}~\bibnamefont {Leoni}}, \bibinfo {author} {\bibfnamefont
  {C.}~\bibnamefont {Mihai}}, \bibinfo {author} {\bibfnamefont {P.~J.~R.}\
  \bibnamefont {Mason}}, \bibinfo {author} {\bibfnamefont {R.}~\bibnamefont
  {M\ifmmode~\u{a}\else \u{a}\fi{}rginean}}, \bibinfo {author} {\bibfnamefont
  {A.}~\bibnamefont {Negret}}, \bibinfo {author} {\bibfnamefont
  {D.}~\bibnamefont {Pantelic\ifmmode~\u{a}\else \u{a}\fi{}}}, \bibinfo
  {author} {\bibfnamefont {Z.}~\bibnamefont {Podolyak}}, \bibinfo {author}
  {\bibfnamefont {P.~H.}\ \bibnamefont {Regan}}, \bibinfo {author}
  {\bibfnamefont {T.}~\bibnamefont {Sava}}, \bibinfo {author} {\bibfnamefont
  {L.}~\bibnamefont {Stroe}}, \bibinfo {author} {\bibfnamefont
  {S.}~\bibnamefont {Toma}}, \bibinfo {author} {\bibfnamefont {C.~A.}\
  \bibnamefont {Ur}}, \ and\ \bibinfo {author} {\bibfnamefont {E.}~\bibnamefont
  {Wilson}},\ }\href {\doibase 10.1103/PhysRevC.89.064314} {\bibfield
  {journal} {\bibinfo  {journal} {Phys. Rev. C}\ }\textbf {\bibinfo {volume}
  {89}},\ \bibinfo {pages} {064314} (\bibinfo {year} {2014})}\BibitemShut
  {NoStop}%
\bibitem [{\citenamefont {Crider}\ \emph {et~al.}(2016)\citenamefont {Crider},
  \citenamefont {Prokop}, \citenamefont {Liddick}, \citenamefont {Al-Shudifat},
  \citenamefont {Ayangeakaa}, \citenamefont {Carpenter}, \citenamefont
  {Carroll}, \citenamefont {Chen}, \citenamefont {Chiara}, \citenamefont
  {David}, \citenamefont {Dombos}, \citenamefont {Go}, \citenamefont
  {Grzywacz}, \citenamefont {Harker}, \citenamefont {Janssens}, \citenamefont
  {Larson}, \citenamefont {Lauritsen}, \citenamefont {Lewis}, \citenamefont
  {Quinn}, \citenamefont {Recchia}, \citenamefont {Spyrou}, \citenamefont
  {Suchyta}, \citenamefont {Walters},\ and\ \citenamefont {Zhu}}]{Cri16}%
  \BibitemOpen
  \bibfield  {author} {\bibinfo {author} {\bibfnamefont {B.}~\bibnamefont
  {Crider}}, \bibinfo {author} {\bibfnamefont {C.}~\bibnamefont {Prokop}},
  \bibinfo {author} {\bibfnamefont {S.}~\bibnamefont {Liddick}}, \bibinfo
  {author} {\bibfnamefont {M.}~\bibnamefont {Al-Shudifat}}, \bibinfo {author}
  {\bibfnamefont {A.}~\bibnamefont {Ayangeakaa}}, \bibinfo {author}
  {\bibfnamefont {M.}~\bibnamefont {Carpenter}}, \bibinfo {author}
  {\bibfnamefont {J.}~\bibnamefont {Carroll}}, \bibinfo {author} {\bibfnamefont
  {J.}~\bibnamefont {Chen}}, \bibinfo {author} {\bibfnamefont {C.}~\bibnamefont
  {Chiara}}, \bibinfo {author} {\bibfnamefont {H.}~\bibnamefont {David}},
  \bibinfo {author} {\bibfnamefont {A.}~\bibnamefont {Dombos}}, \bibinfo
  {author} {\bibfnamefont {S.}~\bibnamefont {Go}}, \bibinfo {author}
  {\bibfnamefont {R.}~\bibnamefont {Grzywacz}}, \bibinfo {author}
  {\bibfnamefont {J.}~\bibnamefont {Harker}}, \bibinfo {author} {\bibfnamefont
  {R.}~\bibnamefont {Janssens}}, \bibinfo {author} {\bibfnamefont
  {N.}~\bibnamefont {Larson}}, \bibinfo {author} {\bibfnamefont
  {T.}~\bibnamefont {Lauritsen}}, \bibinfo {author} {\bibfnamefont
  {R.}~\bibnamefont {Lewis}}, \bibinfo {author} {\bibfnamefont
  {S.}~\bibnamefont {Quinn}}, \bibinfo {author} {\bibfnamefont
  {F.}~\bibnamefont {Recchia}}, \bibinfo {author} {\bibfnamefont
  {A.}~\bibnamefont {Spyrou}}, \bibinfo {author} {\bibfnamefont
  {S.}~\bibnamefont {Suchyta}}, \bibinfo {author} {\bibfnamefont
  {W.}~\bibnamefont {Walters}}, \ and\ \bibinfo {author} {\bibfnamefont
  {S.}~\bibnamefont {Zhu}},\ }\href {\doibase
  https://doi.org/10.1016/j.physletb.2016.10.020} {\bibfield  {journal}
  {\bibinfo  {journal} {Phys. Lett. B}\ }\textbf {\bibinfo {volume} {763}},\
  \bibinfo {pages} {108 } (\bibinfo {year} {2016})}\BibitemShut {NoStop}%
\bibitem [{\citenamefont {Morales}\ \emph {et~al.}(2016)\citenamefont
  {Morales}, \citenamefont {Benzoni}, \citenamefont {Watanabe}, \citenamefont
  {Nishimura}, \citenamefont {Browne}, \citenamefont {Daido}, \citenamefont
  {Doornenbal}, \citenamefont {Fang}, \citenamefont {Lorusso}, \citenamefont
  {Patel}, \citenamefont {Rice}, \citenamefont {Sinclair}, \citenamefont
  {S\"oderstr\"om}, \citenamefont {Sumikama}, \citenamefont {Wu}, \citenamefont
  {Xu}, \citenamefont {Yagi}, \citenamefont {Yokoyama}, \citenamefont {Baba},
  \citenamefont {Avigo}, \citenamefont {Bello~Garrote}, \citenamefont {Blasi},
  \citenamefont {Bracco}, \citenamefont {Camera}, \citenamefont {Ceruti},
  \citenamefont {Crespi}, \citenamefont {de~Angelis}, \citenamefont {Delattre},
  \citenamefont {Dombradi}, \citenamefont {Gottardo}, \citenamefont {Isobe},
  \citenamefont {Kojouharov}, \citenamefont {Kurz}, \citenamefont {Kuti},
  \citenamefont {Matsui}, \citenamefont {Melon}, \citenamefont {Mengoni},
  \citenamefont {Miyazaki}, \citenamefont {Modamio-Hoyborg}, \citenamefont
  {Momiyama}, \citenamefont {Napoli}, \citenamefont {Niikura}, \citenamefont
  {Orlandi}, \citenamefont {Sakurai}, \citenamefont {Sahin}, \citenamefont
  {Sohler}, \citenamefont {Shaffner}, \citenamefont {Taniuchi}, \citenamefont
  {Taprogge}, \citenamefont {Vajta}, \citenamefont {Valiente-Dob\'on},
  \citenamefont {Wieland},\ and\ \citenamefont {Yalcinkaya}}]{Mor16}%
  \BibitemOpen
  \bibfield  {author} {\bibinfo {author} {\bibfnamefont {A.~I.}\ \bibnamefont
  {Morales}}, \bibinfo {author} {\bibfnamefont {G.}~\bibnamefont {Benzoni}},
  \bibinfo {author} {\bibfnamefont {H.}~\bibnamefont {Watanabe}}, \bibinfo
  {author} {\bibfnamefont {S.}~\bibnamefont {Nishimura}}, \bibinfo {author}
  {\bibfnamefont {F.}~\bibnamefont {Browne}}, \bibinfo {author} {\bibfnamefont
  {R.}~\bibnamefont {Daido}}, \bibinfo {author} {\bibfnamefont
  {P.}~\bibnamefont {Doornenbal}}, \bibinfo {author} {\bibfnamefont
  {Y.}~\bibnamefont {Fang}}, \bibinfo {author} {\bibfnamefont {G.}~\bibnamefont
  {Lorusso}}, \bibinfo {author} {\bibfnamefont {Z.}~\bibnamefont {Patel}},
  \bibinfo {author} {\bibfnamefont {S.}~\bibnamefont {Rice}}, \bibinfo {author}
  {\bibfnamefont {L.}~\bibnamefont {Sinclair}}, \bibinfo {author}
  {\bibfnamefont {P.-A.}\ \bibnamefont {S\"oderstr\"om}}, \bibinfo {author}
  {\bibfnamefont {T.}~\bibnamefont {Sumikama}}, \bibinfo {author}
  {\bibfnamefont {J.}~\bibnamefont {Wu}}, \bibinfo {author} {\bibfnamefont
  {Z.~Y.}\ \bibnamefont {Xu}}, \bibinfo {author} {\bibfnamefont
  {A.}~\bibnamefont {Yagi}}, \bibinfo {author} {\bibfnamefont {R.}~\bibnamefont
  {Yokoyama}}, \bibinfo {author} {\bibfnamefont {H.}~\bibnamefont {Baba}},
  \bibinfo {author} {\bibfnamefont {R.}~\bibnamefont {Avigo}}, \bibinfo
  {author} {\bibfnamefont {F.~L.}\ \bibnamefont {Bello~Garrote}}, \bibinfo
  {author} {\bibfnamefont {N.}~\bibnamefont {Blasi}}, \bibinfo {author}
  {\bibfnamefont {A.}~\bibnamefont {Bracco}}, \bibinfo {author} {\bibfnamefont
  {F.}~\bibnamefont {Camera}}, \bibinfo {author} {\bibfnamefont
  {S.}~\bibnamefont {Ceruti}}, \bibinfo {author} {\bibfnamefont {F.~C.~L.}\
  \bibnamefont {Crespi}}, \bibinfo {author} {\bibfnamefont {G.}~\bibnamefont
  {de~Angelis}}, \bibinfo {author} {\bibfnamefont {M.-C.}\ \bibnamefont
  {Delattre}}, \bibinfo {author} {\bibfnamefont {Z.}~\bibnamefont {Dombradi}},
  \bibinfo {author} {\bibfnamefont {A.}~\bibnamefont {Gottardo}}, \bibinfo
  {author} {\bibfnamefont {T.}~\bibnamefont {Isobe}}, \bibinfo {author}
  {\bibfnamefont {I.}~\bibnamefont {Kojouharov}}, \bibinfo {author}
  {\bibfnamefont {N.}~\bibnamefont {Kurz}}, \bibinfo {author} {\bibfnamefont
  {I.}~\bibnamefont {Kuti}}, \bibinfo {author} {\bibfnamefont {K.}~\bibnamefont
  {Matsui}}, \bibinfo {author} {\bibfnamefont {B.}~\bibnamefont {Melon}},
  \bibinfo {author} {\bibfnamefont {D.}~\bibnamefont {Mengoni}}, \bibinfo
  {author} {\bibfnamefont {T.}~\bibnamefont {Miyazaki}}, \bibinfo {author}
  {\bibfnamefont {V.}~\bibnamefont {Modamio-Hoyborg}}, \bibinfo {author}
  {\bibfnamefont {S.}~\bibnamefont {Momiyama}}, \bibinfo {author}
  {\bibfnamefont {D.~R.}\ \bibnamefont {Napoli}}, \bibinfo {author}
  {\bibfnamefont {M.}~\bibnamefont {Niikura}}, \bibinfo {author} {\bibfnamefont
  {R.}~\bibnamefont {Orlandi}}, \bibinfo {author} {\bibfnamefont
  {H.}~\bibnamefont {Sakurai}}, \bibinfo {author} {\bibfnamefont
  {E.}~\bibnamefont {Sahin}}, \bibinfo {author} {\bibfnamefont
  {D.}~\bibnamefont {Sohler}}, \bibinfo {author} {\bibfnamefont
  {H.}~\bibnamefont {Shaffner}}, \bibinfo {author} {\bibfnamefont
  {R.}~\bibnamefont {Taniuchi}}, \bibinfo {author} {\bibfnamefont
  {J.}~\bibnamefont {Taprogge}}, \bibinfo {author} {\bibfnamefont
  {Z.}~\bibnamefont {Vajta}}, \bibinfo {author} {\bibfnamefont {J.~J.}\
  \bibnamefont {Valiente-Dob\'on}}, \bibinfo {author} {\bibfnamefont
  {O.}~\bibnamefont {Wieland}}, \ and\ \bibinfo {author} {\bibfnamefont
  {M.}~\bibnamefont {Yalcinkaya}},\ }\href {\doibase
  10.1103/PhysRevC.93.034328} {\bibfield  {journal} {\bibinfo  {journal} {Phys.
  Rev. C}\ }\textbf {\bibinfo {volume} {93}},\ \bibinfo {pages} {034328}
  (\bibinfo {year} {2016})}\BibitemShut {NoStop}%
\bibitem [{\citenamefont {Morales}\ \emph {et~al.}(2017)\citenamefont
  {Morales}, \citenamefont {Benzoni}, \citenamefont {Watanabe}, \citenamefont
  {Tsunoda}, \citenamefont {Otsuka}, \citenamefont {Nishimura}, \citenamefont
  {Browne}, \citenamefont {Daido}, \citenamefont {Doornenbal}, \citenamefont
  {Fang}, \citenamefont {Lorusso}, \citenamefont {Patel}, \citenamefont {Rice},
  \citenamefont {Sinclair}, \citenamefont {SÃ¶derstrÃ¶m}, \citenamefont
  {Sumikama}, \citenamefont {Wu}, \citenamefont {Xu}, \citenamefont {Yagi},
  \citenamefont {Yokoyama}, \citenamefont {Baba}, \citenamefont {Avigo},
  \citenamefont {Garrote}, \citenamefont {Blasi}, \citenamefont {Bracco},
  \citenamefont {Camera}, \citenamefont {Ceruti}, \citenamefont {Crespi},
  \citenamefont {de~Angelis}, \citenamefont {Delattre}, \citenamefont
  {Dombradi}, \citenamefont {Gottardo}, \citenamefont {Isobe}, \citenamefont
  {Kojouharov}, \citenamefont {Kurz}, \citenamefont {Kuti}, \citenamefont
  {Matsui}, \citenamefont {Melon}, \citenamefont {Mengoni}, \citenamefont
  {Miyazaki}, \citenamefont {Modamio-Hoybjor}, \citenamefont {Momiyama},
  \citenamefont {Napoli}, \citenamefont {Niikura}, \citenamefont {Orlandi},
  \citenamefont {Sakurai}, \citenamefont {Sahin}, \citenamefont {Sohler},
  \citenamefont {Schaffner}, \citenamefont {Taniuchi}, \citenamefont
  {Taprogge}, \citenamefont {Vajta}, \citenamefont {Valiente-DobÃ³n},
  \citenamefont {Wieland},\ and\ \citenamefont {Yalcinkaya}}]{Mor17}%
  \BibitemOpen
  \bibfield  {author} {\bibinfo {author} {\bibfnamefont {A.}~\bibnamefont
  {Morales}}, \bibinfo {author} {\bibfnamefont {G.}~\bibnamefont {Benzoni}},
  \bibinfo {author} {\bibfnamefont {H.}~\bibnamefont {Watanabe}}, \bibinfo
  {author} {\bibfnamefont {Y.}~\bibnamefont {Tsunoda}}, \bibinfo {author}
  {\bibfnamefont {T.}~\bibnamefont {Otsuka}}, \bibinfo {author} {\bibfnamefont
  {S.}~\bibnamefont {Nishimura}}, \bibinfo {author} {\bibfnamefont
  {F.}~\bibnamefont {Browne}}, \bibinfo {author} {\bibfnamefont
  {R.}~\bibnamefont {Daido}}, \bibinfo {author} {\bibfnamefont
  {P.}~\bibnamefont {Doornenbal}}, \bibinfo {author} {\bibfnamefont
  {Y.}~\bibnamefont {Fang}}, \bibinfo {author} {\bibfnamefont {G.}~\bibnamefont
  {Lorusso}}, \bibinfo {author} {\bibfnamefont {Z.}~\bibnamefont {Patel}},
  \bibinfo {author} {\bibfnamefont {S.}~\bibnamefont {Rice}}, \bibinfo {author}
  {\bibfnamefont {L.}~\bibnamefont {Sinclair}}, \bibinfo {author}
  {\bibfnamefont {P.-A.}\ \bibnamefont {SÃ¶derstrÃ¶m}}, \bibinfo {author}
  {\bibfnamefont {T.}~\bibnamefont {Sumikama}}, \bibinfo {author}
  {\bibfnamefont {J.}~\bibnamefont {Wu}}, \bibinfo {author} {\bibfnamefont
  {Z.}~\bibnamefont {Xu}}, \bibinfo {author} {\bibfnamefont {A.}~\bibnamefont
  {Yagi}}, \bibinfo {author} {\bibfnamefont {R.}~\bibnamefont {Yokoyama}},
  \bibinfo {author} {\bibfnamefont {H.}~\bibnamefont {Baba}}, \bibinfo {author}
  {\bibfnamefont {R.}~\bibnamefont {Avigo}}, \bibinfo {author} {\bibfnamefont
  {F.~B.}\ \bibnamefont {Garrote}}, \bibinfo {author} {\bibfnamefont
  {N.}~\bibnamefont {Blasi}}, \bibinfo {author} {\bibfnamefont
  {A.}~\bibnamefont {Bracco}}, \bibinfo {author} {\bibfnamefont
  {F.}~\bibnamefont {Camera}}, \bibinfo {author} {\bibfnamefont
  {S.}~\bibnamefont {Ceruti}}, \bibinfo {author} {\bibfnamefont
  {F.}~\bibnamefont {Crespi}}, \bibinfo {author} {\bibfnamefont
  {G.}~\bibnamefont {de~Angelis}}, \bibinfo {author} {\bibfnamefont {M.-C.}\
  \bibnamefont {Delattre}}, \bibinfo {author} {\bibfnamefont {Z.}~\bibnamefont
  {Dombradi}}, \bibinfo {author} {\bibfnamefont {A.}~\bibnamefont {Gottardo}},
  \bibinfo {author} {\bibfnamefont {T.}~\bibnamefont {Isobe}}, \bibinfo
  {author} {\bibfnamefont {I.}~\bibnamefont {Kojouharov}}, \bibinfo {author}
  {\bibfnamefont {N.}~\bibnamefont {Kurz}}, \bibinfo {author} {\bibfnamefont
  {I.}~\bibnamefont {Kuti}}, \bibinfo {author} {\bibfnamefont {K.}~\bibnamefont
  {Matsui}}, \bibinfo {author} {\bibfnamefont {B.}~\bibnamefont {Melon}},
  \bibinfo {author} {\bibfnamefont {D.}~\bibnamefont {Mengoni}}, \bibinfo
  {author} {\bibfnamefont {T.}~\bibnamefont {Miyazaki}}, \bibinfo {author}
  {\bibfnamefont {V.}~\bibnamefont {Modamio-Hoybjor}}, \bibinfo {author}
  {\bibfnamefont {S.}~\bibnamefont {Momiyama}}, \bibinfo {author}
  {\bibfnamefont {D.}~\bibnamefont {Napoli}}, \bibinfo {author} {\bibfnamefont
  {M.}~\bibnamefont {Niikura}}, \bibinfo {author} {\bibfnamefont
  {R.}~\bibnamefont {Orlandi}}, \bibinfo {author} {\bibfnamefont
  {H.}~\bibnamefont {Sakurai}}, \bibinfo {author} {\bibfnamefont
  {E.}~\bibnamefont {Sahin}}, \bibinfo {author} {\bibfnamefont
  {D.}~\bibnamefont {Sohler}}, \bibinfo {author} {\bibfnamefont
  {H.}~\bibnamefont {Schaffner}}, \bibinfo {author} {\bibfnamefont
  {R.}~\bibnamefont {Taniuchi}}, \bibinfo {author} {\bibfnamefont
  {J.}~\bibnamefont {Taprogge}}, \bibinfo {author} {\bibfnamefont
  {Z.}~\bibnamefont {Vajta}}, \bibinfo {author} {\bibfnamefont
  {J.}~\bibnamefont {Valiente-DobÃ³n}}, \bibinfo {author} {\bibfnamefont
  {O.}~\bibnamefont {Wieland}}, \ and\ \bibinfo {author} {\bibfnamefont
  {M.}~\bibnamefont {Yalcinkaya}},\ }\href {\doibase
  https://doi.org/10.1016/j.physletb.2016.12.025} {\bibfield  {journal}
  {\bibinfo  {journal} {Phys. Lett. B}\ }\textbf {\bibinfo {volume} {765}},\
  \bibinfo {pages} {328 } (\bibinfo {year} {2017})}\BibitemShut {NoStop}%
\bibitem [{\citenamefont {Leoni}\ \emph {et~al.}(2017)\citenamefont {Leoni},
  \citenamefont {Fornal}, \citenamefont {M\ifmmode~\u{a}\else
  \u{a}\fi{}rginean}, \citenamefont {Sferrazza}, \citenamefont {Tsunoda},
  \citenamefont {Otsuka}, \citenamefont {Bocchi}, \citenamefont {Crespi},
  \citenamefont {Bracco}, \citenamefont {Aydin}, \citenamefont {Boromiza},
  \citenamefont {Bucurescu}, \citenamefont
  {Cieplicka-Ory\ifmmode~\grave{n}\else \`{n}\fi{}czak}, \citenamefont
  {Costache}, \citenamefont {C\ifmmode~\u{a}\else \u{a}\fi{}linescu},
  \citenamefont {Florea}, \citenamefont {Ghi\ifmmode \mbox{\c{t}}\else
  \c{t}\fi{}\ifmmode~\u{a}\else \u{a}\fi{}}, \citenamefont {Glodariu},
  \citenamefont {Ionescu}, \citenamefont {Iskra}, \citenamefont {Krzysiek},
  \citenamefont {M\ifmmode~\u{a}\else \u{a}\fi{}rginean}, \citenamefont
  {Mihai}, \citenamefont {Mihai}, \citenamefont {Mitu}, \citenamefont
  {Negre\ifmmode~\mbox{\c{t}}\else \c{t}\fi{}}, \citenamefont {Ni\ifmmode
  \mbox{\c{t}}\else \c{t}\fi{}\ifmmode~\u{a}\else \u{a}\fi{}}, \citenamefont
  {Ol\ifmmode~\u{a}\else \u{a}\fi{}cel}, \citenamefont {Oprea}, \citenamefont
  {Pascu}, \citenamefont {Petkov}, \citenamefont {Petrone}, \citenamefont
  {Porzio}, \citenamefont {\ifmmode~\mbox{\c{S}}\else \c{S}\fi{}erban},
  \citenamefont {Sotty}, \citenamefont {Stan}, \citenamefont
  {\ifmmode~\mbox{\c{S}}\else \c{S}\fi{}tiru}, \citenamefont {Stroe},
  \citenamefont {\ifmmode \mbox{\c{S}}\else \c{S}\fi{}uv\ifmmode \u{a}\else
  \u{a}\fi{}il\ifmmode~\u{a}\else \u{a}\fi{}}, \citenamefont {Toma},
  \citenamefont {Turturic\ifmmode~\u{a}\else \u{a}\fi{}}, \citenamefont
  {Ujeniuc},\ and\ \citenamefont {Ur}}]{Leo17}%
  \BibitemOpen
  \bibfield  {author} {\bibinfo {author} {\bibfnamefont {S.}~\bibnamefont
  {Leoni}}, \bibinfo {author} {\bibfnamefont {B.}~\bibnamefont {Fornal}},
  \bibinfo {author} {\bibfnamefont {N.}~\bibnamefont {M\ifmmode~\u{a}\else
  \u{a}\fi{}rginean}}, \bibinfo {author} {\bibfnamefont {M.}~\bibnamefont
  {Sferrazza}}, \bibinfo {author} {\bibfnamefont {Y.}~\bibnamefont {Tsunoda}},
  \bibinfo {author} {\bibfnamefont {T.}~\bibnamefont {Otsuka}}, \bibinfo
  {author} {\bibfnamefont {G.}~\bibnamefont {Bocchi}}, \bibinfo {author}
  {\bibfnamefont {F.~C.~L.}\ \bibnamefont {Crespi}}, \bibinfo {author}
  {\bibfnamefont {A.}~\bibnamefont {Bracco}}, \bibinfo {author} {\bibfnamefont
  {S.}~\bibnamefont {Aydin}}, \bibinfo {author} {\bibfnamefont
  {M.}~\bibnamefont {Boromiza}}, \bibinfo {author} {\bibfnamefont
  {D.}~\bibnamefont {Bucurescu}}, \bibinfo {author} {\bibfnamefont
  {N.}~\bibnamefont {Cieplicka-Ory\ifmmode~\grave{n}\else \`{n}\fi{}czak}},
  \bibinfo {author} {\bibfnamefont {C.}~\bibnamefont {Costache}}, \bibinfo
  {author} {\bibfnamefont {S.}~\bibnamefont {C\ifmmode~\u{a}\else
  \u{a}\fi{}linescu}}, \bibinfo {author} {\bibfnamefont {N.}~\bibnamefont
  {Florea}}, \bibinfo {author} {\bibfnamefont {D.~G.}\ \bibnamefont
  {Ghi\ifmmode \mbox{\c{t}}\else \c{t}\fi{}\ifmmode~\u{a}\else \u{a}\fi{}}},
  \bibinfo {author} {\bibfnamefont {T.}~\bibnamefont {Glodariu}}, \bibinfo
  {author} {\bibfnamefont {A.}~\bibnamefont {Ionescu}}, \bibinfo {author}
  {\bibfnamefont {L.}~\bibnamefont {Iskra}}, \bibinfo {author} {\bibfnamefont
  {M.}~\bibnamefont {Krzysiek}}, \bibinfo {author} {\bibfnamefont
  {R.}~\bibnamefont {M\ifmmode~\u{a}\else \u{a}\fi{}rginean}}, \bibinfo
  {author} {\bibfnamefont {C.}~\bibnamefont {Mihai}}, \bibinfo {author}
  {\bibfnamefont {R.~E.}\ \bibnamefont {Mihai}}, \bibinfo {author}
  {\bibfnamefont {A.}~\bibnamefont {Mitu}}, \bibinfo {author} {\bibfnamefont
  {A.}~\bibnamefont {Negre\ifmmode~\mbox{\c{t}}\else \c{t}\fi{}}}, \bibinfo
  {author} {\bibfnamefont {C.~R.}\ \bibnamefont {Ni\ifmmode \mbox{\c{t}}\else
  \c{t}\fi{}\ifmmode~\u{a}\else \u{a}\fi{}}}, \bibinfo {author} {\bibfnamefont
  {A.}~\bibnamefont {Ol\ifmmode~\u{a}\else \u{a}\fi{}cel}}, \bibinfo {author}
  {\bibfnamefont {A.}~\bibnamefont {Oprea}}, \bibinfo {author} {\bibfnamefont
  {S.}~\bibnamefont {Pascu}}, \bibinfo {author} {\bibfnamefont
  {P.}~\bibnamefont {Petkov}}, \bibinfo {author} {\bibfnamefont
  {C.}~\bibnamefont {Petrone}}, \bibinfo {author} {\bibfnamefont
  {G.}~\bibnamefont {Porzio}}, \bibinfo {author} {\bibfnamefont
  {A.}~\bibnamefont {\ifmmode~\mbox{\c{S}}\else \c{S}\fi{}erban}}, \bibinfo
  {author} {\bibfnamefont {C.}~\bibnamefont {Sotty}}, \bibinfo {author}
  {\bibfnamefont {L.}~\bibnamefont {Stan}}, \bibinfo {author} {\bibfnamefont
  {I.}~\bibnamefont {\ifmmode~\mbox{\c{S}}\else \c{S}\fi{}tiru}}, \bibinfo
  {author} {\bibfnamefont {L.}~\bibnamefont {Stroe}}, \bibinfo {author}
  {\bibfnamefont {R.}~\bibnamefont {\ifmmode \mbox{\c{S}}\else
  \c{S}\fi{}uv\ifmmode \u{a}\else \u{a}\fi{}il\ifmmode~\u{a}\else \u{a}\fi{}}},
  \bibinfo {author} {\bibfnamefont {S.}~\bibnamefont {Toma}}, \bibinfo {author}
  {\bibfnamefont {A.}~\bibnamefont {Turturic\ifmmode~\u{a}\else \u{a}\fi{}}},
  \bibinfo {author} {\bibfnamefont {S.}~\bibnamefont {Ujeniuc}}, \ and\
  \bibinfo {author} {\bibfnamefont {C.~A.}\ \bibnamefont {Ur}},\ }\href
  {\doibase 10.1103/PhysRevLett.118.162502} {\bibfield  {journal} {\bibinfo
  {journal} {Phys. Rev. Lett.}\ }\textbf {\bibinfo {volume} {118}},\ \bibinfo
  {pages} {162502} (\bibinfo {year} {2017})}\BibitemShut {NoStop}%
\bibitem [{\citenamefont {M\ifmmode~\u{a}\else \u{a}\fi{}rginean}\ \emph
  {et~al.}(2020)\citenamefont {M\ifmmode~\u{a}\else \u{a}\fi{}rginean},
  \citenamefont {Little}, \citenamefont {Tsunoda}, \citenamefont {Leoni},
  \citenamefont {Janssens}, \citenamefont {Fornal}, \citenamefont {Otsuka},
  \citenamefont {Michelagnoli}, \citenamefont {Stan}, \citenamefont {Crespi},
  \citenamefont {Costache}, \citenamefont {Lica}, \citenamefont {Sferrazza},
  \citenamefont {Turturica}, \citenamefont {Ayangeakaa}, \citenamefont
  {Auranen}, \citenamefont {Barani}, \citenamefont {Bender}, \citenamefont
  {Bottoni}, \citenamefont {Boromiza}, \citenamefont {Bracco}, \citenamefont
  {C\ifmmode~\u{a}\else \u{a}\fi{}linescu}, \citenamefont {Campbell},
  \citenamefont {Carpenter}, \citenamefont {Chowdhury}, \citenamefont
  {Ciema\l{}a}, \citenamefont {Cieplicka-Ory\ifmmode~\grave{n}\else
  \`{n}\fi{}czak}, \citenamefont {Cline}, \citenamefont {Clisu}, \citenamefont
  {Crawford}, \citenamefont {Dinescu}, \citenamefont {Dudouet}, \citenamefont
  {Filipescu}, \citenamefont {Florea}, \citenamefont {Forney}, \citenamefont
  {Fracassetti}, \citenamefont {Gade}, \citenamefont {Gheorghe}, \citenamefont
  {Hayes}, \citenamefont {Harca}, \citenamefont {Henderson}, \citenamefont
  {Ionescu}, \citenamefont {Iskra}, \citenamefont {Jentschel}, \citenamefont
  {Kandzia}, \citenamefont {Kim}, \citenamefont {Kondev}, \citenamefont
  {Korschinek}, \citenamefont {K\"oster}, \citenamefont {Krishichayan},
  \citenamefont {Krzysiek}, \citenamefont {Lauritsen}, \citenamefont {Li},
  \citenamefont {M\ifmmode~\u{a}\else \u{a}\fi{}rginean}, \citenamefont
  {Maugeri}, \citenamefont {Mihai}, \citenamefont {Mihai}, \citenamefont
  {Mitu}, \citenamefont {Mutti}, \citenamefont {Negret}, \citenamefont
  {Ni\ifmmode \mbox{\c{t}}\else \c{t}\fi{}\ifmmode~\u{a}\else \u{a}\fi{}},
  \citenamefont {Ol\ifmmode~\u{a}\else \u{a}\fi{}cel}, \citenamefont {Oprea},
  \citenamefont {Pascu}, \citenamefont {Petrone}, \citenamefont {Porzio},
  \citenamefont {Rhodes}, \citenamefont {Seweryniak}, \citenamefont {Schumann},
  \citenamefont {Sotty}, \citenamefont {Stolze}, \citenamefont {\ifmmode
  \mbox{\c{S}}\else \c{S}\fi{}uv\ifmmode \u{a}\else
  \u{a}\fi{}il\ifmmode~\u{a}\else \u{a}\fi{}}, \citenamefont {Toma},
  \citenamefont {Ujeniuc}, \citenamefont {Walters}, \citenamefont {Wu},
  \citenamefont {Wu}, \citenamefont {Zhu},\ and\ \citenamefont
  {Ziliani}}]{Mar20}%
  \BibitemOpen
  \bibfield  {author} {\bibinfo {author} {\bibfnamefont {N.}~\bibnamefont
  {M\ifmmode~\u{a}\else \u{a}\fi{}rginean}}, \bibinfo {author} {\bibfnamefont
  {D.}~\bibnamefont {Little}}, \bibinfo {author} {\bibfnamefont
  {Y.}~\bibnamefont {Tsunoda}}, \bibinfo {author} {\bibfnamefont
  {S.}~\bibnamefont {Leoni}}, \bibinfo {author} {\bibfnamefont {R.~V.~F.}\
  \bibnamefont {Janssens}}, \bibinfo {author} {\bibfnamefont {B.}~\bibnamefont
  {Fornal}}, \bibinfo {author} {\bibfnamefont {T.}~\bibnamefont {Otsuka}},
  \bibinfo {author} {\bibfnamefont {C.}~\bibnamefont {Michelagnoli}}, \bibinfo
  {author} {\bibfnamefont {L.}~\bibnamefont {Stan}}, \bibinfo {author}
  {\bibfnamefont {F.~C.~L.}\ \bibnamefont {Crespi}}, \bibinfo {author}
  {\bibfnamefont {C.}~\bibnamefont {Costache}}, \bibinfo {author}
  {\bibfnamefont {R.}~\bibnamefont {Lica}}, \bibinfo {author} {\bibfnamefont
  {M.}~\bibnamefont {Sferrazza}}, \bibinfo {author} {\bibfnamefont
  {A.}~\bibnamefont {Turturica}}, \bibinfo {author} {\bibfnamefont {A.~D.}\
  \bibnamefont {Ayangeakaa}}, \bibinfo {author} {\bibfnamefont
  {K.}~\bibnamefont {Auranen}}, \bibinfo {author} {\bibfnamefont
  {M.}~\bibnamefont {Barani}}, \bibinfo {author} {\bibfnamefont {P.~C.}\
  \bibnamefont {Bender}}, \bibinfo {author} {\bibfnamefont {S.}~\bibnamefont
  {Bottoni}}, \bibinfo {author} {\bibfnamefont {M.}~\bibnamefont {Boromiza}},
  \bibinfo {author} {\bibfnamefont {A.}~\bibnamefont {Bracco}}, \bibinfo
  {author} {\bibfnamefont {S.}~\bibnamefont {C\ifmmode~\u{a}\else
  \u{a}\fi{}linescu}}, \bibinfo {author} {\bibfnamefont {C.~M.}\ \bibnamefont
  {Campbell}}, \bibinfo {author} {\bibfnamefont {M.~P.}\ \bibnamefont
  {Carpenter}}, \bibinfo {author} {\bibfnamefont {P.}~\bibnamefont
  {Chowdhury}}, \bibinfo {author} {\bibfnamefont {M.}~\bibnamefont
  {Ciema\l{}a}}, \bibinfo {author} {\bibfnamefont {N.}~\bibnamefont
  {Cieplicka-Ory\ifmmode~\grave{n}\else \`{n}\fi{}czak}}, \bibinfo {author}
  {\bibfnamefont {D.}~\bibnamefont {Cline}}, \bibinfo {author} {\bibfnamefont
  {C.}~\bibnamefont {Clisu}}, \bibinfo {author} {\bibfnamefont {H.~L.}\
  \bibnamefont {Crawford}}, \bibinfo {author} {\bibfnamefont {I.~E.}\
  \bibnamefont {Dinescu}}, \bibinfo {author} {\bibfnamefont {J.}~\bibnamefont
  {Dudouet}}, \bibinfo {author} {\bibfnamefont {D.}~\bibnamefont {Filipescu}},
  \bibinfo {author} {\bibfnamefont {N.}~\bibnamefont {Florea}}, \bibinfo
  {author} {\bibfnamefont {A.~M.}\ \bibnamefont {Forney}}, \bibinfo {author}
  {\bibfnamefont {S.}~\bibnamefont {Fracassetti}}, \bibinfo {author}
  {\bibfnamefont {A.}~\bibnamefont {Gade}}, \bibinfo {author} {\bibfnamefont
  {I.}~\bibnamefont {Gheorghe}}, \bibinfo {author} {\bibfnamefont {A.~B.}\
  \bibnamefont {Hayes}}, \bibinfo {author} {\bibfnamefont {I.}~\bibnamefont
  {Harca}}, \bibinfo {author} {\bibfnamefont {J.}~\bibnamefont {Henderson}},
  \bibinfo {author} {\bibfnamefont {A.}~\bibnamefont {Ionescu}}, \bibinfo
  {author} {\bibfnamefont {L.~W.}\ \bibnamefont {Iskra}}, \bibinfo {author}
  {\bibfnamefont {M.}~\bibnamefont {Jentschel}}, \bibinfo {author}
  {\bibfnamefont {F.}~\bibnamefont {Kandzia}}, \bibinfo {author} {\bibfnamefont
  {Y.~H.}\ \bibnamefont {Kim}}, \bibinfo {author} {\bibfnamefont {F.~G.}\
  \bibnamefont {Kondev}}, \bibinfo {author} {\bibfnamefont {G.}~\bibnamefont
  {Korschinek}}, \bibinfo {author} {\bibfnamefont {U.}~\bibnamefont
  {K\"oster}}, \bibinfo {author} {\bibnamefont {Krishichayan}}, \bibinfo
  {author} {\bibfnamefont {M.}~\bibnamefont {Krzysiek}}, \bibinfo {author}
  {\bibfnamefont {T.}~\bibnamefont {Lauritsen}}, \bibinfo {author}
  {\bibfnamefont {J.}~\bibnamefont {Li}}, \bibinfo {author} {\bibfnamefont
  {R.}~\bibnamefont {M\ifmmode~\u{a}\else \u{a}\fi{}rginean}}, \bibinfo
  {author} {\bibfnamefont {E.~A.}\ \bibnamefont {Maugeri}}, \bibinfo {author}
  {\bibfnamefont {C.}~\bibnamefont {Mihai}}, \bibinfo {author} {\bibfnamefont
  {R.~E.}\ \bibnamefont {Mihai}}, \bibinfo {author} {\bibfnamefont
  {A.}~\bibnamefont {Mitu}}, \bibinfo {author} {\bibfnamefont {P.}~\bibnamefont
  {Mutti}}, \bibinfo {author} {\bibfnamefont {A.}~\bibnamefont {Negret}},
  \bibinfo {author} {\bibfnamefont {C.~R.}\ \bibnamefont {Ni\ifmmode
  \mbox{\c{t}}\else \c{t}\fi{}\ifmmode~\u{a}\else \u{a}\fi{}}}, \bibinfo
  {author} {\bibfnamefont {A.}~\bibnamefont {Ol\ifmmode~\u{a}\else
  \u{a}\fi{}cel}}, \bibinfo {author} {\bibfnamefont {A.}~\bibnamefont {Oprea}},
  \bibinfo {author} {\bibfnamefont {S.}~\bibnamefont {Pascu}}, \bibinfo
  {author} {\bibfnamefont {C.}~\bibnamefont {Petrone}}, \bibinfo {author}
  {\bibfnamefont {C.}~\bibnamefont {Porzio}}, \bibinfo {author} {\bibfnamefont
  {D.}~\bibnamefont {Rhodes}}, \bibinfo {author} {\bibfnamefont
  {D.}~\bibnamefont {Seweryniak}}, \bibinfo {author} {\bibfnamefont
  {D.}~\bibnamefont {Schumann}}, \bibinfo {author} {\bibfnamefont
  {C.}~\bibnamefont {Sotty}}, \bibinfo {author} {\bibfnamefont {S.~M.}\
  \bibnamefont {Stolze}}, \bibinfo {author} {\bibfnamefont {R.}~\bibnamefont
  {\ifmmode \mbox{\c{S}}\else \c{S}\fi{}uv\ifmmode \u{a}\else
  \u{a}\fi{}il\ifmmode~\u{a}\else \u{a}\fi{}}}, \bibinfo {author}
  {\bibfnamefont {S.}~\bibnamefont {Toma}}, \bibinfo {author} {\bibfnamefont
  {S.}~\bibnamefont {Ujeniuc}}, \bibinfo {author} {\bibfnamefont {W.~B.}\
  \bibnamefont {Walters}}, \bibinfo {author} {\bibfnamefont {C.~Y.}\
  \bibnamefont {Wu}}, \bibinfo {author} {\bibfnamefont {J.}~\bibnamefont {Wu}},
  \bibinfo {author} {\bibfnamefont {S.}~\bibnamefont {Zhu}}, \ and\ \bibinfo
  {author} {\bibfnamefont {S.}~\bibnamefont {Ziliani}},\ }\href {\doibase
  10.1103/PhysRevLett.125.102502} {\bibfield  {journal} {\bibinfo  {journal}
  {Phys. Rev. Lett.}\ }\textbf {\bibinfo {volume} {125}},\ \bibinfo {pages}
  {102502} (\bibinfo {year} {2020})}\BibitemShut {NoStop}%
\bibitem [{\citenamefont {Porzio}\ \emph {et~al.}()\citenamefont {Porzio},
  \citenamefont {Michelagnoli}, \citenamefont {Cieplicka-Ory\'nczak},
  \citenamefont {Sferrazza}, \citenamefont {Leoni}, \citenamefont {Fornal},
  \citenamefont {Tsunoda}, \citenamefont {Otsuka}, \citenamefont {Bottoni},
  \citenamefont {Costache}, \citenamefont {Crespi}, \citenamefont {Iskra},
  \citenamefont {Jentschel}, \citenamefont {Kandzia}, \citenamefont {Kim},
  \citenamefont {K\"oester}, \citenamefont {M\u{a}rginean}, \citenamefont
  {Mihai}, \citenamefont {Mutti},\ and\ \citenamefont {Turturica}}]{Por20}%
  \BibitemOpen
  \bibfield  {author} {\bibinfo {author} {\bibfnamefont {C.}~\bibnamefont
  {Porzio}}, \bibinfo {author} {\bibfnamefont {C.}~\bibnamefont
  {Michelagnoli}}, \bibinfo {author} {\bibfnamefont {N.}~\bibnamefont
  {Cieplicka-Ory\'nczak}}, \bibinfo {author} {\bibfnamefont {M.}~\bibnamefont
  {Sferrazza}}, \bibinfo {author} {\bibfnamefont {S.}~\bibnamefont {Leoni}},
  \bibinfo {author} {\bibfnamefont {B.}~\bibnamefont {Fornal}}, \bibinfo
  {author} {\bibfnamefont {Y.}~\bibnamefont {Tsunoda}}, \bibinfo {author}
  {\bibfnamefont {T.}~\bibnamefont {Otsuka}}, \bibinfo {author} {\bibfnamefont
  {S.}~\bibnamefont {Bottoni}}, \bibinfo {author} {\bibfnamefont
  {C.}~\bibnamefont {Costache}}, \bibinfo {author} {\bibfnamefont {F.~C.~L.}\
  \bibnamefont {Crespi}}, \bibinfo {author} {\bibfnamefont {L.~W.}\
  \bibnamefont {Iskra}}, \bibinfo {author} {\bibfnamefont {M.}~\bibnamefont
  {Jentschel}}, \bibinfo {author} {\bibfnamefont {F.}~\bibnamefont {Kandzia}},
  \bibinfo {author} {\bibfnamefont {Y.-H.}\ \bibnamefont {Kim}}, \bibinfo
  {author} {\bibfnamefont {U.}~\bibnamefont {K\"oester}}, \bibinfo {author}
  {\bibfnamefont {N.}~\bibnamefont {M\u{a}rginean}}, \bibinfo {author}
  {\bibfnamefont {C.}~\bibnamefont {Mihai}}, \bibinfo {author} {\bibfnamefont
  {P.}~\bibnamefont {Mutti}}, \ and\ \bibinfo {author} {\bibfnamefont
  {A.}~\bibnamefont {Turturica}},\ }\href@noop {} {\bibinfo  {journal}
  {accepted in Phys. Rev. C}\ }\BibitemShut {NoStop}%
\bibitem [{\citenamefont {Uozumi}\ \emph {et~al.}(1994)\citenamefont {Uozumi},
  \citenamefont {Iwamoto}, \citenamefont {Widodo}, \citenamefont {Nohtomi},
  \citenamefont {Sakae}, \citenamefont {Matoba}, \citenamefont {Nakano},
  \citenamefont {Maki},\ and\ \citenamefont {Koori}}]{Uoz94}%
  \BibitemOpen
\bibfield  {journal} {  }\bibfield  {author} {\bibinfo {author} {\bibfnamefont
  {Y.}~\bibnamefont {Uozumi}}, \bibinfo {author} {\bibfnamefont
  {O.}~\bibnamefont {Iwamoto}}, \bibinfo {author} {\bibfnamefont
  {S.}~\bibnamefont {Widodo}}, \bibinfo {author} {\bibfnamefont
  {A.}~\bibnamefont {Nohtomi}}, \bibinfo {author} {\bibfnamefont
  {T.}~\bibnamefont {Sakae}}, \bibinfo {author} {\bibfnamefont
  {M.}~\bibnamefont {Matoba}}, \bibinfo {author} {\bibfnamefont
  {M.}~\bibnamefont {Nakano}}, \bibinfo {author} {\bibfnamefont
  {T.}~\bibnamefont {Maki}}, \ and\ \bibinfo {author} {\bibfnamefont
  {N.}~\bibnamefont {Koori}},\ }\href {\doibase
  https://doi.org/10.1016/0375-9474(94)90740-4} {\bibfield  {journal} {\bibinfo
   {journal} {Nucl. Phys. A}\ }\textbf {\bibinfo {volume} {576}},\ \bibinfo
  {pages} {123 } (\bibinfo {year} {1994})}\BibitemShut {NoStop}%
\bibitem [{\citenamefont {Bocchi}\ \emph {et~al.}(2016)\citenamefont {Bocchi},
  \citenamefont {Leoni}, \citenamefont {Fornal}, \citenamefont {Col\`{o}},
  \citenamefont {Bortignon}, \citenamefont {Bottoni}, \citenamefont {Bracco},
  \citenamefont {Michelagnoli}, \citenamefont {Bazzacco}, \citenamefont
  {Blanc}, \citenamefont {de~France}, \citenamefont {Jentschel}, \citenamefont
  {K\"{o}ster}, \citenamefont {Mutti}, \citenamefont {Regis}, \citenamefont
  {Simpson}, \citenamefont {Soldner}, \citenamefont {Ur}, \citenamefont
  {Urban}, \citenamefont {Fraile}, \citenamefont {Lozeva}, \citenamefont
  {Belvito}, \citenamefont {Benzoni}, \citenamefont {Bruce}, \citenamefont
  {Carroll}, \citenamefont {Cieplicka-Orynczak}, \citenamefont {Crespi},
  \citenamefont {Didierjean}, \citenamefont {Jolie}, \citenamefont {Korten},
  \citenamefont {Kr\"{o}ll}, \citenamefont {Lalkovski}, \citenamefont {Mach},
  \citenamefont {Marginean}, \citenamefont {Melon}, \citenamefont {Mengoni},
  \citenamefont {Million}, \citenamefont {Nannini}, \citenamefont {Napoli},
  \citenamefont {Olaizola}, \citenamefont {Paziy}, \citenamefont {Podolyak},
  \citenamefont {Regan}, \citenamefont {Saed-Samii}, \citenamefont {Szpak},\
  and\ \citenamefont {Vedia}}]{Boc16}%
  \BibitemOpen
  \bibfield  {author} {\bibinfo {author} {\bibfnamefont {G.}~\bibnamefont
  {Bocchi}}, \bibinfo {author} {\bibfnamefont {S.}~\bibnamefont {Leoni}},
  \bibinfo {author} {\bibfnamefont {B.}~\bibnamefont {Fornal}}, \bibinfo
  {author} {\bibfnamefont {G.}~\bibnamefont {Col\`{o}}}, \bibinfo {author}
  {\bibfnamefont {P.}~\bibnamefont {Bortignon}}, \bibinfo {author}
  {\bibfnamefont {S.}~\bibnamefont {Bottoni}}, \bibinfo {author} {\bibfnamefont
  {A.}~\bibnamefont {Bracco}}, \bibinfo {author} {\bibfnamefont
  {C.}~\bibnamefont {Michelagnoli}}, \bibinfo {author} {\bibfnamefont
  {D.}~\bibnamefont {Bazzacco}}, \bibinfo {author} {\bibfnamefont
  {A.}~\bibnamefont {Blanc}}, \bibinfo {author} {\bibfnamefont
  {G.}~\bibnamefont {de~France}}, \bibinfo {author} {\bibfnamefont
  {M.}~\bibnamefont {Jentschel}}, \bibinfo {author} {\bibfnamefont
  {U.}~\bibnamefont {K\"{o}ster}}, \bibinfo {author} {\bibfnamefont
  {P.}~\bibnamefont {Mutti}}, \bibinfo {author} {\bibfnamefont
  {J.}~\bibnamefont {Regis}}, \bibinfo {author} {\bibfnamefont
  {G.}~\bibnamefont {Simpson}}, \bibinfo {author} {\bibfnamefont
  {T.}~\bibnamefont {Soldner}}, \bibinfo {author} {\bibfnamefont
  {C.}~\bibnamefont {Ur}}, \bibinfo {author} {\bibfnamefont {W.}~\bibnamefont
  {Urban}}, \bibinfo {author} {\bibfnamefont {L.}~\bibnamefont {Fraile}},
  \bibinfo {author} {\bibfnamefont {R.}~\bibnamefont {Lozeva}}, \bibinfo
  {author} {\bibfnamefont {B.}~\bibnamefont {Belvito}}, \bibinfo {author}
  {\bibfnamefont {G.}~\bibnamefont {Benzoni}}, \bibinfo {author} {\bibfnamefont
  {A.}~\bibnamefont {Bruce}}, \bibinfo {author} {\bibfnamefont
  {R.}~\bibnamefont {Carroll}}, \bibinfo {author} {\bibfnamefont
  {N.}~\bibnamefont {Cieplicka-Orynczak}}, \bibinfo {author} {\bibfnamefont
  {F.}~\bibnamefont {Crespi}}, \bibinfo {author} {\bibfnamefont
  {F.}~\bibnamefont {Didierjean}}, \bibinfo {author} {\bibfnamefont
  {J.}~\bibnamefont {Jolie}}, \bibinfo {author} {\bibfnamefont
  {W.}~\bibnamefont {Korten}}, \bibinfo {author} {\bibfnamefont
  {T.}~\bibnamefont {Kr\"{o}ll}}, \bibinfo {author} {\bibfnamefont
  {S.}~\bibnamefont {Lalkovski}}, \bibinfo {author} {\bibfnamefont
  {H.}~\bibnamefont {Mach}}, \bibinfo {author} {\bibfnamefont {N.}~\bibnamefont
  {Marginean}}, \bibinfo {author} {\bibfnamefont {B.}~\bibnamefont {Melon}},
  \bibinfo {author} {\bibfnamefont {D.}~\bibnamefont {Mengoni}}, \bibinfo
  {author} {\bibfnamefont {B.}~\bibnamefont {Million}}, \bibinfo {author}
  {\bibfnamefont {A.}~\bibnamefont {Nannini}}, \bibinfo {author} {\bibfnamefont
  {D.}~\bibnamefont {Napoli}}, \bibinfo {author} {\bibfnamefont
  {B.}~\bibnamefont {Olaizola}}, \bibinfo {author} {\bibfnamefont
  {V.}~\bibnamefont {Paziy}}, \bibinfo {author} {\bibfnamefont
  {Z.}~\bibnamefont {Podolyak}}, \bibinfo {author} {\bibfnamefont
  {P.}~\bibnamefont {Regan}}, \bibinfo {author} {\bibfnamefont
  {N.}~\bibnamefont {Saed-Samii}}, \bibinfo {author} {\bibfnamefont
  {B.}~\bibnamefont {Szpak}}, \ and\ \bibinfo {author} {\bibfnamefont
  {V.}~\bibnamefont {Vedia}},\ }\href {\doibase
  https://doi.org/10.1016/j.physletb.2016.06.065} {\bibfield  {journal}
  {\bibinfo  {journal} {Phys. Lett. B}\ }\textbf {\bibinfo {volume} {760}},\
  \bibinfo {pages} {273 } (\bibinfo {year} {2016})}\BibitemShut {NoStop}%
\bibitem [{\citenamefont {Bottoni}\ \emph {et~al.}(2019)\citenamefont
  {Bottoni}, \citenamefont {Iskra}, \citenamefont {Leoni}, \citenamefont
  {Fornal}, \citenamefont {Col\`o}, \citenamefont {Bazzacco}, \citenamefont
  {Gatti}, \citenamefont {Benzoni}, \citenamefont {Blanc}, \citenamefont
  {Bocchi}, \citenamefont {Bracco}, \citenamefont {Cieplicka-Ory\'nczak},
  \citenamefont {Crespi}, \citenamefont {Jentschel}, \citenamefont {K\"oster},
  \citenamefont {Michelagnoli}, \citenamefont {Million}, \citenamefont {Mutti},
  \citenamefont {Soldner}, \citenamefont {Ur},\ and\ \citenamefont
  {Urban}}]{Bot19Zako}%
  \BibitemOpen
  \bibfield  {author} {\bibinfo {author} {\bibfnamefont {S.}~\bibnamefont
  {Bottoni}}, \bibinfo {author} {\bibfnamefont {L.}~\bibnamefont {Iskra}},
  \bibinfo {author} {\bibfnamefont {S.}~\bibnamefont {Leoni}}, \bibinfo
  {author} {\bibfnamefont {B.}~\bibnamefont {Fornal}}, \bibinfo {author}
  {\bibfnamefont {G.}~\bibnamefont {Col\`o}}, \bibinfo {author} {\bibfnamefont
  {D.}~\bibnamefont {Bazzacco}}, \bibinfo {author} {\bibfnamefont
  {L.}~\bibnamefont {Gatti}}, \bibinfo {author} {\bibfnamefont
  {G.}~\bibnamefont {Benzoni}}, \bibinfo {author} {\bibfnamefont
  {A.}~\bibnamefont {Blanc}}, \bibinfo {author} {\bibfnamefont
  {G.}~\bibnamefont {Bocchi}}, \bibinfo {author} {\bibfnamefont
  {A.}~\bibnamefont {Bracco}}, \bibinfo {author} {\bibfnamefont
  {N.}~\bibnamefont {Cieplicka-Ory\'nczak}}, \bibinfo {author} {\bibfnamefont
  {F.}~\bibnamefont {Crespi}}, \bibinfo {author} {\bibfnamefont
  {M.}~\bibnamefont {Jentschel}}, \bibinfo {author} {\bibfnamefont
  {U.}~\bibnamefont {K\"oster}}, \bibinfo {author} {\bibfnamefont
  {C.}~\bibnamefont {Michelagnoli}}, \bibinfo {author} {\bibfnamefont
  {B.}~\bibnamefont {Million}}, \bibinfo {author} {\bibfnamefont
  {P.}~\bibnamefont {Mutti}}, \bibinfo {author} {\bibfnamefont
  {T.}~\bibnamefont {Soldner}}, \bibinfo {author} {\bibfnamefont
  {C.}~\bibnamefont {Ur}}, \ and\ \bibinfo {author} {\bibfnamefont
  {W.}~\bibnamefont {Urban}},\ }\href
  {http://www.actaphys.uj.edu.pl/fulltext?series=Reg&vol=50&page=285}
  {\bibfield  {journal} {\bibinfo  {journal} {Acta. Phys. Pol. B}\ }\textbf
  {\bibinfo {volume} {50}},\ \bibinfo {pages} {285} (\bibinfo {year}
  {2019})}\BibitemShut {NoStop}%
\end{thebibliography}%

\end{document}